\documentclass[aps,prd,10pt,tightenlines,notitlepage,showpacs,nofootinbib,superscriptaddress,twocolumn,eqsecnum,raggedbottom,preprintnumbers,longbibliography]{revtex4-1}

\usepackage{bbm,lmodern,amsmath,graphicx,epsfig,amssymb,amsfonts,dsfont,mathtools}
\usepackage{amsmath,amssymb,mathrsfs}
\usepackage{rotating}
\usepackage{bigstrut}
\usepackage{morefloats}
\usepackage{multirow}
\usepackage{longtable}
\usepackage{dcolumn}
\usepackage{placeins}
\usepackage[normalem]{ulem}
\usepackage[dvipsnames]{xcolor}
\usepackage[utf8]{inputenc}
\usepackage{nicefrac} 
\usepackage{physics}
\usepackage{braket}
\usepackage{siunitx}

\usepackage{hyperref}
\hypersetup{
	colorlinks	=true,
	urlcolor	=blue,
	linkcolor	=blue,
	citecolor	=blue,
	pdftitle	={Coupled-channel contributions to the Gerasimov-Drell-Hearn sum rule from the J\"ulich-Bonn approach},
	pdfauthor	={C.~Schneider, D.~R\"onchen, C.~Hanhart, U.-G.~Mei\ss ner, },
	pdfsubject	={GDH sum rule}
	}

\allowdisplaybreaks[1]

\newcommand{\be}{\begin{equation}}
\newcommand{\ee}{\end{equation}}
\newcommand{\ba}{\begin{eqnarray}}
\newcommand{\ea}{\end{eqnarray}}

\newcommand{\npo}{{\rm NP}}
\begin{document}


\title{Coupled-channel contributions to the GDH sum rule from the J\"ulich-Bonn approach}

\author{C.~Schneider}
\email{c.schneider@fz-juelich.de}
\affiliation{Institute for Advanced Simulation (IAS-4), Forschungszentrum J\"ulich, 
52425 J\"ulich, Germany}

\author{D.~R\"onchen}
\email{d.roenchen@fz-juelich.de}
\affiliation{Institute for Advanced Simulation (IAS-4), Forschungszentrum J\"ulich, 
52425 J\"ulich, Germany}

\author{C.~Hanhart}
\email{c.hanhart@fz-juelich.de}
\affiliation{Institute for Advanced Simulation (IAS-4), Forschungszentrum J\"ulich, 
52425 J\"ulich, Germany}

\author{Ulf-G.~Mei\ss ner}
\email{meissner@hiskp.uni-bonn.de}
\affiliation{Institute for Advanced Simulation (IAS-4), Forschungszentrum J\"ulich, 
52425 J\"ulich, Germany}
\affiliation{Helmholtz-Institut f\"ur Strahlen- und Kernphysik (Theorie) and Bethe Center for Theoretical
Physics,  Universit\"at Bonn, 53115 Bonn, Germany}
\affiliation{Peng Huanwu Collaborative Center for Research and Education, International Institute for Interdisciplinary and Frontiers, Beihang University, Beijing 100191, China}

\begin{abstract}
We study the Gerasimov-Drell-Hearn (GDH) sum rule within a dynamical coupled-channel approach, the Jülich-Bonn model for light baryon resonances based on fits to an extensive data base of pion and photon induced data. Recently published photoproduction data for different observables with $\pi N$ and $\eta N$ final states are analyzed simultaneously with older data for the reactions $\pi N\to \pi N$,  $\eta N$, $K\Lambda$, $K\Sigma$ and  $\gamma p\to\pi N$, $\eta N$, $K\Lambda$, $K\Sigma$. The impact of the new data on the resonance spectrum is investigated and the contribution of the individual channels to the GDH integral is determined. 
\end{abstract}


\maketitle

\section{Introduction}

The Gerasimov-Drell-Hearn (GDH) sum rule~\cite{Gerasimov:1965et,Drell:1966jv}, derived from fundamental principles of Quantum Field Theory, allows to probe the widely discussed spin structure of the nucleon experimentally~\cite{Drechsel:2000ct,Drechsel:2004ki,Schumacher:2005an,Helbing:2006zp,Deur:2018roz}. While the value of the nucleon spin is known, quantifying the individual contributions from quarks and gluons remains a challenge. The GDH sum rule relates the anomalous magnetic moment $\kappa_N$ of the nucleon to the integrated difference of the total helicity-dependent photoproduction cross-sections $\Delta \sigma=\sigma_{3/2}-\sigma_{1/2}$, which involves all possible photon-induced final states.
The experimental confirmation of the GDH sum rule poses a considerable  challenge: the measurement of  $\Delta \sigma$ requires a circularly polarized photon beam and a longitudinally polarized nucleon target and the data have to be taken over a large energy range. Moreover, an inclusive measurement of all possible outgoing channels is not feasible in photoproduction, in contrast to electroproduction processes. In Refs.~\cite{CLAS:2017ozc,CLAS:2021apd}, e.g., the proton spin structure was studied with polarized electron beams by the CLAS Collaboation. Those measurements can be extrapolated to the photon point at zero momentum transfer using generalized GDH integrals obtained from chiral effective field theory~\cite{Bernard:1992nz,Ji:1999pd,Bernard:2002pw,Bernard:2012hb,Alarcon:2020icz}. Dedicated experimental programs to confirm the GDH sum rule directly in photoproduction processes were carried out especially by the GDH Collaboration at MAMI and ELSA~\cite{GDH:2001zzk,GDH:2003xhc,Dutz:2004zz}.
To date, photoproduction data for $\Delta \sigma$, which is directly related to the double-polarization $E$, are available for $\pi N$, $\pi\pi N$ and $\eta N$ final states. Predictions from theory or phenomenological models can fill the gap of missing channels or energy regions not covered by experiment. The single-pion contributions to the  GDH integral were calculated, e.g. within the GWU/SAID~\cite{Arndt:2005wk,Strakovsky:2022tvu} or the MAID~\cite{Drechsel:2007if} frameworks, the $K\Sigma$ contribution in Ref.~\cite{Mart:2019fau} using an isobar model for $K\Sigma$ photoproduction off proton and neutron targets.

Especially valuable with regard to contributions of different hadronic final states are predictions from coupled-channel approaches where multiple initial and final states are analyzed simultaneously. Examples for such approaches are the Bonn-Gatchina~\cite{Anisovich:2011fc,CBELSATAPS:2019ylw,Sarantsev:2025lik}, Kent State~\cite{Hunt:2018wqz},  ANL/Osaka~\cite{Kamano:2013iva,Kamano:2016bgm} and Jülich-Bonn~\cite{Ronchen:2012eg,Ronchen:2022hqk} models that extract the spectrum of light baryon resonances in combined studies of pion- and photon-induced reactions. The latter two models fall into the class of so-called dynamical coupled-channel approaches that employ the hadron exchange picture and involve an integration over off-shell momenta in the scattering equation. See Ref.~\cite{Doring:2025sgb} for a recent review on dynamical coupled-channel approaches.

In this work, we determine the contributions of the channels $\pi^0 p$, $\pi^+ n$, $\eta p$, $K^+\Lambda^0$, $K^0 \Sigma^+$, and $K^+\Sigma^0$ to the GDH sum rule for photoproduction processes off a proton target within the Jülich-Bonn framework, including contributions from channels for which no data on $\Delta\sigma$ are available. The predictions are based on fits to an extensive pion- and photoproduction data base, including recently published data sets. In addition, we provide updated values for the $N^*$ and $\Delta$ resonance parameters.

The paper is organized in the following way: In Sec.~\ref{sec:background} we give a short overview of the theoretical formalism used in the J\"ulich-Bonn model. In Sec.~\ref{sec:new_data} we list the newly included data sets and describe the numerical details of the fit. In Sec.~\ref{Sec:Fit_results} we present the updated fit results and discuss changes in pole position of the extracted resonance spectrum. In Sec.~\ref{sec:GDH_sum_rule} we give a short introduction to the GDH sum rule and discuss the different contributions of the two-body channels included in our approach. Additional information is gathered in the Appendix.

\section{Theoretical Framework}
\label{sec:background}

In this section, we give an overview of the J\"ulich-Bonn dynamical coupled-channel (DCC) approach. For more details on the theoretical framework, we refer to Refs.~\cite{Ronchen:2012eg,Ronchen:2014cna,Ronchen:2015vfa, Wang:2022osj}. 
\subsection{Hadronic processes}
\label{sec:Hadronic_part}
The hadronic meson-baryon scattering is described by the $T$-matrix $T_{\mu\nu}$ and its dynamics is given by a Lippmann-Schwinger-like equation which involves an integration over intermediate off-shell momenta as shown in Eq.~\eqref{eq:LS}, 
\begin{align}
&T_{\mu\nu}(p,p',W)=V_{\mu\nu}(p,p',W) \label{eq:LS}\\
&+\sum_{\kappa}\int\limits_0^\infty dq\,
 q^2\,V_{\mu\kappa}(p,q,W)G^{}_\kappa(q,W)\,T_{\kappa\nu}(q,p',W) ~ .\nonumber
\end{align}
Here, $W$ is the center-of-mass energy, $q$ ($p',p$) are the intermediate (incoming, outgoing) momenta and $V_{\mu\nu}$ denotes the interaction potential for the incoming (outgoing) meson-baryon channel $\nu$ ($\mu$). 

The current model includes the channels $\pi N, \eta N, K\Lambda, K\Sigma,\sigma N, \rho N, \pi \Delta$. The latter three channels are used to effectively parameterize the three-body channel $\pi\pi N$ consistent with the corresponding $\pi\pi$ and $\pi N$ phase shifts~\cite{Schutz:1998jx,Krehl:1999km}. The sum in Eq.~\eqref{eq:LS} runs over all intermediate channels $\kappa$ with
 $G_{\kappa}(q,W)$ being the meson-baryon propagator.
The channel space was recently extended to include the process $\pi N\to\omega N$ in an analysis restricted to pion-induced reactions~\cite{Wang:2022osj}. The extension of the full model, comprising pion- and photon-induced processes, to the $\omega N$-channel is in progress.

We emphasize that the real, dispersive parts of the amplitude are taken into account, a prerequisite for retaining analyticity. Eq.~\eqref{eq:LS} is formulated in isospin and partial-wave basis (corresponding  indices suppressed for better readability), where we include angular momenta up to $J=9/2$.

The Lippmann-Schwinger equation is consistent with two-body unitarity. Three-body unitarity is approximately fulfilled in our approach, see, e.g., Ref.~\cite{Mai:2017vot} for a manifestly three-body unitary framework.
The scattering potential $V_{\mu\nu}$ in Eq.~\eqref{eq:LS} is derived from an effective Lagrangian by using time-ordered perturbation theory (TOPT).  It is constructed of $t$- and $u$-channel exchanges of known mesons and baryons and $s$-channel diagrams, which represent genuine resonances, as well as phenomenological contact diagrams, which are used to absorb physics beyond the explicit processes. For further details on the explicit form of the scattering potential see Refs.~\cite{Ronchen:2012eg,Wang:2022osj}.

The $T$-matrix can be decomposed into a pole part and a non-pole part 
\begin{equation}
    T=T^{\text{P}}+T^{\text{NP}}~.
\end{equation}
The non-pole part $T^{\text{NP}}$ is build of the potentials from the $t$- and $u$-channel which constitute the non-pole part of the potential $V^{\text{NP}}$:
\begin{align}
&T^{\text{NP}}_{\mu\nu}(p,p',W)=V^{\text{NP}}_{\mu\nu}(p,p',W) \label{eq:T_NP}\\
&+\sum_{\kappa}\int\limits_0^\infty dq\,
 q^2\,V^{\text{NP}}_{\mu\kappa}(p,q,W)G^{}_\kappa(q,W)\,T^{\text{NP}}_{\kappa\nu}(q,p',W) ~ .\nonumber
\end{align}

Note that the separation into a pole and non-pole part is purely due to technical advantages and we define resonance states as poles in the full $T$-matrix as further specified in Sec.~\ref{sec:res_spec}. The scattering equation Eq.~\eqref{eq:T_NP} for the non-pole part can dynamically generate poles which are not included as genuine poles via $s$-channel terms. The pole part $T^{\text{P}}$, on the other hand, includes the $s$-channel pole terms and is also evaluated from $T^\npo$ using the prescription 
\begin{equation}
    T^{\text{P}}_{\mu\nu}(p,p',W)=\sum_{i,j}\Gamma^{a}_{\mu,i}(p)\, (D^{-1})_{ij}(W)\,\Gamma^{c}_{\nu,j}(p')~,
    \label{eq:tpo}
\end{equation}
where $\Gamma^{a}_{\mu,i}$ is the dressed vertex function which describes the annihilation of the $i$-th resonance into channel $\mu$, and $\Gamma^{c}_{\nu,j}$ is the dressed vertex function for the creation of the $j$-th resonance from the channel $\nu$. The dressed resonance vertex functions are constructed using $T^\npo$ as shown in Eq.~\eqref{eq:propagator_s_channel_dressed_vf} below,

\begin{widetext}
\begin{eqnarray}
    \Gamma^{a}_{\mu,i}(p)&=&\gamma^{a}_{\mu,i}(p)+\sum_{\kappa}\int_0^\infty \dd{q} q^2 \, T^{\text{NP}}_{\mu\nu}(p,q,W)\, G_\kappa(q,W)\, \gamma_{\kappa,i}^{a}(q)\nonumber\\
    \Gamma^{c}_{\nu,j}(p')&=&\gamma^{c}_{\nu,j}(p')+\sum_{\kappa}\int_0^\infty \dd{q} q^2\,   \gamma_{\kappa,j}^{c}(q) \, G_\kappa(q,W) \, T^{\text{NP}}_{\mu\nu}(q,p',W)\nonumber\\
    D_{ij}(W)&=&\delta_{ij}(W-m_i^b)-\Sigma_{ij}(W)\nonumber\\
    \Sigma_{ij}(W)&=&\sum_{\kappa} \int_0^\infty \dd{q} q^2 \, \gamma_{\kappa,j}^{c}(q)\, G_\kappa(q,W)\, \Gamma^{a}_{\mu,i}(q)~.
    \label{eq:propagator_s_channel_dressed_vf}
\end{eqnarray}
\end{widetext}
In Eq.~\eqref{eq:tpo}, $D^{-1}$ is the propagator for $s$-channel resonances related to the self-energies $\Sigma_{ij}$ as listed in Eq.~\eqref{eq:propagator_s_channel_dressed_vf}. Here the $\gamma$'s are the bare vertex functions with explicit forms given in Ref.~\cite{Doring:2010ap,Ronchen:2015vfa} and $m_i^b$ are the bare mass parameters of the $i$-th resonance. 

The indices $i,j=1,2,3$ characterize the $s$-channel state ($i,j=1,2$) or phenomenological contact term ($i,j=3$) in a given partial wave. Whether one or two bare $s$-channel states are included per partial wave is chosen as demanded by the fit. The contact terms are included on the same footing as the genuine resonance terms due to technical reasons, yet without inducing a pole in the amplitude. 
In the current study we include a maximum of two $s$-channel states and one contact term per partial wave. Note that relating a pole in the amplitude unambiguously to a given $s$-channel term is not possible, because of the highly nonlinear nature of the approach. 
Related to that, a classification of ``$s$-channel poles" as genuine three-quark states and, in contrast, associating a purely molecular nature to dynamically generated poles is too simplistic. 
Instead one has to employ compositeness or elementariness criteria as done in Refs.~\cite{Wang:2023snv,Wang:2025ecf}  in the framework of the JüBo model.   

The $T$-matrix can be used to evaluate observables which can then be fitted to experimental data. See Refs.~\cite{Doring:2010ap,Ronchen:2014cna,Ronchen:2015vfa} for explicit expressions of cross section and polarization observables. The normalised, dimensionless partial-wave amplitude $\tau$ is directly related to $T$ by
\begin{equation}
    \tau_{\mu\nu}=-\pi \sqrt{\rho_\mu\rho_\nu}\, T_{\mu\nu}~,
\end{equation}
where $\rho$ is the kinematical phase factor
\begin{equation*}
    \rho_\kappa=\frac{p_\kappa}{W}E_{b,\kappa}E_{m,\kappa}~,
\end{equation*}
with $E_{b/m,\kappa}=\sqrt{\vec p^{~2}+M_{b/m,\kappa}^2}$ the on-shell energy of the baryon/meson in channel $\kappa$ and $p_\kappa$ is the corresponding on-shell three-momentum.

\subsection{Photoproduction processes}

To include photoproduction processes, the J\"ulich-Bonn DCC approach was extended in a semi-phenomenological way in Ref.~\cite{Ronchen:2014cna}. Here the electric and magnetic photoproduction multipole amplitudes $\cal{M}$ are given by
\begin{multline}
{\cal M}_{\mu\gamma}(p,W)=V_{\mu\gamma}(p,W) \\+\sum_{\kappa}\int\limits_0^\infty dq\,q^2\,
T_{\mu\kappa}(p,q,W)G^{}_\kappa(q,W)V_{\kappa\gamma}(q,W)\ .
\label{eq:photo}
\end{multline}
The channel index $\gamma$ denotes the initial channel of $\gamma N$ with a real photon ($Q^2=0$), and $\mu$ $(\kappa)$ are the final (intermediate) meson-baryon channel. Note that $T_{\mu\kappa}$ in Eq.~\eqref{eq:photo} is the hadronic $T$-matrix of Eq.~\eqref{eq:LS} with the off-shell momentum $q$ and the on-shell momentum $p$ and $G_\kappa$ is the same hadronic two-body propagator as in Eq.~\eqref{eq:LS}. The channel space for meson-baryon pairs currently includes $\kappa=(\pi N, \eta N, K\Lambda, K\Sigma, \pi \Delta)$ and will be complemented once the approach is extended to two pion or vector meson photoproduction. A similar parameterization was recently applied to virtual photons ($Q^2\neq 0$) using the Jülich-Bonn-Washington (JBW) framework for electroproduction reactions~\cite{Mai:2021vsw,Mai:2021aui,Mai:2023cbp,Wang:2024byt}, which includes the Jülich-Bonn amplitude as input at the photon point at $Q^2=0$.

In Eq.~\eqref{eq:photo} the photoproduction potential $V_{\mu\gamma}$ is given by 

\begin{equation}
    V_{\mu\gamma}(p,W)=\alpha^{\text{NP}}_{\mu\gamma}(p,W)+\sum_i\frac{\gamma^a_{\mu,i}(p)\gamma_{\gamma,i}(W)}{W-m^b_i}\,,
\label{eq:Vphoto1}
\end{equation}
where $\gamma_{\mu, i}^{a}$ denotes the bare meson-baryon-to-resonance vertex function (same as in Sec.~\ref{sec:Hadronic_part}) and $\gamma_{\gamma,i}$ is the photon-to-resonance vertex function. The photon-vertex $\gamma_{\gamma,i}$ is parameterized phenomenologically via a polynomial function in the energy $W$ and includes free parameters for each genuine $s$-channel state. The non-pole part of Eq.~\eqref{eq:photo} $\alpha^{\text{NP}}_{\mu\gamma}$ is also parameterized by energy-dependent polynomials which introduce additional fit parameters depending on the partial wave and the final hadronic state. This polynomial parameterization is numerically advantageous to a field-theoretical description (as done e.g. in Ref.~\cite{Huang:2011as}). Further details on the explicit forms of $\gamma_{\gamma,i}$ and $\alpha^{\text{NP}}_{\mu\gamma}$ are given in Ref.~\cite{Ronchen:2014cna}.

\section{Determination of the free model parameters}
\label{sec:new_data}

\subsection{Database}
The data included in this study are listed in Tab.~\ref{tab:data}. References to all considered pion and photon induced data  can be found online~\cite{Juelichmodel:online}. 
Note that for the elastic $\pi N$ channel we do not fit directly to data but use the partial-wave amplitudes of GWU/SAID WI08 analysis~\cite{Workman:2012hx}. We use the energy-dependent solution in steps of $\SI{5}{\mega\electronvolt}$  from $\pi N$ threshold up to $W=\SI{2400}{\mega\electronvolt}$ which leads to the number of fitted points for the elastic $\pi N$ channel quoted in Tab.~\ref{tab:data}.
The photoproduction data sets were mainly obtained from the GWU/SAID~\cite{SAID} and BnGa webpages~\cite{BnGa_web}. 

For the process ${\gamma} {p} \rightarrow \pi^0 p$ we include recent data for the double polarization observable ${E}$, which was published by CLAS \cite{CLAS:2023ddn}. Since the observable ${E}$ is closely related to $\Delta \sigma$ by the relation
\begin{equation}
    \dv{\Delta \sigma}{\Omega}=-2\dv{\sigma_0}{\Omega}{E}~,
\end{equation}
its inclusion is important for the determination  of the contribution of the $\pi^0p$ channel to the GDH sum rule. Previously published data sets on $E$ or $\Delta \sigma$ in $\gamma p\to \pi^0p$, $\pi^+n$, $\eta p$ from MAMI~\cite{GDH:2002pkk,GDH:2004ydy,Ahrens:2006gp,A2:2024ydg}, CBELSA~\cite{CBELSATAPS:2013btn,CBELSATAPS:2019hhr,CBELSATAPS:2019ylw}, 
and CLAS~\cite{CLAS:2015ykk,CLAS:2015pjm} are also included in the fit.

In this study we also include new data on the double polarization observable $G$ for single pion photoproduction off the proton published by the CLAS Collaboration \cite{CLAS:2021udy}. This increases the data base for this observable considerably, especially for the $\pi^+n$ channel (from 86 to 303 data points).

New data on ${d\Delta \sigma}/{d\Omega}$ were also recently published by the A2 Collaboration at MAMI ~\cite{A2CollaborationatMAMI:2023twj}. This data together with the solution of this study is shown in Figs.~\ref{fig:delta13_A2Mami} and \ref{fig:delta13_A2Mami2} in the Appendix. Although we did not include this dataset in the current fit because it was published shortly after the major part of the simulations was completed, we achieve a good data description.

Furthermore, we included recent $\eta$-photoproduction data from the LEPS2/BGOegg collaboration for the differential cross section ${d\sigma}/{d\Omega}$ and the photon beam asymmetry $\Sigma$ \cite{LEPS2BGOegg:2022dop}. These data cover a large polar angle region $-1<\cos\theta_{\text{c.m.}}<0.6$. Note that the beam asymmetry data for center of mass energies above $\SI{2.1}{\giga\electronvolt}$ was covered for the first time in that work.

In addition also the recent polarization data for $\gamma p\to K^0\Sigma^+$ by CLAS~\cite{CLAS:2024bzi} were included. 
An earlier JüBo fit to those data was already presented in Ref.~\cite{CLAS:2024bzi}.

\begin{table*}
\caption{Data included in the fit with new data sets highlighted in red. A full list of references to the different experimental publications can be found online~\cite{Juelichmodel:online}.} 
\begin{center}
\renewcommand{\arraystretch}{1.9}
\begin {tabular}{l|l|r} 
\hline\hline
Reaction & Observables ($\#$ data points) & $\#$ data p./channel \\ \hline
$\pi N\to\pi N$ & {PWA  GW-SAID WI08 \cite{Workman:2012hx} (ED solution) } & {8,396}\\
$\pi^-p\to\eta n$ &{$d\sigma/d\Omega$ (676), $P$ (79) }
& 755\\
$\pi^-p\to K^0 \Lambda$ &{$d\sigma/d\Omega$ (814), $P$ (472), $\beta$ (72) }
& 1,358\\
$\pi^-p\to K^0 \Sigma^0$ &{$d\sigma/d\Omega$ (470), $P$ (120) }
& 590\\
$\pi^-p\to K^+ \Sigma^-$ &{$d\sigma/d\Omega$ (150)}
& 150\\
$\pi^+p\to K^+ \Sigma^+$ &{$d\sigma/d\Omega$ (1124), $P$ (551) , $\beta$ (7)}
& 1,682\\ 
 \hline
$\gamma p\to \pi^0p$ & $d\sigma/d\Omega$ (18721), $\Sigma$ (3287), $P$ (768), $T$ (1404), $\Delta\sigma_{31}$ (140),& \\
& $G$ (393\textcolor{red}{+198})~\cite{CLAS:2021udy}, $H$ (225), $E$ (1227\textcolor{red}{+495})~\cite{CLAS:2023ddn}, $F$ (397), $C_{x^\prime_\text {L}}$ (74), $C_{z^\prime_\text{L}}$ (26)
&27,355 \\
$\gamma p\to \pi^+n$ & $d\sigma/d\Omega$ (5670), $\Sigma$ (1456), $P$ (265), $T$ (718), $\Delta\sigma_{31}$ (231), &\\
& $G$ (86\textcolor{red}{+217})~\cite{CLAS:2021udy}, $H$ (128), $E$ (903)
& 9,674 \\
$\gamma p\to \eta p$ &$d\sigma/d\Omega$ (9112\textcolor{red}{+320})~\cite{LEPS2BGOegg:2022dop}, $\Sigma$ (535\textcolor{red}{+80})~\cite{LEPS2BGOegg:2022dop}, $P$ (63), $T$ (291), $F$ (144), $E$ (306), $G$ (47), $H$ (56)
& 10,954\\
$\gamma p\to K^+\Lambda$ & 
{\footnotesize $d\sigma/d\Omega$ (2563), 
$P$ (1663), 
$\Sigma$ (459), 
$T$ (383), 
$C_{x^\prime}$ (121), 
$C_{z^\prime}$ (123), 
$O_{x^\prime}$ (66), 
$O_{z^\prime}$ (66), 
$O_x$ (314), 
$O_z$ (314) 
}
&6,072 \\ 
$\gamma p\to K^+\Sigma^0$ & {\footnotesize $d\sigma/d\Omega$ (4381), $P$ (402), $\Sigma$ (280) $T$ (127), $C_{x'}$ (94), $C_{z'}$ (94), $O_{x}$ (127), $O_{z}$ (127)  } & 5,632\\
$\gamma p\to K^0\Sigma^+$ & {\footnotesize $d\sigma/d\Omega$ (281), $P$ (188) , $\Sigma$ (21), $T$ (21), $O_x$ (21), $O_z$ (21)} & 553 \\ \hline
& \multicolumn{1}{r}{in total} & {73,171}  
\\
\hline\hline
\end {tabular}
\end{center}
\label{tab:data}
\end{table*}

\subsection{Numerical details}
\label{sec:numerics}
We perform a $\chi^2$ minimization to fit the free model parameters to the experimental data using MINUIT~\cite{James:1975dr} on the JURECA-DC supercomputer at the J\"ulich Supercomputing Center~\cite{JURECA}. 
The free parameters are the same as in our previous analysis JüBo2022~\cite{Ronchen:2022hqk}: 
\begin{itemize}
    \item 134 for the 21 $s$-channel diagrams of Eq.~\eqref{eq:propagator_s_channel_dressed_vf} (bare masses and couplings to the pertinent channels)
    \item 61 from phenomenological contact terms of Eq.~\eqref{eq:propagator_s_channel_dressed_vf} (bare couplings)
    \item 764 parameters directly connected to the photon interaction of Eq.~\eqref{eq:Vphoto1} (coefficients of the polynomials).
\end{itemize} 
More details on the free parameters can be found in Ref.~\cite{Ronchen:2022hqk}.
It should be noted that even though we always use the full data base to determine the parameter values, we cannot fit all parameters simultaneously due to the complexity of the model and numerical limitations. The majority of the large number of parameters originates from the polynomial parameterization of the photoproduction amplitude and does not induce resonance poles in the scattering matrix $T_{\mu\nu}$. While the number of parameters of this class is not predetermined by the model and likely not all of them are indispensable, this flexibility can be regarded as an advantage because it helps to keep the number of genuine $s$-channel states at a minimum. A possible way to reduce the number of parameters systematically could be by using model selection tools such as the LASSO method~\cite{Tibshirani:1996fxl,Landay:2016cjw,Landay:2018wgf} which is planned for the future. 

Experimental systematic errors are usually only available for the more recent data sets. They are included as angle-independent normalization factors, as done in the GWU/SAID analysis~\cite{Doring:2016snk}. 
We consider an additional $5\%$ uncertainty for older data sets on top of the statistical one to account for systematic errors.

As can be seen in Tab.~\ref{tab:data}, the number of data points for different channels and observables varies significantly. This leads to small data sets being mostly ignored in the $\chi^2$ minimization. To allow smaller data sets to have an impact, we introduce weights in the $\chi^2$. This procedure is typical for the kind of analyses of this type~\cite{Arndt:1995bj, Anisovich:2011fc, Briscoe:2020qat, CLAS:2015ykk,ParticleDataGroup:2024cfk}.

To perform a proper statistical error analysis, one would have to study the propagation of systematic and statistical uncertainties from the experimental data to the extracted baryon resonance parameters, by also taking into account covariance matrices, which poses a considerable numerical challenge that is beyond the scope of the current work. We note that until now this has not been carried out in a rigorous way by any of the coupled-channel analysis groups.

Instead, we follow the procedure of our previous studies~\cite{Ronchen:2018ury,Ronchen:2022hqk} to qualitatively estimate the uncertainties of the resonance parameters from re-fits with a modified parameterization of our model by including additional $s$-channel states. 
For example in each of the 16 partial waves with zero or one $s$-channel resonance in the original fit, we include an additional genuine state and perform re-fits of all free parameters as given above. We use the maximal deviation from the original resonance parameter values in these re-fits as our estimated uncertainties. As there was no significant improvement of the data description in these re-fits, we can conclude that none of the additional $s$-channel states has to be included in the original parameterization of the model. 
This procedure gives us a qualitative estimation of relative uncertainties for the resonance parameters. 
We consider this a compromise since a rigorous error analysis is not possible in the present study.

\section{Results}
\label{Sec:Fit_results}
\subsection{Fit results for new data sets}
\begin{figure*}
\begin{center}
\includegraphics[width=1.\linewidth]{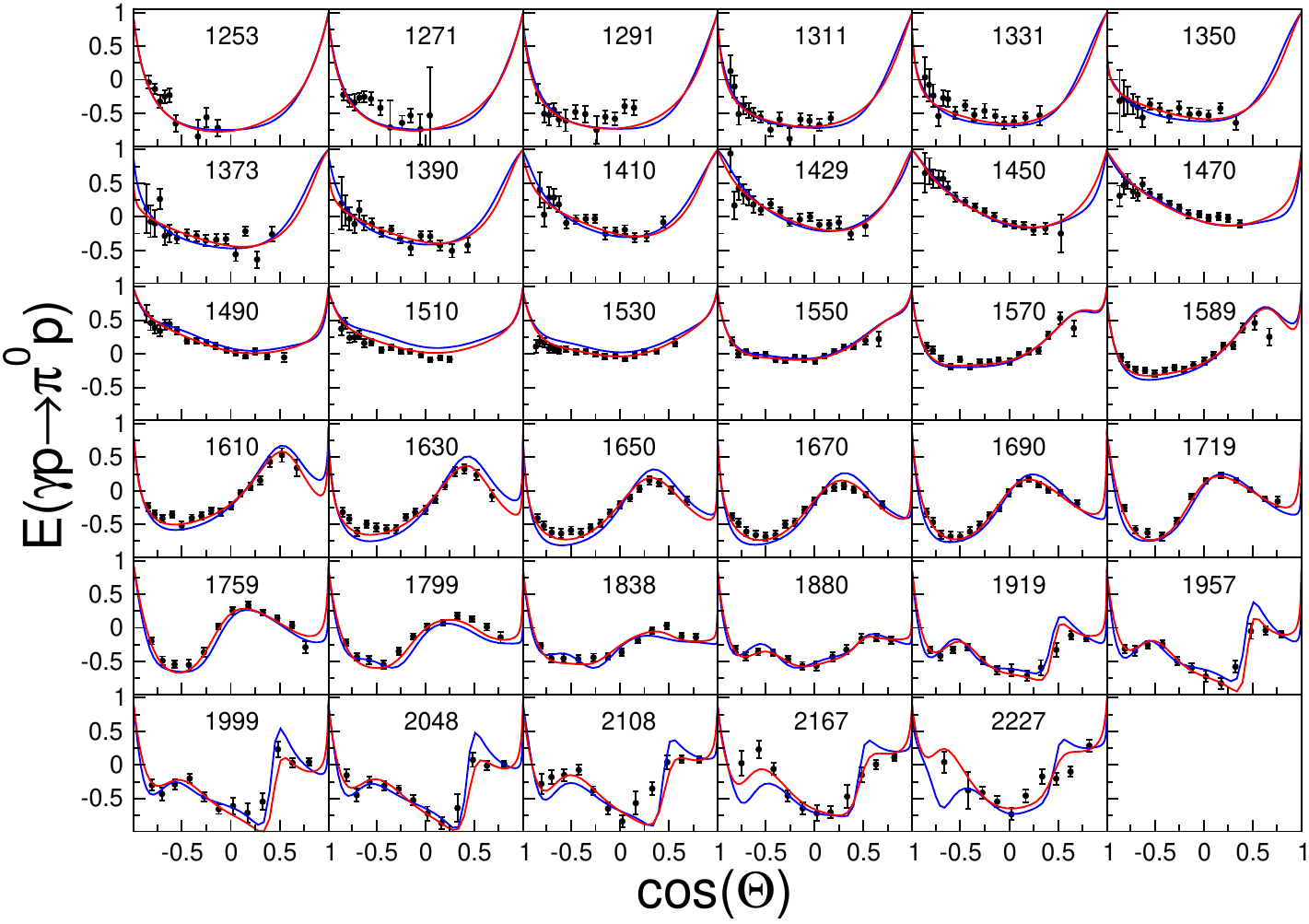} 
\end{center}
\caption{Current fit results (red) and 2022 fit~\cite{Ronchen:2022hqk} (blue) for comparison for the double spin polarization observable ${E}$ for the process $\gamma p\to \pi^0p$. Data  from \cite{CLAS:2023ddn}. The numbers in each plot denote the center of mass energy in MeV.}
\label{fig:Epi0pCLAS2021}
\end{figure*}

\begin{figure*}
\begin{center}
\includegraphics[width=1.\linewidth]{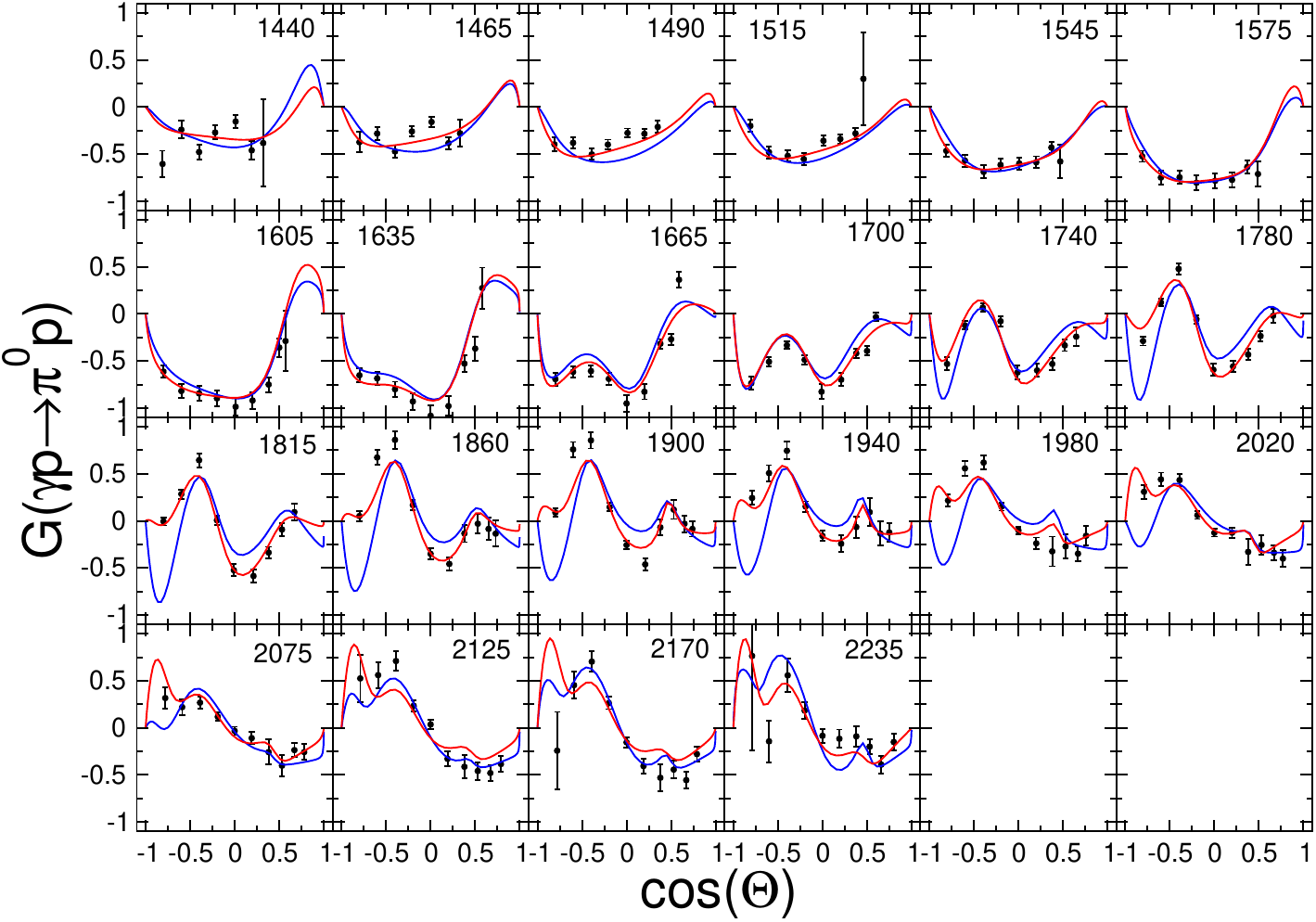} 
\end{center}
\caption{Current fit results (red) and 2022 fit~\cite{Ronchen:2022hqk} (blue) for comparison for the double polarization observable ${G}$ for the process $\gamma p\to \pi^0p$. Data  from \cite{CLAS:2021udy}. The numbers in each plot denote the center of mass energy in MeV.}
\label{fig:Gpi0pCLAS2021}
\end{figure*}

\begin{figure*}
\begin{center}
\includegraphics[width=1.\linewidth]{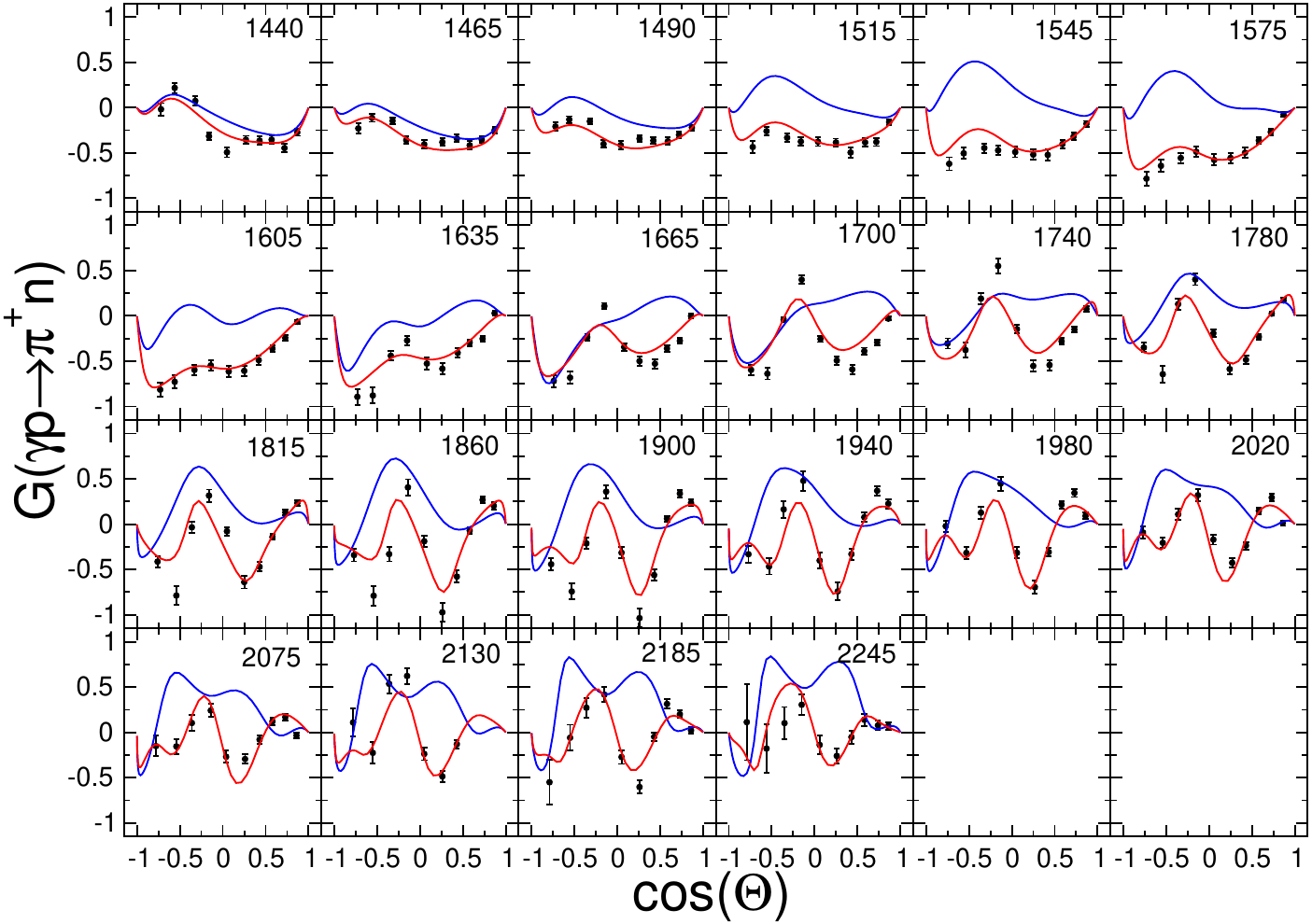} 
\end{center}
\caption{Current fit results (red) and 2022 fit~\cite{Ronchen:2022hqk} (blue) for comparison for the double polarization observable ${G}$ for the process $\gamma p\to \pi^+n$. Data from \cite{CLAS:2021udy}. The numbers in each plot denote the center of mass energy in MeV.}
\label{fig:GpipnCLAS2021}
\end{figure*}

\begin{figure*}
\begin{center}
\includegraphics[width=1.\linewidth]{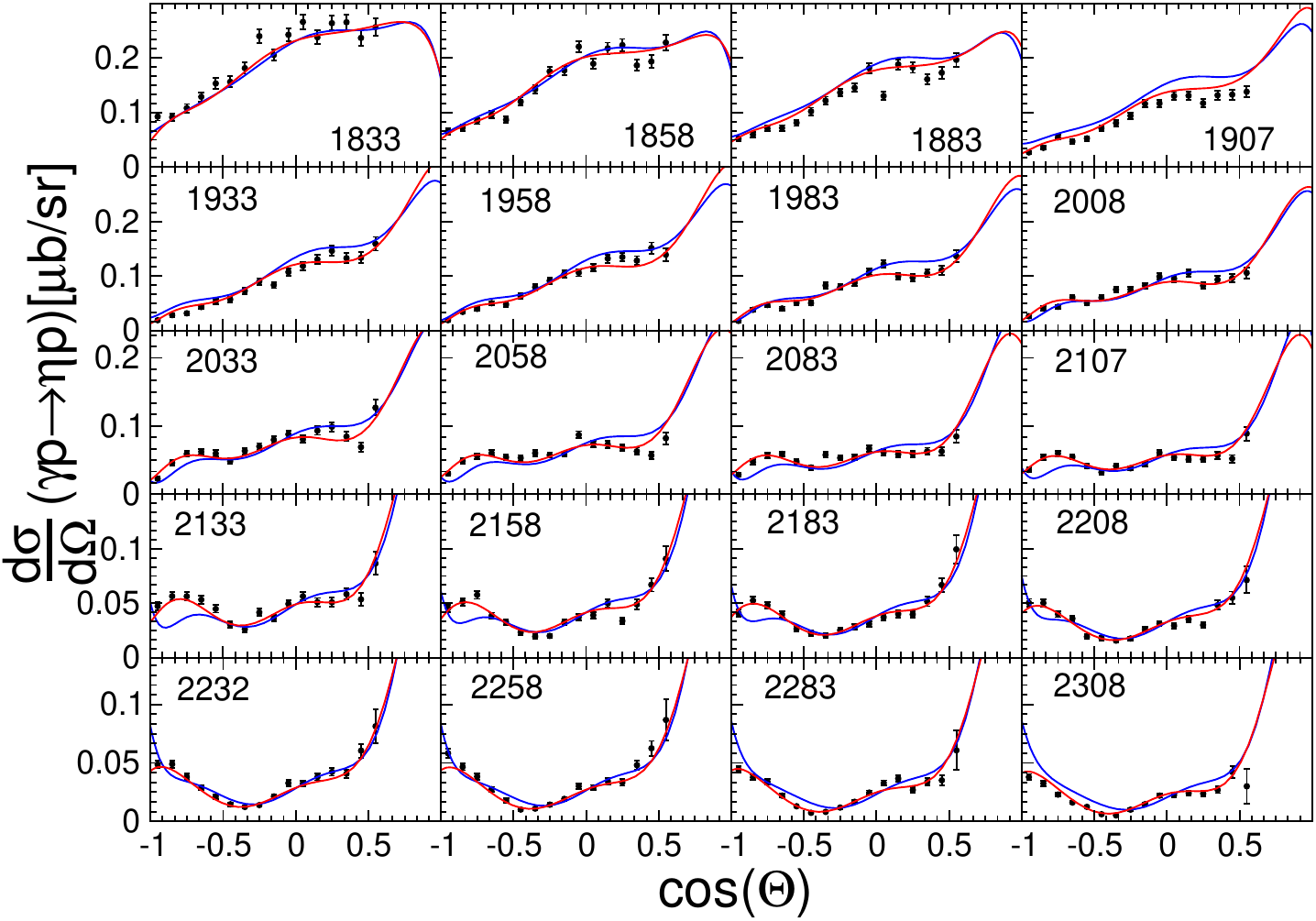} 
\end{center}
\caption{Current fit results (red) and 2022 fit~\cite{Ronchen:2022hqk} (blue) for comparison for the differential cross section for the process $\gamma p\to \eta p$. Data from \cite{LEPS2BGOegg:2022dop}. The numbers in each plot denote the center of mass energy in MeV.}
\label{fig:dsdoetapLEPS2022}
\end{figure*}

\begin{figure*}
\begin{center}
\includegraphics[width=1.\linewidth]{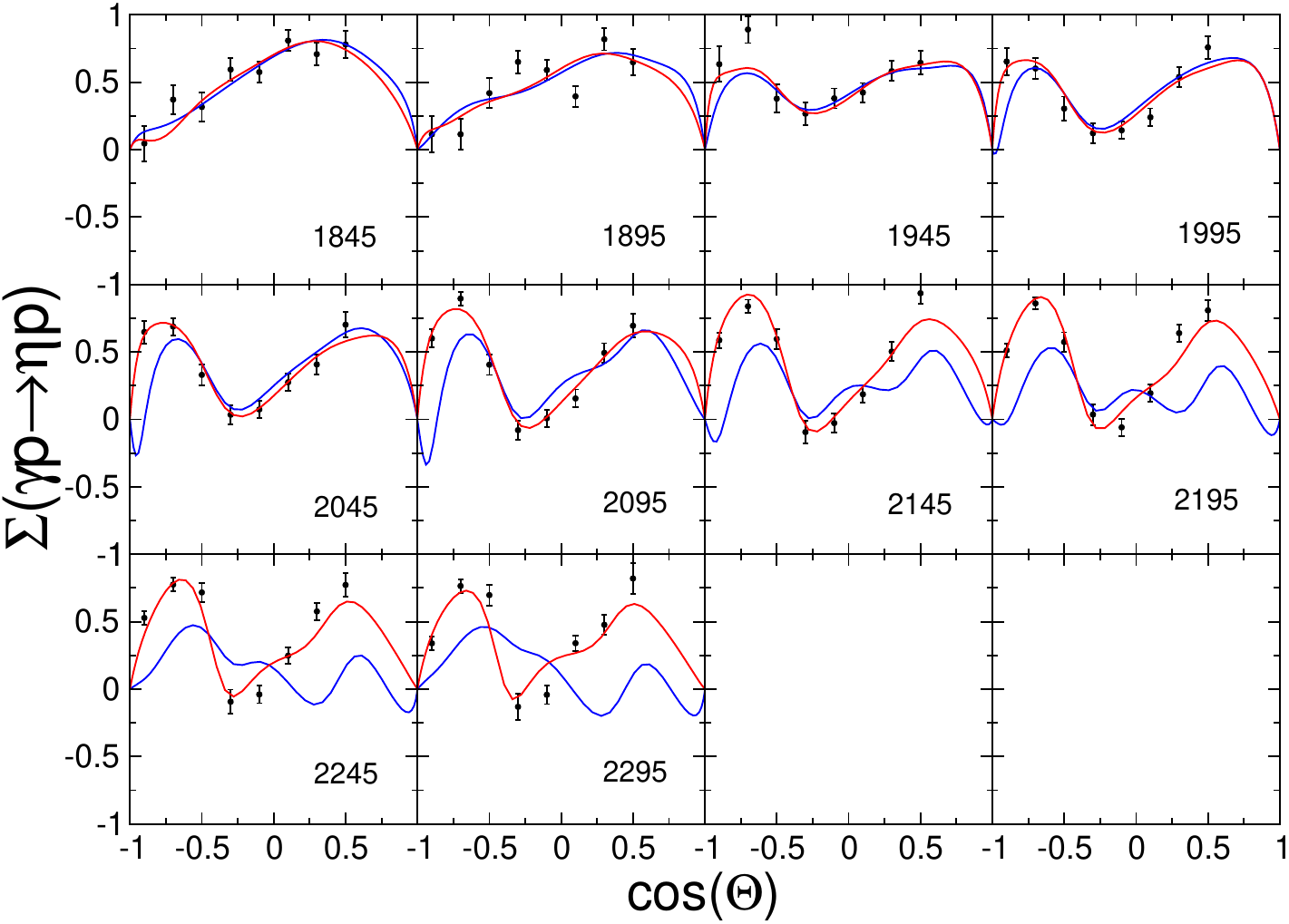} 
\end{center}
\caption{Current fit results (red) and 2022 fit~\cite{Ronchen:2022hqk} (blue) for comparison for the photon beam asymmetry $\Sigma$ for the process $\gamma p\to \eta p$. Data from \cite{LEPS2BGOegg:2022dop}. The numbers in each plot denote the center of mass energy in MeV.}
\label{fig:SetapLEPS2022}
\end{figure*}

In Figs.~\ref{fig:Epi0pCLAS2021} to~\ref{fig:SetapLEPS2022}
we present the current fit result, solution ``JüBo2025", for the newly included data sets, in comparison to the result of our previous study JüBo2022~\cite{Ronchen:2022hqk}. We list the $\chi^2$-values of these new data sets in Tab.~\ref{tab:chi2newdata}.

\begin{table}
\caption{$\chi^2$-values per number of data points ($\#$) and weights used in the weighted fit for the newly included datasets presented in Figs.~\ref{fig:Epi0pCLAS2021} to~\ref{fig:SetapLEPS2022}.}
\begin{center}
\renewcommand{\arraystretch}{1.9}
\begin {tabular}{|l|c|c|c|} 
\hline
Reaction & Observable ($\#$) & $\chi^2$/$\#$ & weight\\ \hline
$\gamma p\to \pi^0 p$  & ${E}$ (495) \cite{CLAS:2023ddn} & 1.62 & 60\\
 & ${G}$ (198) \cite{CLAS:2021udy} & 1.57 & 100\\ \hline
$\gamma p\to \pi^+ n$  &  ${G}$ (217) \cite{CLAS:2021udy} & 3.02 & 130\\ \hline
$\gamma p\to \eta N$ & $d\sigma/d\Omega$ (320) \cite{LEPS2BGOegg:2022dop} &  0.85 & 60\\
  & $\Sigma$ (80) \cite{LEPS2BGOegg:2022dop} & 0.63 & 90 \\ 
\hline 

\end{tabular}
\end{center}
\label{tab:chi2newdata}
\end{table}

The new data on the double polarization observable ${E}$ for the process $\gamma p\to \pi^0p$ only lead to minor improvements, since the previous fit result was already in good agreement (Fig.~\ref{fig:Epi0pCLAS2021}).

For the double polarization observable ${G}$, we find slight improvements for the process $\gamma p \rightarrow \pi^0 p$ , especially at higher energies, see Fig.~\ref{fig:Gpi0pCLAS2021}. But for the process  $\gamma p \rightarrow \pi^+ n$ we achieve strong improvements, c.f Fig.~\ref{fig:GpipnCLAS2021}. This can be explained by the considerable increase in number of data points, which was more than tripled.

The new data for the reaction $\gamma p\rightarrow \eta p$ lead to an improvement in the description of backward angles at energies above $\SI{2}{\giga\electronvolt}$ as presented in Figs.~\ref{fig:dsdoetapLEPS2022} and \ref{fig:SetapLEPS2022}. Note that for the observable $\Sigma$ the data for energies above $\SI{2.1}{\giga\electronvolt}$ was covered for the first time, which explains the large overall improvement with respect to  the 2022 solution. 

\subsection{Resonance spectrum}
\label{sec:res_spec}
Resonances are defined as poles on the unphysical Riemann sheet of the full scattering matrix $T$ in the complex energy plane. The analytic properties, sheet structure, cuts and the analytic continuation of the amplitude to the second sheet within our model are described in detail in Ref.~\cite{Doring:2009yv}. 

We use the normalised residues to quantify the coupling strengths of individual states to hadronic channels. The definition of normalised residues within our framework is in agreement with that of the Particle Data Group (PDG)~\cite{ParticleDataGroup:2024cfk} and can be found in Ref.~\cite{Ronchen:2012eg}. In Ref.~\cite{Heuser:2024biq} a novel approach is presented to relate the complex residues to the more intuitive branching fractions. A determination of branching ratios for baryon resonances along those lines is planned for the future.  Our method to calculate the residues of the complex poles can be found in the appendix of Ref.~\cite{Doring:2010ap}. For the coupling of the $\gamma N$ channel to the resonances we use the PDG convention to define the so-called photocouplings at the pole. The explicit definition can be found in Ref.~\cite{Ronchen:2014cna}.

In Tab.~\ref{tab:poles1} and \ref{tab:poles2} we present the pole positions $W_0$ and the residues of the states found in this study. The photocouplings at the pole are listed in Tab.~\ref{tab:photo}. We compare the values with the results from the J\"uBo2022 study \cite{Ronchen:2022hqk}. We also list PDG~\cite{ParticleDataGroup:2024cfk} values, if an estimate is given. We find all 4-star resonances for $I=1/2$ and $I=3/2$ up to $J=9/2$, except for the $N(1895)1/2^+$. This resonance is not needed in our present study. As mentioned in Ref.~\cite{Tiator:2018heh}, this resonance was found to be important for the description of the near threshold $\eta'N$ data. This suggests that once the J\"uBo analysis is extended to the $\eta'N$ channel, it will be seen if the $N(1895)1/2^+$ is needed. Such an extension is planned for the future. There are also a few resonance states with a lower star rating that we find within our analysis.

Compared to our previous analysis JüBo2022~\cite{Ronchen:2022hqk} we find no new resonance states but observe some significant changes in the resonance parameters. Those differences are discussed in the following. The different partial waves $L_{2I\, 2J}$ are labeled with reference to the $\pi N$ channel as specified in Tab.~11 of Ref.~\cite{Ronchen:2012eg}. For the individual resonance states we follow the naming scheme of the PDG~\cite{ParticleDataGroup:2024cfk}.

\subsection*{Changes in the $N^*$ resonance spectrum}

The two $S_{11}$ resonance states $N(1535)1/2^-$ and $N(1650)1/2^-$ originate from bare $s$-channel states. Compared to the J\"uBo2022 values, the normalized residue of the $N(1535)1/2^-$ for the channel $K\Sigma$ decreased significantly. However, since the $N(1535)1/2^-$ lies relatively far below the $K\Sigma$ threshold, this normalized residue is difficult to determine. 
For the $N(1650)1/2^-$ the uncertainties for the residues are smaller than before and also the pole position is similar as in JüBo2022. 

In the $P_{11}$ partial wave we find the poles of the $N(1440) 1/2^+$ and $N(1710) 1/2^+$ resonances. The former state is the famous Roper resonance, which is dynamically generated in the Jülich model, predominantly by correlated $2\pi$ exchange with $\rho$ quantum numbers combined with strong contributions from the $\sigma N$ channel~\cite{Schutz:1998jx,Krehl:1999km}.  The main difference in this partial wave to the 2022 values is that the $N(1710) 1/2^+$ shifted in the real part further away from the PDG estimate. The nucleon ground state is also included in this partial wave as a bare $s$-channel state with its bare mass and $\pi N$ coupling normalized so that the dressed quantities are fixed to the physical values~\cite{Ronchen:2015vfa}. 

The $P_{13}$ partial wave has two $s$-channel induced poles, the $N(1720)3/2^+$ and $N(1900)3/2^+$. For the $N(1720)3/2^+$ we find a smaller width and all residues are lower than in J\"uBo2022. The $N(1900)3/2^+$ is found to have a significantly broader width and a larger magnitude of the photocoupling $A^{3/2}$ at the pole in the present study. In the JüBo2022 study~\cite{Ronchen:2022hqk}, we observed that these two resonances have a high impact on the $\gamma p\to \eta p$ data by turning off the corresponding couplings. This is still the case in the present study, the $N(1900)3/2^+$ residue into $\eta N$ even increased. This is also visualized in the Appendix in Fig.~\ref{fig:LEPS_data_without_P13}.

The 4-star resonance $N(1520)3/2^-$ is observed in the $D_{13}$ partial wave. Given that this well-established state had small uncertainties in previous JüBo analyses, the change in the  pole position in the current study is noticeable, especially for the imaginary part. 
As in the J\"uBo2022 \cite{Ronchen:2022hqk} analysis we see further indications for a dynamically generated $N(1875)3/2^-$ at $1914 (1) {-} i\, 343 (2)\, \si{\mega\electronvolt}$, but its width of $686\,\si{\mega\electronvolt}$ is still significantly larger than the estimate of the PDG~\cite{ParticleDataGroup:2024cfk} of $160\pm 60 \,\si{\mega\electronvolt}$. The influence of such a broad state on the physical axis, and for that the data, is very limited.

We observe the $N(1675)5/2^-$ in the $D_{15}$ partial wave which originates from an $s$-channel state. For this resonance the real part of the pole position changed and the elastic residue as well as the normalised residue into the $\eta N$ channel decreased.

In the $F_{15}$ partial wave we find the $s$-channel resonance $N(1680)5/2^+$ but without any significant change in the resonance parameters.

The $N(1990) 7/2^+$ can be observed in the $F_{17}$ wave. The changes compared to the 2022 values are within uncertainties except for the elastic $\pi N$ residue, which is, however, very small and therefore difficult to determine. We also find that this resonance has a large influence on the $\eta p$ data, which is in accordance with its large $\eta N$ residue. This is shown in the appendix in Fig.~\ref{fig:LEPS_data_without_F17}. Although this state has only a 2-star rating by the PDG, we can confirm our observation in the JüBo2022 analysis that the $N(1990) 7/2^+$ has  significant influence in the energy range of its pole position and also confirm the relatively small imaginary part. Based on this observation we propose an upgrade to a 3-star rating.  

The $G_{17}$ features the $N(2190) 7/2^-$ resonance. We notice a significant decrease in the width compared to the JüBo2022 result while still getting an acceptable fit to the $\pi N$ partial waves of the GWU/SAID \cite{Workman:2012hx}. We already observed this change after the inclusion of the new polarization data for $K^0\Sigma^+$ photoproduction~\cite{CLAS:2024bzi}, the resonance parameters remain stable with regard to the latter analysis. Our width is in clear disagreement with the PDG estimate. We also find a smaller photocoupling $A^{3/2}$ for this resonance. 

In the $G_{19}$ partial wave we have one pole, the $N(2250)9/2^-$. Here the pole position changed in real and imaginary part but all pole parameters have a high uncertainty in our current analysis. We also notice a significant shift in the photocoupling $A^{1/2}$.

For the $H_{19}$ we can find again only one pole, the $N(2220)9/2^+$. For this one the real part decreased significantly, but also carries a large uncertainty on the pole position. We also notice a significant smaller photocoupling $A^{1/2}$.

\subsection*{Changes in the $\Delta$ resonance spectrum}

The $S_{31}$ partial wave features the well-established resonance $\Delta(1620)1/2^-$. The moderate  changes in the parameters are mostly accompanied by higher uncertainties of the residues and agree with the 2022 values.

The $P_{31}$ features only the $\Delta(1910)1/2^+$. The large width increased significantly in the present study compared to 2022 and is further away from the PDG estimate. The photocoupling at the pole $A^{1/2}$ also changed significantly but has a higher uncertainty.

In the $P_{33}$ we have the $\Delta(1232)$ as well as the two resonances $\Delta(1600)$ and the $\Delta(1920)$. 
While the pole parameters of the $\Delta(1232)$ are very similar to our previous study, the width of the $\Delta(1600)$ changed more clearly. 
As in 2022 the $\Delta(1920)$ is very broad and we cannot claim much evidence for this state.

In the $D_{33}$ we find the $\Delta(1700)3/2^-$. Here the real part of the pole position and the width both increased and have a lower uncertainty. Also the $\pi N$ residue as well as all normalised residues increased significantly. 

For the $D_{35}$ we have the $\Delta(1930)5/2^-$ as the only $s$-channel pole. Here the only significant change is in the width in its current analysis.

The $F_{35}$ features the $\Delta(1905)5/2^+$ as the only resonance pole. For this, the real part of the pole position changed slightly but its width significantly.

In the $F_{37}$ and $G_{37}$ partial waves we observe two poles each without significant changes compared to the JüBo2022 analysis.

One $s$-channel diagram is included in the $G_{39}$ partial wave which induces the $\Delta(2400)9/2^-$ pole. Here the real part of the pole position was found to be higher than $2.5\,\si{\giga\electronvolt}$ and the width was significantly smaller than in 2022. Also all the residues increased significantly in the current analysis.

\begin{table*} \caption{We list extracted resonance parameters of the $I=1/2$ resonances: Pole positions $W_0$ ($\Gamma_{\rm tot}$ defined as -2Im$W_0$), elastic $\pi N$ residues $(|r_{\pi N}|,\theta_{\pi N\to\pi N})$, and the normalized residues $(\sqrt{\Gamma_{\pi N}\Gamma_\mu}/\Gamma_{\rm tot},\theta_{\pi N\to \mu})$ for the inelastic reactions $\pi
N\to \mu$ with $\mu=\eta N$, $K\Lambda$, $K\Sigma$. 
We show the results of the present study (``2025") and the J\"uBo2022 results (``2022") for comparison~\cite{Ronchen:2022hqk} and the estimates from the Particle Data Group~\cite{ParticleDataGroup:2024cfk} (``PDG"), if available, as well as the PDG star rating.}
\begin{center}
\renewcommand{\arraystretch}{1.3}
\resizebox{\textwidth}{!}{
\begin {tabular}{ll|ll|ll|ll|ll|ll} 
\hline\hline
&&\multicolumn{1}{|l}{Re $W_0$ \hspace*{0.5cm} }
& \multicolumn{1}{l|}{$-$2Im $W_0$\hspace*{0.1cm} }
& \multicolumn{1}{l}{$|r_{\pi N}|$\hspace*{0.2cm}} 
& \multicolumn{1}{l}{$\theta_{\pi N\to\pi N}$ } 
& \multicolumn{1}{|l}{$\displaystyle{\frac{\Gamma^{1/2}_{\pi N}\Gamma^{1/2}_{\eta N}}{\Gamma_{\rm tot}}}$}
& \multicolumn{1}{l|}{$\theta_{\pi N\to\eta N}$\hspace*{0.1cm}}
& \multicolumn{1}{l}{$\displaystyle{\frac{\Gamma^{1/2}_{\pi N}\Gamma^{1/2}_{K\Lambda}}{\Gamma_{\rm tot}}}$} 
& \multicolumn{1}{l|}{$\theta_{\pi N\to K\Lambda}$\hspace*{0.1cm}}
& \multicolumn{1}{l}{$\displaystyle{\frac{\Gamma^{1/2}_{\pi N}\Gamma^{1/2}_{K\Sigma}}{\Gamma_{\rm tot}}}$} 
& \multicolumn{1}{l}{$\theta_{\pi N\to K\Sigma}$}
\bigstrut[t]\\[0.2cm]
&&\multicolumn{1}{|l}{[MeV]} & \multicolumn{1}{l|}{[MeV]} & \multicolumn{1}{l}{[MeV]} & \multicolumn{1}{l}{[deg]} 
& \multicolumn{1}{|l}{[\%]}  & \multicolumn{1}{l|}{[deg]} & \multicolumn{1}{l}{[\%]}  & \multicolumn{1}{l|}{[deg]} &\multicolumn{1}{l}{[\%]} & \multicolumn{1}{l}{[deg]} \\
		  & fit &&&&&&&&
\bigstrut[t]\\
\hline

$N (1535)$ 1/2$^-$ &  2025 & $1504$ (1) & $78$ (2) & $20$ (1) & $-27$ (1) & $55$ (1) & $128$ (0) & $22$ (1) & $-56$ (1) & $6.7$ (1.3) & $79$ (9) \\  
& 2022 & 1504 (0)& 74 (1)& 18 (1)& $-37$ (3) & 50 (3) & 118 (3) & 26 (2) & $-67$ (3) & 28 (2) & 92 (3) \\
%
 ****	&PDG& $1510\pm 10$ & $110^{+20}_{-30}$ & $25\pm 10$& $-20\pm 20$	&---&---&---&---&---&---		\\ \hline
	
$N(1650)$ $1/2^-$ & 2025 & $1671$ (2) & $127$ (3) & $42$ (1) & $-54$ (2) & $24$ (1) & $39$ (3) & $17$ (0) & $-73$ (2) & $27$ (1) & $-61$ (2) \\  
 & 2022 & 1678 (3) & 127 (3) &59 (21) & $-18$ (46) & 34 (12) & 71 (45) & 26 (10) &  $-40$ (46) & 41 (15) & $-21$ (47)\\
%
  ****&PDG &$1665\pm 15$ &$135\pm 35 $ & $45^{+10}_{-20} $ & $-70^{+20}_{-10}$ & ---&---&---&---&---&---\\
				\hline
				
 $N (1440)$ 1/2$^+$ & 2025 & $1359$ (1) & $213$ (2) & $61$ (1) & $-99$ (1) & $8.4$ (0.4) & $-32$ (2) & $6.2$ (1.8) & $140$ (4) & $0.9$ (1.4) & $-23$ (9) \\  
 & 2022 & 1353 (1) & 203 (3) &  59 (2) & $-104$ (4) & 8.4 (0.4) & $-28$ (4)& 2.5 (0.9) & $-92$ (86) & 0.2 (0.5) & $-32$ (154)\\
%
  **** &PDG & $1370\pm 10$ &$175\pm 15$ &$50\pm 4$ & $-90\pm 10$ &---&---&---&---&---&---\\
				\hline


$N (1710)$  1/2$^+$ & 2025 & $1588$ (5) & $113$ (9) & $3.4$ (0.7) & $-115$ (4) & $21$ (1) & $86$ (2) & $14$ (2) & $-157$ (4) & $4.7$ (0.4) & $161$ (3) \\  

& 2022 &1605 (14) & 115 (9) & 5.5 (4.7) & $-114$ (57) & 28 (26) & 91 (63) & 20 (19)& $-144$ (77) & 5.5 (4.8) & 162 (305) \\
%
****  &PDG &$1700\pm 20$ & $ 120\pm 40$ &$7^{+3}_{-4}$ &$190^{+80}_{-70}$ &--- &---&---&---&---&---\\
				\hline
 
%

 $N (1720)$ 3/2$^+$ & 2025 & $1720$ (2) & $166$ (4) & $4.8$ (3.7) & $-16$ (43) & $2.5$ (1.6) & $99$ (32) & $1.2$ (0.3) & $-71$ (31) & $1.2$ (0.9) & $107$ (37) \\  
& 2022 & 1726 (8) & 185 (12) & 15 (2) & $-60$ (5) & 4.9 (0.9) & 64 (10) & 3.4 (0.4) & $-101$ (8) & 5.9 (1) & 82 (6) \\
%
 **** &PDG &$1680^{+30}_{-20}$ &$200^{+200}_{-50} $ &$15\pm 10$ & $-110\pm 50$ &--- &---&--- &---&---&---\\
				\hline

 $N (1900)$ 3/2$^+$ & 2025 & $1904$ (1) & $141$ (1) & $1.1$ (0.3) & $-93$ (4) & $2.2$ (0.2) & $-2$ (4) & $4.2$ (0.3) & $-62$ (3) & $1.5$ (0.2) & $-79$ (8) \\  
 & 2022 & 1905 (3) & 93 (4) & 1.6 (0.3) & 44 (21) & 1.0 (0.3) & 55 (29)&2.9 (0.6) & 5.4 (18.6)&1.3 (0.3) & $-40(18)$\\
%
****  &PDG &$1920\pm 20$ & $130^{+30}_{-40}$ &$4\pm 2$ &$-10\pm 30$&---&---&---&---&---&---\\
				\hline
 
 $N (1520)$ 3/2$^-$ & 2025 & $1496$ (1) & $102$ (6) & $26$ (5.2) & $-17$ (4) & $1.8$ (0.8) & $68$ (9) & $2.9$ (7.4) & $140$ (15) & $3.1$ (20) & $-36$ (4) \\  
 & 2022 & 1482 (6) & 126 (18)& 27 (21) & $-36$ (48) & 2.1 (1.8) &34 (53) & 2.6 (1.9) & 127 (47) & 1.0 (1.2) & 94 (68) \\
%
 ****	 &PDG &$1510\pm 5$ &$110^{+10}_{-5}$ &$35\pm 3$ &$-10\pm 5$ &--- &--- &---&---&---&---\\
				 \hline

 $N (1675)$ 5/2$^-$ & 2025 & $1644$ (1) & $117$ (2) & $14$ (5) & $13$ (16) & $3.5$ (1.0) & $-9$ (16) & $<0.1$ (0.0) & $-65$ (49) & $2.2$ (0.5) & $-144$ (16) \\  
 & 2022 & 1652 (3) & 119 (1) & 22 (1) & $-17$ (2) & 6.3 (0.9) & $-39$ (2) & $<0.1$ (0.2) & 174 (161) & 2.4 (0.2) &$-166$ (5) \\
%
****  &PDG &$ 1655\pm 5$ &$135\pm15$ &$26^{+6}_{-4} $ &$-22\pm 5 $ &--- &--- &---&---&---&---\\
				\hline

 
 $N (1680)$ 5/2$^+$ & 2025 &  $1663$ (5) & $114$ (3) & $46$ (18) & $-41$ (12) & $0.1$ (0.1) & $-44$ (110) & $0.9$ (0.3) & $-121$ (15) & $0.1$ (0.2) & $-49$ (196) \\  
 & 2022 & 1657 (3) & 120 (2) &36 (1) & $-31$ (1) & 0.6 (0.7)& 118 (2)& 0.6 (0.1)& $-119$ (3) & $<0.1$ (0.2) & $-46$ (29)\\
%
****  &PDG &$1670\pm10$ &$120^{+15}_{-10}$ &$40\pm 5 $ &$-20\pm 10 $ &--- &--- &---&---&---&---\\
				\hline

 $N (1990)$ 7/2$^+$ & 2025 & $1851$ (6) & $82$ (10) & $0.18$ (0.02) & $-99$ (4) & $5.2$ (0.4) & $-50$ (3) & $0.1$ (0.1) & $-62$ (190) & $0.7$ (0.3) & $-64$ (4) \\  
 & 2022 & 1861 (9) & 72 (5) & 0.16 (0.01) & $-119 (4)$ & 4.8 (0.2) & $-43$ (4)& 0.4 (0.1) &133 (4)&  1.0 (0.3) & $-54$ (4) \\
 %
**&PDG &---&--- &--- &--- &--- &--- &---&---&---&---\\
				\hline

 $N (2190)$  7/2$^-$ & 2025 & $1943$ (8) & $161$ (2) & $15$ (1) & $-45$ (1) & $2.9$ (0.3) & $-52$ (1) & $3.6$ (0.5) & $-62$ (1) & $1.1$ (0.2) & $-71$ (2) \\  
 & 2022 & 1965 (12) & 287 (66) & 18 (7) & $-45$ (27)& 2.1 (1) & $-65$ (29) & 2.6 (1.4) & $-78$ (30) & 0.5 (0.2) & $-92$ (31) \\
%
****  &PDG &$2050\pm 100 $ &$400\pm 100 $ &$40\pm20$ &$0\pm 30$ &--- &--- &---&---&---&---\\
				\hline
 
 $N (2250)$ 9/2$^-$ & 2025 & $2194$ (43) & $374$ (65) & $21$ (10) & $-45$ (12) & $3.8$ (1.7) & $-62$ (14) & $0.1$ (0.1) & $-76$ (204) & $1.1$ (0.7) & $-76$ (13) \\  
 & 2022 & 2095 (20) & 422 (26) & 14 (2) & $-67$ (17) & 1.8 (0.2) & $-89$ (9) & 0.3 (0.1) & 80 (9) & 0.4 (0.4) & $-111$ (9) \\
%
****  &PDG &$2150\pm 50$ & $420^{+80}_{-70} $ &$25^{+5}_{-10} $&$-40\pm 20 $ &--- &--- &---&---&---&---\\
				\hline
 
 $N (2220)$ 9/2$^+$ & 2025 & $2009$ (46) & $367$ (41) & $30$ (4) & $-52$ (16) & $2.3$ (0.2) & $-89$ (13) & $1.4$ (0.3) & $-106$ (15) & $0.2$ (0.2) & $-122$ (199) \\  
 & 2022 & 2131 (12) & 388 (12) & 48 (10)  & $-13$ (3) & 4.2 (1.1)   & $-48$ (4)   & 2.0 (0.5)  & $-60$ (4)  & 0.3 (1.6) &$-70$ (4) \\
%
 ****  &PDG &$2150^{+50}_{-20}$ &$400^{+ 80}_{-40}$ &$45^{+15}_{-10} $ &$-40^{+30}_{-20} $ &--- &--- &---&---&---&---\\
\hline\hline
\end {tabular}
}
\end{center}
\label{tab:poles1}
\end{table*}

\begin{table*}
\caption{We list extracted resonance parameters of the $I=3/2$ resonances: Pole positions $W_0$ ($\Gamma_{\rm tot}$ defined as -2Im$W_0$), elastic $\pi N$ residues $(|r_{\pi N}|,\theta_{\pi N\to\pi N})$, and the normalized residues $\pi N\to K\Sigma$
and $\pi N\to\pi\Delta$ with the number in brackets indicating $L$ of the $\pi\Delta$ state. 
We show the results of the present study (``2025") and the J\"uBo2022 results (``2022") for comparison~\cite{Ronchen:2022hqk} and the estimates from the Particle Data Group~\cite{ParticleDataGroup:2024cfk} (``PDG"), if available, as well as the PDG star rating. }

\begin{center}
\renewcommand{\arraystretch}{1.3}
\resizebox{\textwidth}{!}{
\begin {tabular}{ll|ll|ll|ll|ll|ll}  \hline\hline
&&\multicolumn{2}{|l}{Pole position}
 &\multicolumn{2}{|l}{$\pi N$ Residue}
 &\multicolumn{2}{|l}{$K\Sigma$ channel} 
 &\multicolumn{2}{|l|}{$\pi\Delta$, channel (6)}
 &\multicolumn{2}{|l}{$\pi\Delta$, channel (7)}
\bigstrut[t]\\[0.1cm]
&&\multicolumn{1}{|l}{Re $W_0$ \hspace*{0.5cm} }
& \multicolumn{1}{l|}{$-$2Im $W_0$\hspace*{0.1cm} }
& \multicolumn{1}{l}{$|r_{\pi N}|$\hspace*{0.cm}} 
& \multicolumn{1}{l}{$\theta_{\pi N\to\pi N}$ } 
& \multicolumn{1}{|l}{$\displaystyle{\frac{\Gamma^{1/2}_{\pi N}\Gamma^{1/2}_{K\Sigma}}{\Gamma_{\rm tot}}}$}
& \multicolumn{1}{l|}{$\theta_{\pi N\to K\Sigma}$\hspace*{0.1cm}}
& \multicolumn{1}{l}{$\displaystyle{\frac{\Gamma^{1/2}_{\pi N}\Gamma^{1/2}_{\pi\Delta}}{\Gamma_{\rm tot}}}$} 
& \multicolumn{1}{l|}{$\theta_{\pi N\to \pi\Delta}$\hspace*{0.1cm}}
& \multicolumn{1}{l}{$\displaystyle{\frac{\Gamma^{1/2}_{\pi N}\Gamma^{1/2}_{\pi\Delta}}{\Gamma_{\rm tot}}}$} 
& \multicolumn{1}{l}{$\theta_{\pi N\to \pi\Delta}$}
\\
&&\multicolumn{1}{|l}{[MeV]} & \multicolumn{1}{l|}{[MeV]} & \multicolumn{1}{l}{[MeV]} & \multicolumn{1}{l}{[deg]} 
& \multicolumn{1}{|l}{[\%]}  & \multicolumn{1}{l|}{[deg]} & \multicolumn{1}{l}{[\%]}  & \multicolumn{1}{l|}{[deg]} &\multicolumn{1}{l}{[\%]} & \multicolumn{1}{l}{[deg]} \\
		  & fit &&&&&&&&
\bigstrut[t]\\
\hline

 $\Delta(1620)$	1/2$^-$ & 2025 & $1601$ (2) & $79$ (1) & $17$ (3) & $-72$ (19) & $18$ (3) & $-65$ (20) & --- & --- & $51$ (9) {\footnotesize (D)} & $133$ (19) \\  
  & 2022 & 1607 (4) & 85 (5) & 12 (2) & $126$ (4) & 11 (2) & $-120$ (5) & --- &---& 32 (2) {\footnotesize (D)} & 81 (2) \\
%
****& PDG &$1600\pm 10$ &$ 110\pm 30$ & $15\pm5$ &$-100\pm 20$& ---&---&---&---&---&---\\
				\hline
 	 			
$\Delta(1910)$ 1/2$^+$  & 2025 & $1813$ (6) & $653$ (25) & $27$ (21) & $129$ (306) & $0.5$ (1.4) & $-100$ (40) & $15$ (12) {\footnotesize (P)} & $-3$ (63) & --- & --- \\  
&2022 & 1802 (11) & 550 (22) &35 (25) & 93 (14) &0.2 (0.4) & 138 (19)&  24 (18) {\footnotesize (P)}& $-42(14)$ &--- &---\\
%
 ****	& PDG & $1850\pm 50$& $350\pm 150 $ &$25\pm 5$&$-90^{+180}_{-90}$&---&---&---&---&---&--- \\
				\hline
 	 			

$\Delta(1232)$ 3/2$^+$ & 2025 & $1215$ (1) & $92$ (0) & $48$ (0) & $-39$ (0) &   &   &   &   &   &  \\  
&2022 & 1215 (2) & 93 (1)& 50 (2) & $-39$ (1) &&& & \\
%
 ****	& PDG &$1210\pm1$ &$ 100\pm 2$ &$50^{+2}_{-1}$ & $-46^{+1}_{-2}$&&&&& \\
				\hline

 $\Delta(1600)$ 3/2$^+$ & 2025 & $1598$ (1) & $111$ (1) & $6.0$ (0.8) & $-110$ (4) & $9.5$ (1.6) & $8$ (4) & $22$ (4) {\footnotesize (P)} & $87$ (5) & $0.6$ (0.1) {\footnotesize (F)} & $-31$ (9) \\  
&2022 & 1590 (1) & 136 (1) & 11 (1) & $-106$ (2) & 14 (1) & 14 (2) & 30 (3) {\footnotesize (P)}  & 87 (3) & 0.4 (0.04) {\footnotesize (F)}  &$-62$ (9)\\
%
 ****	& PDG &$1520^{+70}_{-50} $ &$280^{+40}_{-30} $ &$25\pm15 $ &$210^{+40}_{-30} $&---&---&---&---&
$1\pm 0.5$&---\\
				\hline

$\Delta(1920)$	3/2$^+$ & 2025 & $1857$ (15) & $875$ (4) & $60$ (8) & $-28$ (19) & $25$ (2.1) & $82$ (13) & $8.0$ (0.8) {\footnotesize(P)}  & $-74$ (13) & $3.3$ (0.5) {\footnotesize(F)}  & $96$ (18) \\  
&2022 & 1883 (4) & 844 (10) & 41 (5) & 11 (7) & 20 (2) & 104 (4) & 5.7 (0.5) {\footnotesize(P)} & $-48$ (5) & 2.0 (0.3) {\footnotesize (F)} & 147 (7)\\
***	& PDG &$1900\pm 50$ &$300\pm 100 $ &$25\pm 10$&$-100\pm 50$&---&---&---&---&---&--- \\
				\hline

$\Delta(1700)$ 3/2$^-$ & 2025 & $1663$ (4) & $354$ (13) & $47$ (13) & $-11$ (20) & $3.1$ (0.7) & $-164$ (15) & $7.7$ (1.7)  {\footnotesize (D)} & $148$ (19) & $59$ (15) {\footnotesize (S)} & $146$ (19) \\  
&2022 & 1637 (64) & 295 (58) & 15 (23) & $-13$ (147) & 0.7 (1.5) & $-176$ (320) &3.8 (7.8)  {\footnotesize (D)}& 127 (254) & 20 (29) {\footnotesize (S)}  & 146 (266)  \\
%
 ****& PDG &$1665\pm 25 $ &$250\pm 50 $ &$25\pm 15$&$-20\pm20 $&---&---&---&---&---&--- \\
				\hline

 $\Delta(1930)$	5/2$^-$ & 2025 & $1835$ (11) & $485$ (20) & $20$ (2) & $-123$ (7) & $0.7$ (0.1) & $28$ (6) & $11$ (2.6) {\footnotesize (D)} & $47$ (6) & $1.2$ (0.1) {\footnotesize (G)} & $147$ (6) \\  
&2022 & 1821 (4)& 447 (13) & 15 (3)  & $-108$ (9) & 1.0 (0.2)  & 49 (9)  & 12 (3)  {\footnotesize (D)}  & 64 (7)  & 0.8 (0.2) {\footnotesize (G)} & 148 (4) \\
%
 *** & PDG &$1850\pm 30 $ &$320^{+130}_{-20}$ &$14\pm 6 $&$-50^{+40}_{-50} $&---&---&---&---&---&--- \\
				\hline
 	 			
 $\Delta(1905)$	5/2$^+$ & 2025 & $1722$ (4) & $264$ (11) & $5.1$ (3.5) & $-65$ (18) & $0.3$ (0.2) & $-179$ (356) & $1.8$ (1.3)  {\footnotesize (F)}  & $47$ (19) & $9.4$ (3.2) {\footnotesize (P)} & $-105$ (22) \\  
&2022 & 1707 (1) & 127 (8) & 3.7 (1.0) & $-92$ (12) & 0.2 (0.03) & 154 (11) & 1.7 (0.3)  {\footnotesize (F)} & 18 (15) & 10 (1) {\footnotesize (P)}& $-109$ (14)\\
%
 ****	& PDG &$ 1770^{+30}_{-20}$ &$300 \pm 40 $ &$20\pm 5$&$-45^{+15}_{-75} $&---&---&---&---&---&--- \\
				 \hline

 $\Delta(1950)$	7/2$^+$ &2025& $1871$ (1) & $170$ (7) & $38$ (3.3) & $8$ (9) & $1.9$ (0.5) & $-43$ (8) & $45$ (4) {\footnotesize (F)}  & $168$ (8) & $2.9$ (0.2) {\footnotesize (H)} & $47$ (9)\\  
&2022 & 1875 (1) & 166 (3) &  27 (2) & 1.1 (2.0) & 2.0 (0.3) & $-40$ (7) & 30 (54) {\footnotesize (F)} & $166$ (2) & 5.1 (0.7) {\footnotesize (H)}& $-11$ (2) \\
%
**** 	& PDG &$1880\pm 10$ &$240\pm 20 $ &$52\pm 8 $ &$ -32\pm8$ &---&---&---&---&---&--- \\
				\hline

$\Delta (2200)$ 7/2$^-$ & 2025 & $1978$ (7) & $327$ (12) & $7.3$ (0.5) & $-76$ (3) & $0.0$ (0.0) & $64$ (3) & $0.4$ (0.1) {\footnotesize (G)}  & $100$ (18) & $18$ (1.1) {\footnotesize (D)} & $100$ (4) \\  
&2022 & 1963 (2) & 328 (3) & 6.8 (0.6) & $-80$ (2) & $<0.1$ (0.03) & $-123$ (2) & 0.3 (0.1) {\footnotesize (G)} & $152$ (5) & 16 (1) {\footnotesize (D)}& 100 (2)\\
%
 ***		& PDG &$2100\pm 50$ &$340\pm 80$ &---&---&---&---&---&---&---&--- \\
				 \hline

 $\Delta (2400)$ 9/2$^-$ & 2025 & $2517$ (49) & $86$ (37) & $25$ (2) & $6$ (6) & $4.5$ (3.4) & $15$ (7) & $65$ (29) {\footnotesize (G)}  & $16$ (7) & $12$ (8) {\footnotesize (I)} & $154$ (11) \\  
& 2022 & 2458 (3) & 280 (2) & 5.4 (5) & 8.4 (33) & 0.4 (0.6) & 17 (30) & 10 (11) {\footnotesize (G)} & 17 (23) & 1.9 (0.5) {\footnotesize (I)}& $-120$ (49) \\
%
** 	& PDG &---&--- &---&---&---&---&---&---&---&--- \\

\hline\hline
\end {tabular}
}
\end{center}
\label{tab:poles2}
\end{table*}

\begin{table*}
\caption{We list the photocouplings at the pole ($A^h_{pole}$, $\vartheta^h$) of the $I=1/2$ (left) and $I=3/2$ resonances (right). 
We show the results of the present study (``2025") and the J\"uBo2022 results (``2022") for comparison~\cite{Ronchen:2022hqk}.
The uncertainties quoted in parentheses provide a rather rough estimate as explained in the text.  
}
\begin{center}
\renewcommand{\arraystretch}{1.5}
\resizebox{2.07\columnwidth}{!}{
\begin {tabular}{ll| cc|cc || l l |cc|cc} 
\hline\hline
& & \multicolumn{1}{c}{$\mathbf{A^{1/2}_{pole}}$\hspace*{0.2cm}} 
& \multicolumn{1}{c|}{$\mathbf{\vartheta^{1/2}}$ } 
& \multicolumn{1}{c}{$\mathbf{A^{3/2}_{pole}}$\hspace*{0.2cm}} 
& \multicolumn{1}{c||}{$\mathbf{\vartheta^{3/2}}$ } 
& & & \multicolumn{1}{c}{$\mathbf{A^{1/2}_{pole}}$\hspace*{0.2cm}} 
& \multicolumn{1}{c|}{$\mathbf{\vartheta^{1/2}}$ } 
& \multicolumn{1}{c}{$\mathbf{A^{3/2}_{pole}}$\hspace*{0.2cm}} 
& \multicolumn{1}{c}{$\mathbf{\vartheta^{3/2}}$ } 
\bigstrut[t]\\[0.2cm]
&& \multicolumn{1}{c}{{\footnotesize[$10^{-3}$ GeV$^{-\nicefrac{1}{2}}$]}} &\multicolumn{1}{c|}{\footnotesize[deg]} & \multicolumn{1}{c}{\footnotesize[$10^{-3}$ GeV$^{-\nicefrac{1}{2}}$]} & \multicolumn{1}{c||}{\footnotesize[deg]} 
&&& \multicolumn{1}{c}{\footnotesize[$10^{-3}$ GeV$^{-\nicefrac{1}{2}}$]} &\multicolumn{1}{c|}{\footnotesize[deg]} & \multicolumn{1}{c}{\footnotesize[$10^{-3}$ GeV$^{-\nicefrac{1}{2}}$]} & \multicolumn{1}{c}{\footnotesize[deg]} 
 \\
		  & fit 		&&&&& &fit &	&
\bigstrut[t]\\
\hline
$N (1535)$ 1/2$^-$ & 2025 & $90$ (2) & $-1$ (2) &  &  & $\Delta(1620)$	1/2$^-$ & 2025 & $30$ (11) & $58$ (12) &  &  \\ 
& 2022 & 84 (5) &$-12$ (3) &&&  & 2022 & 11 (4)& $57$ (24) \\
 \hline
 
 $N (1650)$ 1/2$^-$  & 2025 & $34$ (3) & $-2$ (5) & & & $\Delta(1910)$ 1/2$^+$  &  2025 &  $-164$ (128) & $-5$ (352) & & \\
 & 2022 &39 (10) &$-0.2$ (27)&&&	& 2022 & $-446$ (72) & $-70$ (21)\\ 			
 \hline

 $N (1440)$ 1/2$^+_{(a)}$ & 2025 &$-99$ (15) & $-18$ (4) & & &  $\Delta(1232)$ 3/2$^+$  & 2025 &  $-128$ (5) & $-18$ (3) & $-246$ (3) & $1$ (1)  \\
 & 2022 & $-90$ (13) & $-30$ (5) &&&    & 2022& $-126$ (4)  & $-18$ (3)  & $-245$ (7) & $-0.7$ (1.7)  \\
  \hline

 $N (1710)$  1/2$^+$ & 2025 &$-18$ (2) & $37$ (10) & & & $\Delta(1600)$ 3/2$^+$ & 2025 &  $9$ (4) & $11$ (45) & $-10$ (6) & $105$ (43) \\
 & 2022 & $-18$ (19) & $40$ (109)  && & & 2022  & 25 (10)& 0.5 (5.9)& $-6.0$ (2.6) & $62$ (63) \\ 				 
  \hline

 $N (1720)$ 3/2$^+$ & 2025 & $24$ (10) & $48$ (13) & $-26$ (8) & $-24$ (13) & $\Delta(1920)$	3/2$^+_{(a)}$ & 2025 &  $245$ (30) & $-28$ (9) & $475$ (52) & $10$ (10)\\
 & 2022 & 39 (7) & 60 (10) & $-25$ (7) & $-5.7$ (13) &
  & 2022 & 138 (12) & $-8.9$ (3.9) & 252 (14) & 14 (3) \\
   \hline

$N (1900)$ 3/2$^+$ & 2025 & $6$ (1) & $49$ (21) & $-39$ (2) & $-28$ (3) & $\Delta(1700)$ 3/2$^-$ & 2025 & $244$ (53) & $17$ (11) & $337$ (67) & $-4$ (13)\\
& 2022 & 9.1 (2.7)& 80 (23)& $-7.7$ (3.4)  & $-42$ (23)  &
& 2022 & 163 (120) & $-4.4$ (78)  & 221 (185)  & $-12$ (79)   \\			
\hline

$N (1520)$ 3/2$^-$ &2025 &  $-19$ (3) & $-34$ (8) & $102$ (12) & $17$ (4)& $\Delta(1930)$	5/2$^-$  & 2025 & $218$ (18) & $162$ (24) & $240$ (31) & $173$ (10)\\
& 2022 & $-43$ (25) & $-47$ (20) & 112 (64) &	1.8 (37)	&  			& 2022 &104 (18)& 129 (16) & 322 (44) & 142 (7) \\ 
\hline

 $N (1675)$ 5/2$^-$ & 2025 & $35$ (5) & $17$ (6) & $36$ (4) & $15$ (8) & $\Delta(1905)$	5/2$^+$ & 2025 & $75$ (61) & $-67$ (40) & $-373$ (308) & $94$ (16) \\
 & 2022& 25 (4) & $-1.2$ (7.8) & 51 (4) & $-1.0$ (3.7)&		& 2022 & 55 (8)& $-159$ (3) & $-168$ (40) & $172$ (1.7)  \\
\hline

 $N (1680)$ 5/2$^+$ & 2025 & $-8$ (2) & $119$ (23) & $133$ (34) & $-31$ (8) &  $\Delta(1950)$	7/2$^+$ & 2025 & $-33$ (7) & $-67$ (6) & $-59$ (8) & $-70$ (5) \\
 & 2022 & $-17$ (6) & 70 (14) & 95 (6)& $-57$ (7) & 				  & 2022& $-31$ (4) & $-81$ (7) & $-45$ (4) & $-89$ (4) \\
\hline


  $N (1990)$ 7/2$^+$ & 2025 & $-14$ (3) & $-74$ (20) & $-14$ (3) & $107$ (15)  &  $\Delta (2200)$ 7/2$^-$ &  2025 &  $72$ (17) & $-134$ (8) & $65$ (13) & $-174$ (7) \\
  & 2022 & $-30$ (16) & $-135$ (25) & $-18$ (11) & 53 (32)  &  		 
& 2022 & 104 (22) & $-139$ (3) & 21 (25) & $-180$ (39)\\	 
 \hline

 $N (2190)$  7/2$^-$ & 2025 & $-13$ (2) & $11$ (15) & $20$ (7) & $161$ (15) & $\Delta (2400)$ 9/2$^-$ & 2025 & $15$ (9) & $-102$ (52) & $22$ (22) & $156$ (23) \\
 & 2022 & $-15$ (8) & 111 (17)& 62 (22) & 179 (26) &
  & 2022 & 21 (14) & $-67$ (23) & 22 (14) & 122 (14) \\	
  \hline

				 
 $N (2250)$ 9/2$^-$ &  2025 & $-33$ (8) & $2$ (19) & $58$ (20) & $121$ (20)\\
 & 2022 & $-108$ (14) & 112 (7) & 50 (22) & 69 (16)  \\
 \hline
				 
 $N (2220)$ 9/2$^+$ & 2025 & $74$ (46) & $-78$ (17) & $-185$ (37) & $-14$ (16) \\
 & 2022 & 357 (39) & $-91$ (7) & $-273$ (50) & $-102$ (6) \\

\hline\hline
\end {tabular}
}
\end{center}
\label{tab:photo}
\end{table*}

\section{GDH sum rule}
\label{sec:GDH_sum_rule}

Based on fundamental physics principles such as Lorentz invariance, crossing symmetry, gauge invariance, unitarity, causality and rotational invariance, the Gerasimov-Drell-Hearn (GDH) sum rule (originally formulated in Refs.~\cite{Gerasimov:1965et,Drell:1966jv}) provides a general relation between the difference of the helicity-dependent photoproduction cross sections $\Delta \sigma=\sigma_{3/2}-\sigma_{1/2}$ and the anomalous magnetic moment $\kappa$ of the target particle:
\begin{equation}
    I_{\text{GDH}}=\int_{E_\gamma^{0}}^\infty \dd{E_\gamma}\frac{\Delta \sigma(E_\gamma)}{E_\gamma}=\frac{4\pi^2 S\alpha\kappa^2}{M^2}
    \label{GDH_sum_rule}
\end{equation}
where $M$ is the mass of the target particle, $S$ is its spin, $E_\gamma^0$ the pion-photoproduction threshold and $\alpha=e^2/4\pi$ is the fine-structure constant in terms of the electromagnetic coupling constant $e$. 

In the current study we only consider proton targets, the extension of the J\"uBo model to include neutron photoproduction is planned for the future. The anomalous magnetic moment for the proton is given by $\kappa_p=\mu_p - 1\approx 1.793\mu_N$~\cite{ParticleDataGroup:2024cfk} with the nuclear magneton $\mu_N$. With that the right-hand side of Eq.~\eqref{GDH_sum_rule} evaluates  to $I^p_{\text{GDH}}=\SI{204.78}{\micro\barn}$.

The J\"uBo coupled channel approach allows to extract the contributions to this sum rule of the $\pi N$, $\eta N$, $K\Lambda$ and $K\Sigma$ channels individually. Similar analyses were done in Ref.~\cite{Strakovsky:2022tvu} for single pion photoproduction channels, and there are also analyses considering $\eta N$, $\pi\pi N$ or $KY$ channels such as  Refs.~\cite{Mart:2019fau,Drechsel:2004ki,Zhao:2002wf}. While the J\"uBo approach includes effective $\pi\pi N$ channels in the purely hadronic amplitude, no $\pi\pi N$ photoproduction data are taken into account. Therefore we cannot determine the contribution of this channel to the GDH sum rule directly. 

Since a numerical evaluation of the left-hand side of the integral in Eq.~\eqref{GDH_sum_rule} with an upper limit of infinity is not possible, one defines the so-called {\it running} GDH integral as
\begin{equation}
       I^p_{\text{GDH}}(E_\gamma)=\int_{E_\gamma^{0}}^{E_\gamma} \dd{E'_\gamma}\frac{\Delta \sigma(E'_\gamma)}{E'_\gamma}~.
    \label{Running_GDH_sum_rule}
\end{equation}
where $I^p_{\text{GDH}}$ is now a function of the upper limit of the integral $E_\gamma$.

\begin{figure*}
\begin{center}
\includegraphics[width=0.8\linewidth]{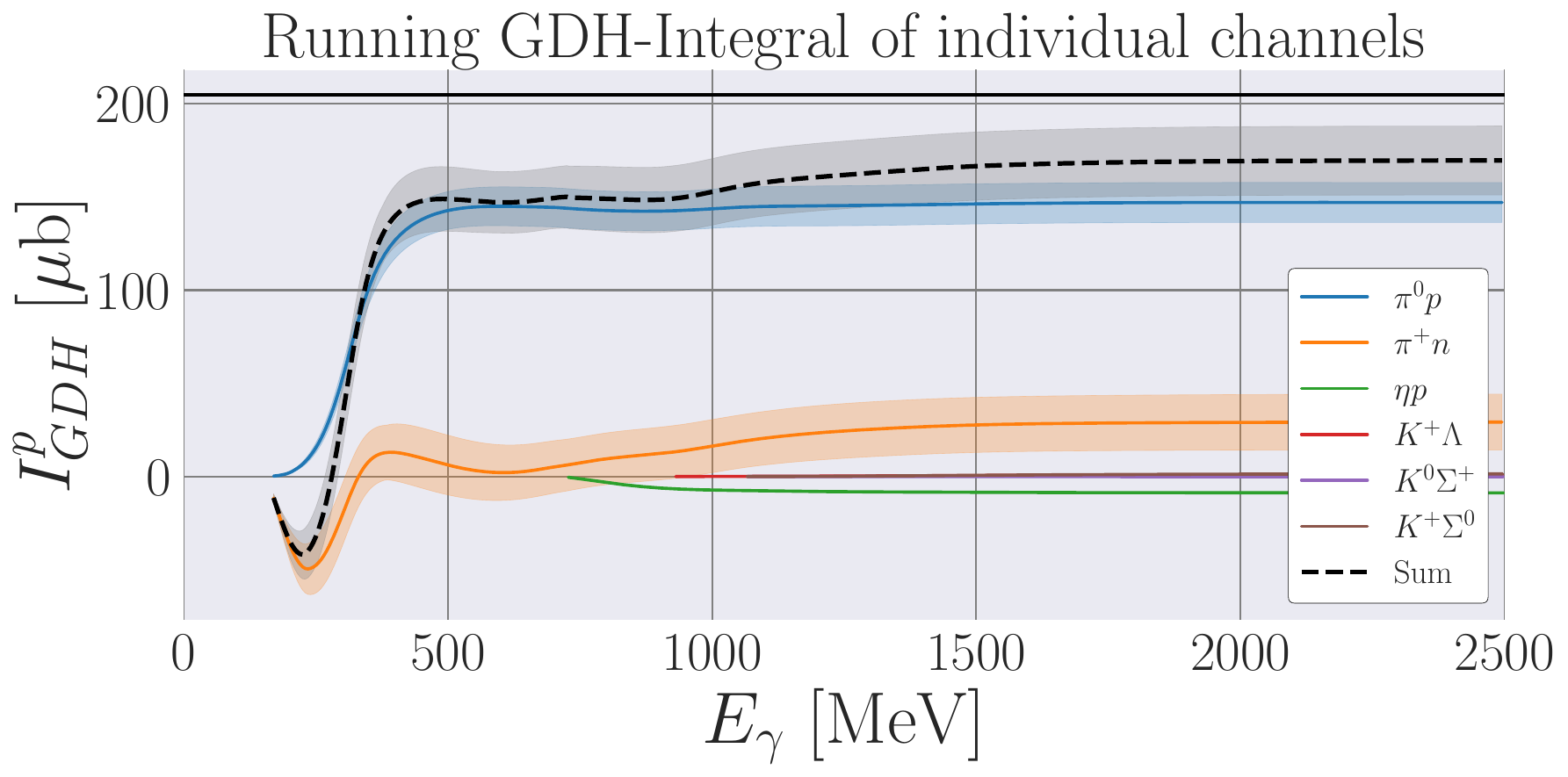} 
\end{center}
\caption{Running GDH integral $I^p_{\text{GDH}}$ as in Eq.~\eqref{Running_GDH_sum_rule} for the different channels with their respective errorbands. The black horizontal line shows the value of $I^p_{\text{GDH}}=\SI{204.78}{\micro\barn}$ calculated from PDG values. The black dashed line with the grey errorband shows the sum of all channels.}
\label{Fig:Running_GDH}
\end{figure*}

In Fig.~\ref{Fig:Running_GDH} we show the individual channel contributions to this running GDH integral from the present analysis with the corresponding uncertainties extracted from the refits described in Sec.~\ref{sec:numerics}. The black dashed line represents the sum of all channels, and the horizontal dotted line the right hand side value of the GDH sum rule of $I^p_{\text{GDH}}=\SI{204.78}{\micro\barn}$. The main contribution is given by the $\pi^0 p$ channel, followed by  $\pi^+ n$. 

All of the 6 channel contributions to the GDH sum rule saturate for high energies to the following values:
\begin{eqnarray}
    I_{\text{GDH}}^p(\pi^0 p)&=&147\pm 7~\si{\micro\barn}~,\nonumber\\
    I_{\text{GDH}}^p(\pi^+ n)&=&29\pm 15~\si{\micro\barn}~,\nonumber\\
    I_{\text{GDH}}^p(\eta p)&=&-8.8\pm 0.1~\si{\micro\barn}~,\nonumber\\
    I_{\text{GDH}}^p(K^+\Lambda^0)&=&0.80\pm 0.05~\si{\micro\barn}~,\nonumber\\
    I_{\text{GDH}}^p(K^0\Sigma^+)&=&-0.12\pm 0.05~\si{\micro\barn}~,\nonumber\\
    I_{\text{GDH}}^p(K^+\Sigma^0)&=&1.42\pm 0.05~\si{\micro\barn}~,\nonumber\\
    I_{\text{GDH}}^p(\text{all})&=&170\pm 19~\si{\micro\barn}~.
\end{eqnarray}

As for the resonance parameters we determine the uncertainties of the contributions to the GDH sum rule for the different channels as explained in \ref{sec:numerics}. As a consequence of this, the uncertainties of the individual channels are necessarily correlated. Therefore, the uncertainty of the sum of all channels is determined by the respective spread of the sum of the integrals.

The contribution of the two single pion channels determined in Ref.~\cite{Strakovsky:2022tvu} was $183.4~\si{\micro\barn}$. The $\pi^0 p$-channel contribution was $\sim 155~\si{\micro\barn}$ and the $\pi^+ n$-channel contributed $\sim 30~\si{\micro\barn}$. Comparing this to our calculated contributions, we observe that our $\pi^+ n$ contribution is in good agreement while our $\pi^0p$ channel is giving a slightly smaller contribution. We also observe a relatively large uncertainty in these two channels, especially for $\pi^+ n$ where less data is available.

\begin{figure*}
\begin{center}
\includegraphics[width=0.8\linewidth]{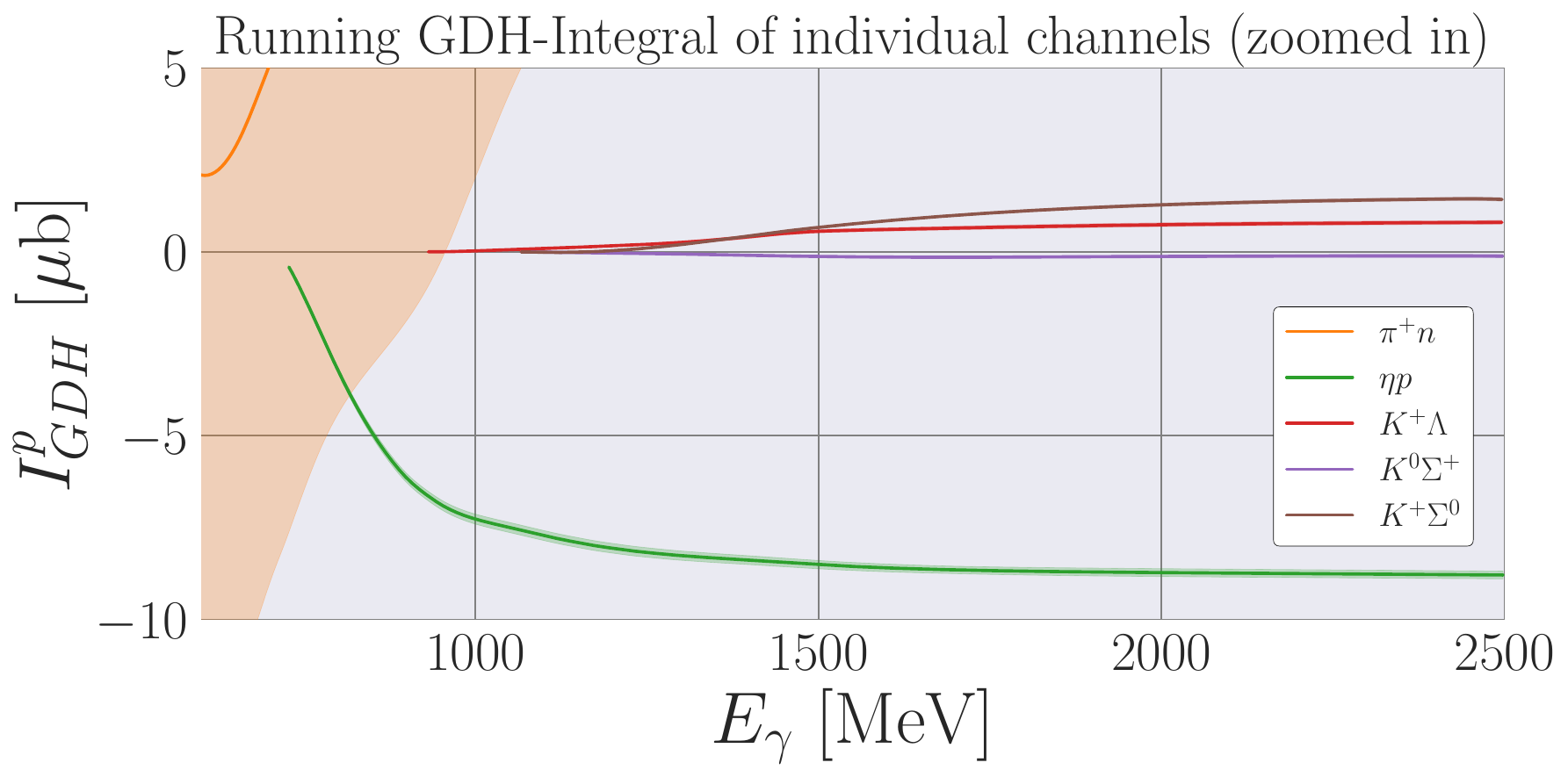} 
\end{center}
\caption{Zoom into the running GDH integral $I^p_{\text{GDH}}$ as in Eq.~\eqref{Running_GDH_sum_rule} for the different channels for better visualization of the higher lying channels.}
\label{Fig:Running_GDH_Zoom}
\end{figure*}

To better visualize the small contributions of the channels $\eta p$, $K^+\Lambda^0$, $K^0\Sigma^+$ and $K^+\Sigma^0$, we add a zoomed-in version in Fig.~\ref{Fig:Running_GDH_Zoom} and present all channels separately in Fig.~\ref{fig:GDH_individual_channel_with_errorband} in the Appendix. We can observe that the $\eta p$ channel is negative and provides the main contribution to these higher lying channels. However, compared to the $\pi N$ channels the other four have only a marginal contribution of $\sim \SI{-6.69}{\micro\barn}$ and even lower the sum of all channels. In some other studies as, e.g., Refs.~\cite{Mart:2019fau,Drechsel:2004ki,Zhao:2002wf} all of those four channels give negative contributions. 
This is not the case in our study, as can be seen for $K^+ \Lambda^0$ and $K^+ \Sigma^0$. Note that no data on $\Delta\sigma$ or $E$ is available for the $KY$ channels yet. The corresponding predicted values for the GDH integral may therefore be subject to even greater uncertainties than suggested by the error band.

The full sum of all channels saturates to the value $I_{\text{GDH}}^p(\text{all})=170\pm 19~\si{\micro\barn}$. This leaves a difference to the right hand side value of $\Delta I^p_\text{GDH}=35\pm 19~\si{\micro\barn}$ to the central value of our sum (black dashed line in Fig.~\ref{Fig:Running_GDH}), which likely corresponds to contributions of channels not considered explicitly in the JüBo model, first and foremost the $\pi\pi N$ channel that exhibits a large total cross section at energies beyond the $\Delta(1232)$ region. Contributions from other missing channels like $\omega N$ or $\eta^\prime N$ are likely less significant.

This hypothesis is supported by other analyses. In Ref.~\cite{Strakovsky:2022tvu} the contributions of $\pi^0 p$ and $\pi^+ n$ to the GDH-sum rule were determined and they found a missing contribution of $\sim \SI{21}{\micro\barn}$. 
In a direct measurement of two-pion photoproduction in Ref.~\cite{GDH:2003fjc}, the contribution of the channel $\gamma p \rightarrow \pi^0 \pi^+ n$ was found to be only  $(11.3\pm0.7\pm0.7)~\si{\micro\barn}$, but only for photon energies up to $E_\gamma=800~\si{\mega\electronvolt}$. 
As mentioned by the authors, this contribution did not saturate yet at that energy. In contrast, in the analyses of Refs.~\cite{Drechsel:2004ki,Hirata:2002tp} a contribution of the $\pi\pi N$ channel over the full energy range was found to be $I_{\text{GDH}}^p(\pi\pi N)=\SI{28}{\micro\barn}$.
Considering our uncertainties on $I_{\text{GDH}}^p(\text{all})$ such values would be enough to reproduce the right hand side of the GDH sum rule of $I^p_{\text{GDH}}=\SI{204.78}{\micro\barn}$ in our analysis. Thus, we conclude that our missing contribution is indeed originating from the missing $\pi\pi N$-channel. Once the model is extended to 2$\pi$ photoproduction off the nucleon, we will be able to verify this assumption quantitatively.

\section{Summary \& Outlook}

In this study, we presented an updated fit result with new data sets for $\gamma p\rightarrow \pi^0 p,\, \pi^+ n$ and $\eta p$ using the J\"ulich-Bonn dynamical coupled-channel model. The current analysis uses more than $73,000$ data points and we fit pion- and photon-induced reactions simultaneously. The spectrum of $N^*$ and $\Delta$ resonances was extracted as complex poles on the unphysical Riemann sheet. 

The new $\eta p$ data from LEPS2/BGOegg collaboration led to an improved description of the backwards peak above $2$\,GeV. 
The current fit also improved a lot for the double polarization ${G}$ for the reaction $\gamma p \rightarrow \pi^+ n$, since our database for this observable was more than tripled by the new dataset from CLAS.

Based on this fit result, we extracted the individual channel contributions to the GDH sum rule for the proton. We found that the channels $\eta p,\, K^+\Lambda^0,\, K^0\Sigma^+,$ and $ K^+\Sigma^0$ all together only contribute marginally compared to the $\pi N$ contributions. The channels considered in this analysis saturates the GDH sum rule to 83\%. The missing part is most probably from the $\pi\pi N$ channels not included in this evaluation.


We plan to extend our model to include photoproduction processes on neutron targets in the near future and also extract the GDH sum rule contributions for these processes.

\section*{Acknowledgements}


The authors gratefully acknowledge computing time on the supercomputer JURECA~\cite{JURECA} at Forschungszentrum Jülich under grant no. {\it baryonspectro}.
This work is supported by the MKW NRW under the funding code No.~NW21-024-A and by the Deutsche Forschungsgemeinschaft (DFG, German Research Foundation) as part of the CRC 1639 NuMeriQS – project no. 511713970.
This work is further supported in part by the EU Horizon 2020 research and innovation program, STRONG-2020 project, and
also funded by the Deutsche Forschungsgemeinschaft (DFG, German Research Foundation) – 491111487. 
In addition, UGM and CH thank the CAS President’s International Fellowship Initiative (PIFI) under Grant Nos.
2025PD0022 and 2025PD0087, respectively, for partial
support.
\appendix

\section{New Data for $\Delta \sigma$}

\begin{figure}[htbp]
\begin{center}
\includegraphics[width=1.\linewidth]{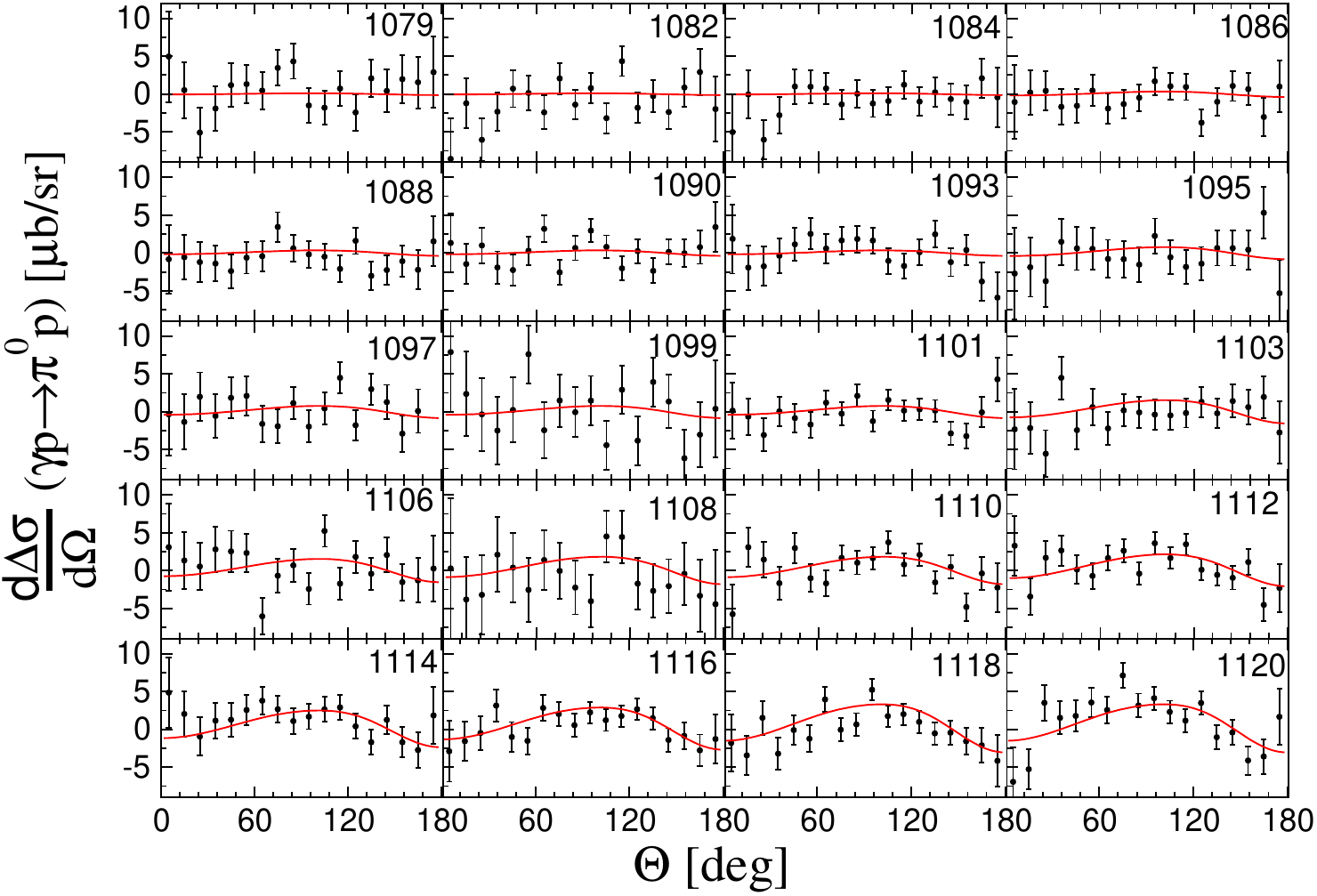} 
\includegraphics[width=1.\linewidth]{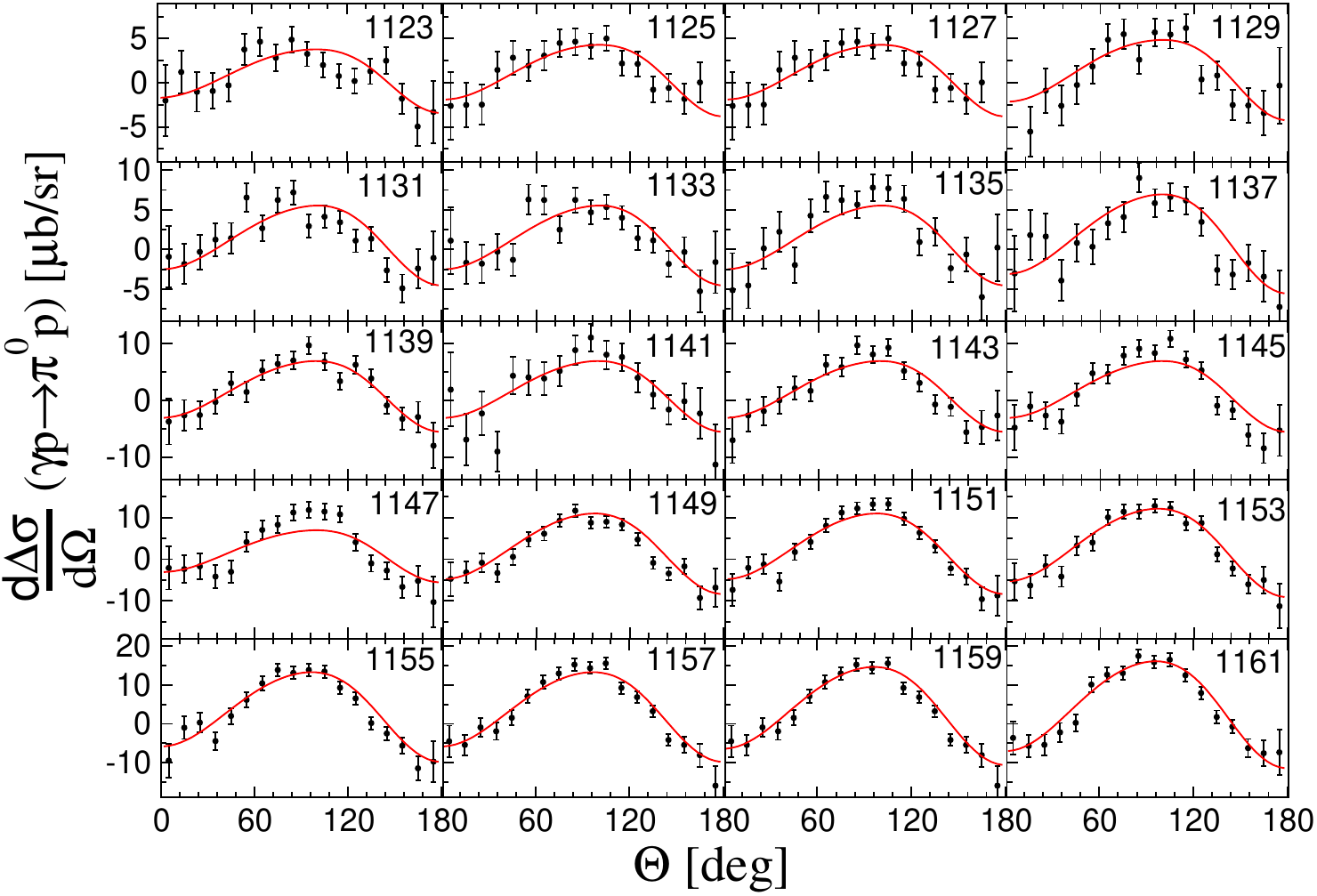} 
\includegraphics[width=1.\linewidth]{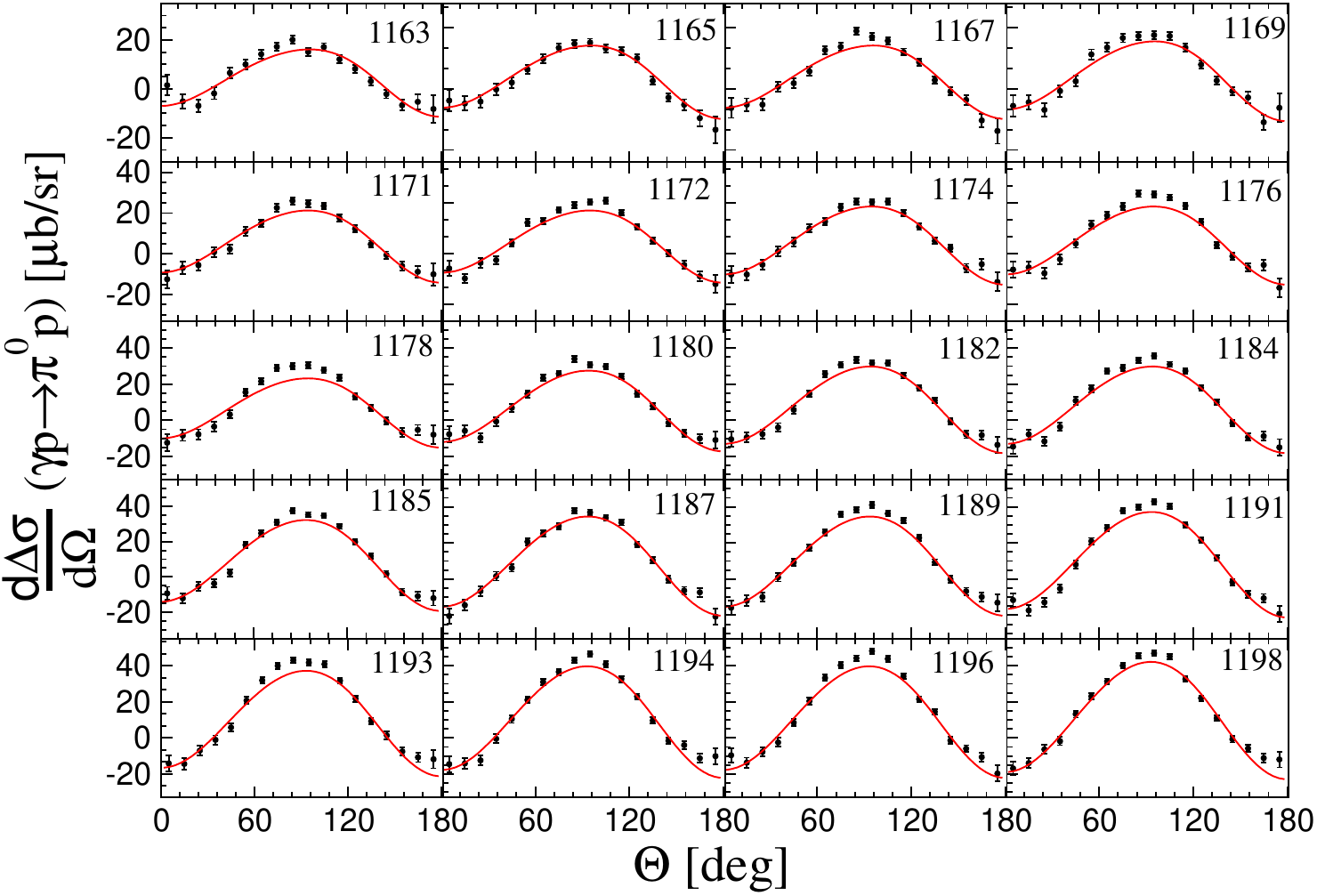} 
\end{center}
\caption{Solution JüBo2025 (red)  for the helicity dependent differential cross section from Ref.~\cite{A2CollaborationatMAMI:2023twj} for the process $\gamma p\to \pi^0 p$. (Data not included in the fit.) The numbers in each plot denote the center of mass energy in MeV.}
\label{fig:delta13_A2Mami}
\end{figure}

\begin{figure}[htbp]
\begin{center} 
\includegraphics[width=1.\linewidth]{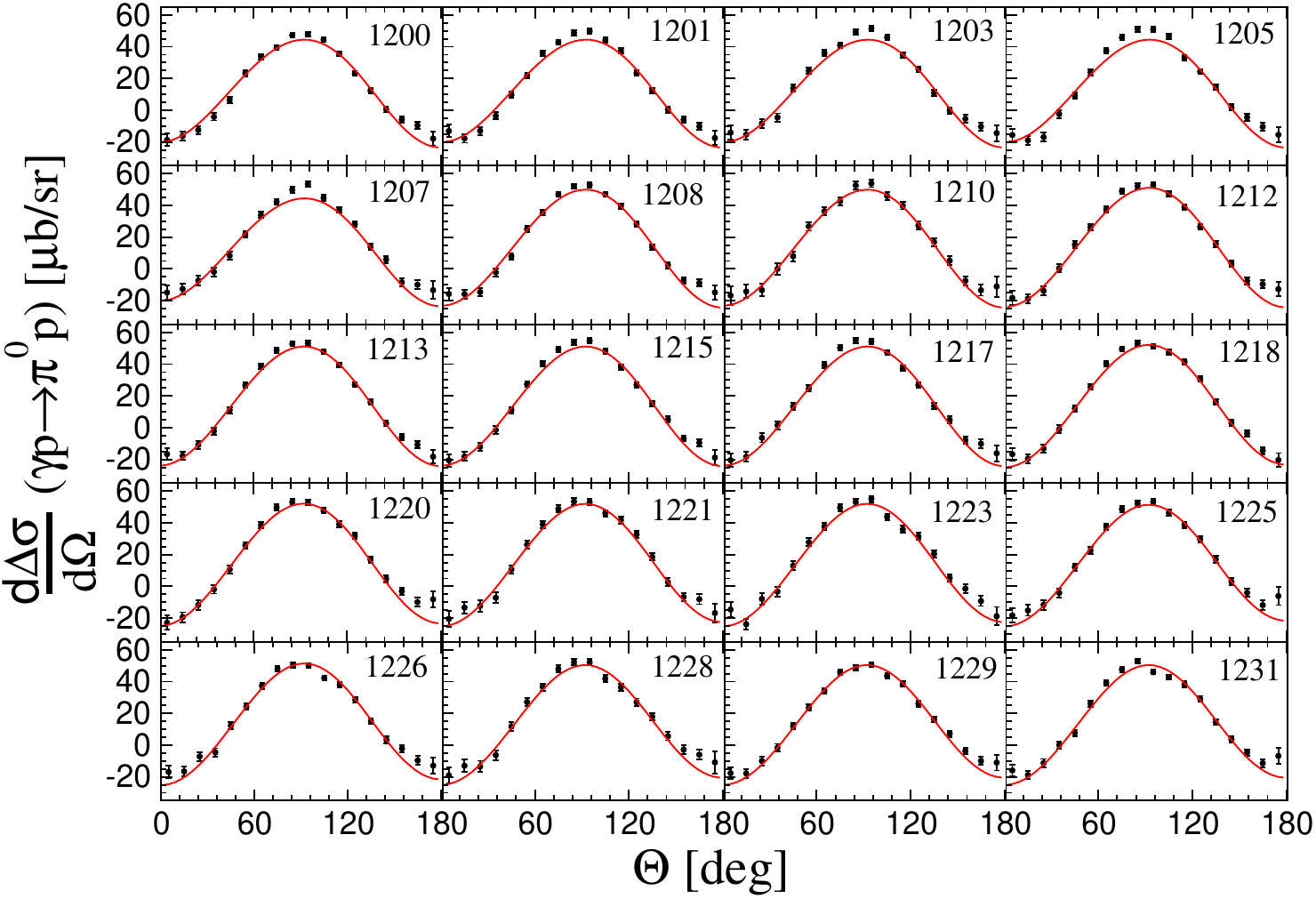} 
\includegraphics[width=1.\linewidth]{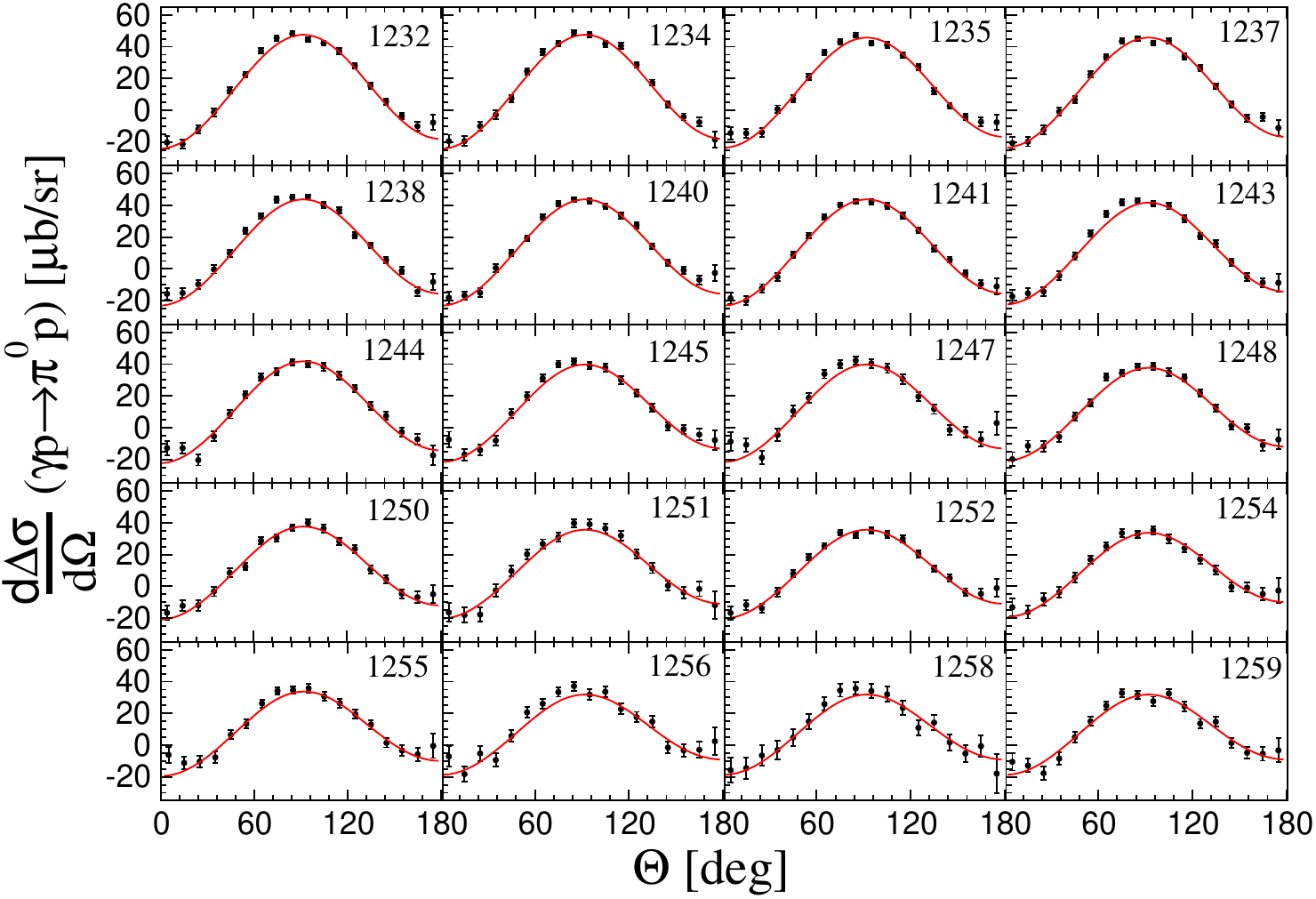} 
\includegraphics[width=1.\linewidth]{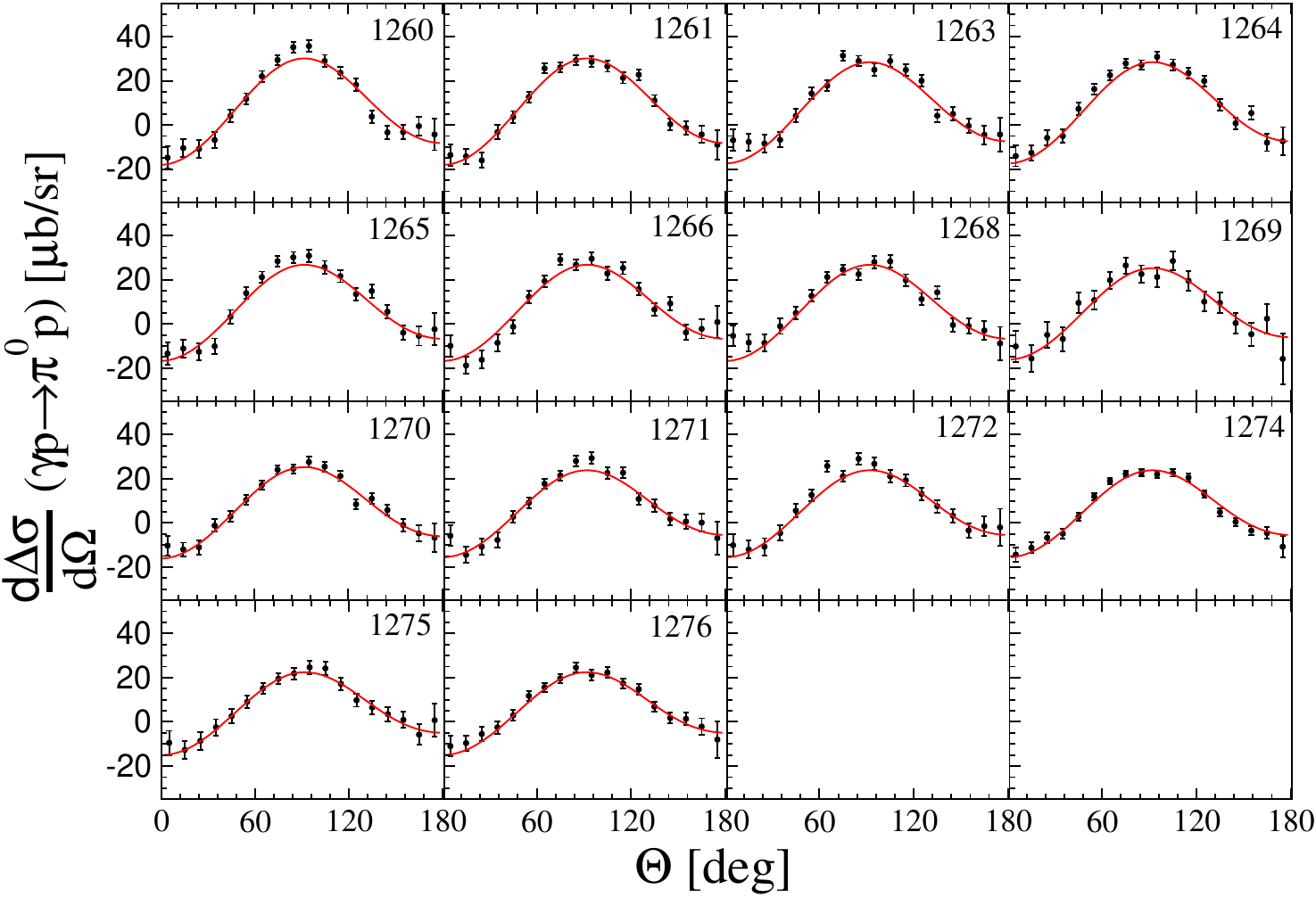}
\end{center}
\caption{Solution JüBo2025 (red)  for the helicity dependent differential cross section from Ref.~\cite{A2CollaborationatMAMI:2023twj} for the process $\gamma p\to \pi^0 p$. (Data not included in the fit.) The numbers in each plot denote the center of mass energy in MeV.}
\label{fig:delta13_A2Mami2}
\end{figure}

In Fig.~\ref{fig:delta13_A2Mami} and \ref{fig:delta13_A2Mami2} one can see the new data from the A2/MAMI-collaboration~\cite{A2CollaborationatMAMI:2023twj} and the solution of the present study. Note that this dataset was not included in the fit. Still, the data are described well.

\section{Individual channel contributions for running GDH-integral}

\begin{figure}[htbp]
\begin{center}
\includegraphics[width=1.\linewidth]{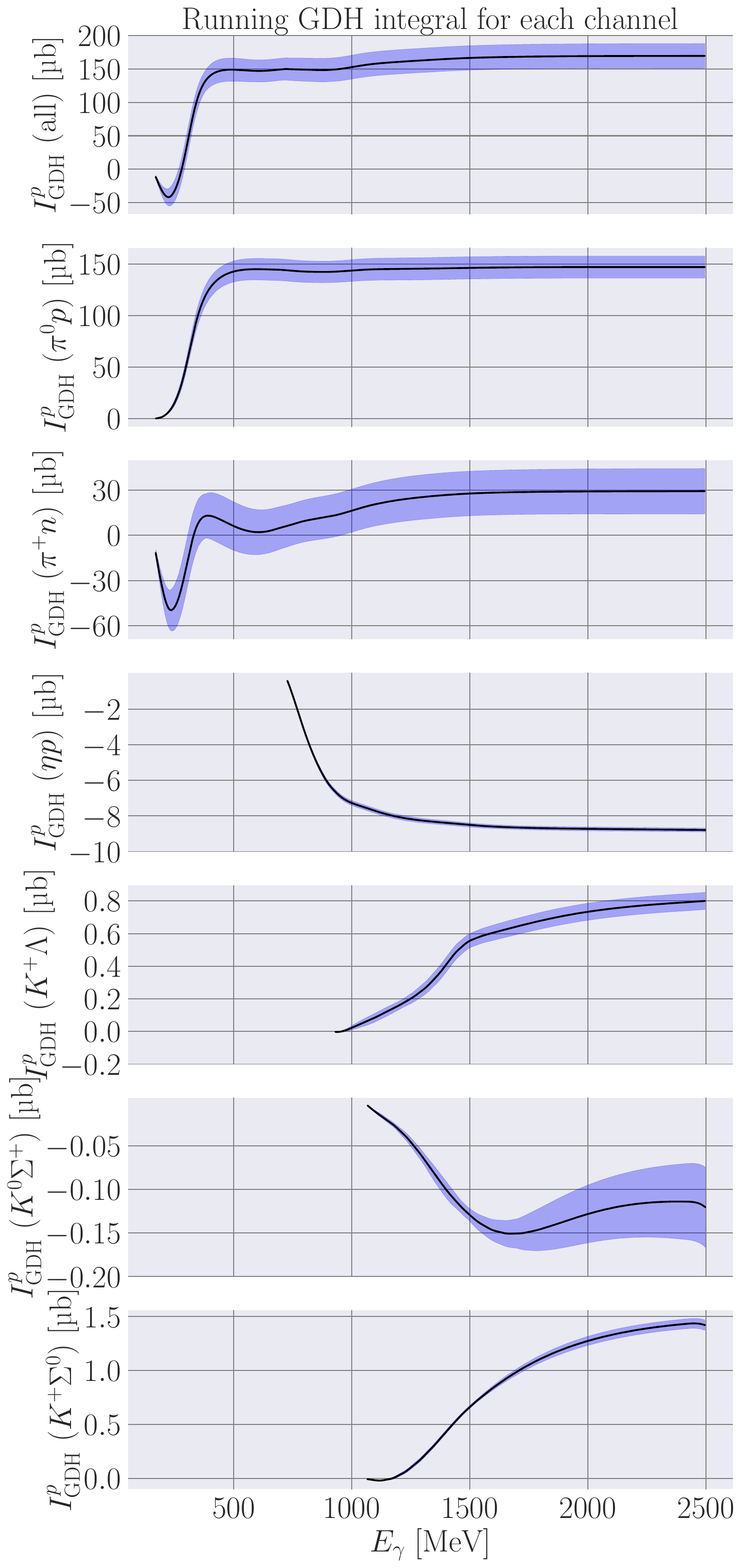} 
\end{center}
\caption{Individual channel contributions to  the running GDH integral $I^p_{\text{GDH}}$ as in Eq.~\eqref{Running_GDH_sum_rule} with errorbands.}
\label{fig:GDH_individual_channel_with_errorband}
\end{figure}

In Fig.~\ref{fig:GDH_individual_channel_with_errorband} we show the individual channel contributions to the running GDH-integral for the six final channels $\pi^0p,\pi^+n,\eta p, K\Lambda,K^0\Sigma^+,K^+\Sigma^0$ together with their individual errorbands. The $y$-axis is scaled such that the errorband is better visible compared to Fig.~\ref{Fig:Running_GDH} and \ref{Fig:Running_GDH_Zoom}.

\section{Influence of specific poles on newly included $\eta p$ data sets}
\begin{figure}[htbp]
\begin{center}
\includegraphics[width=1.\linewidth]{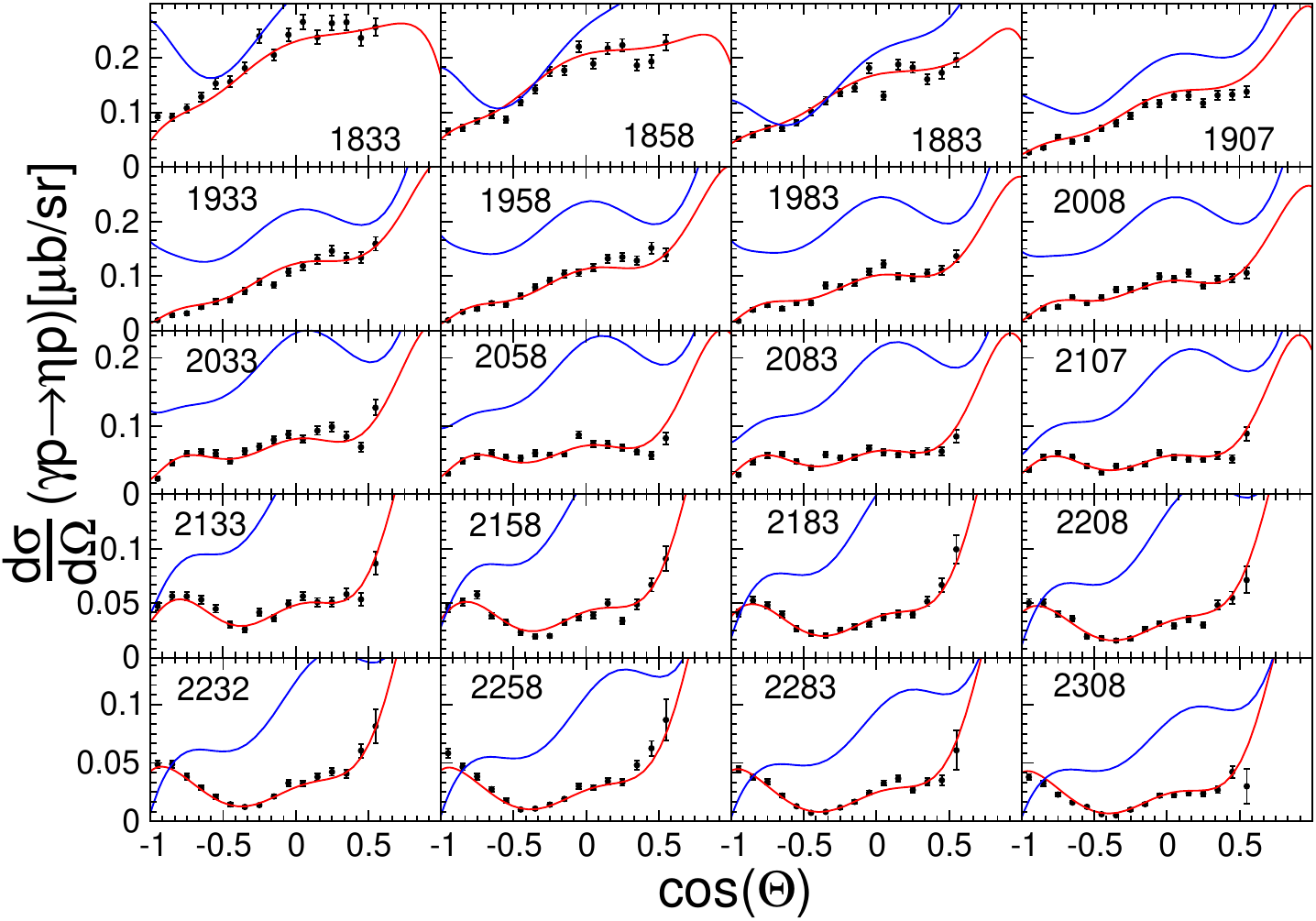} 
\includegraphics[width=1.\linewidth]{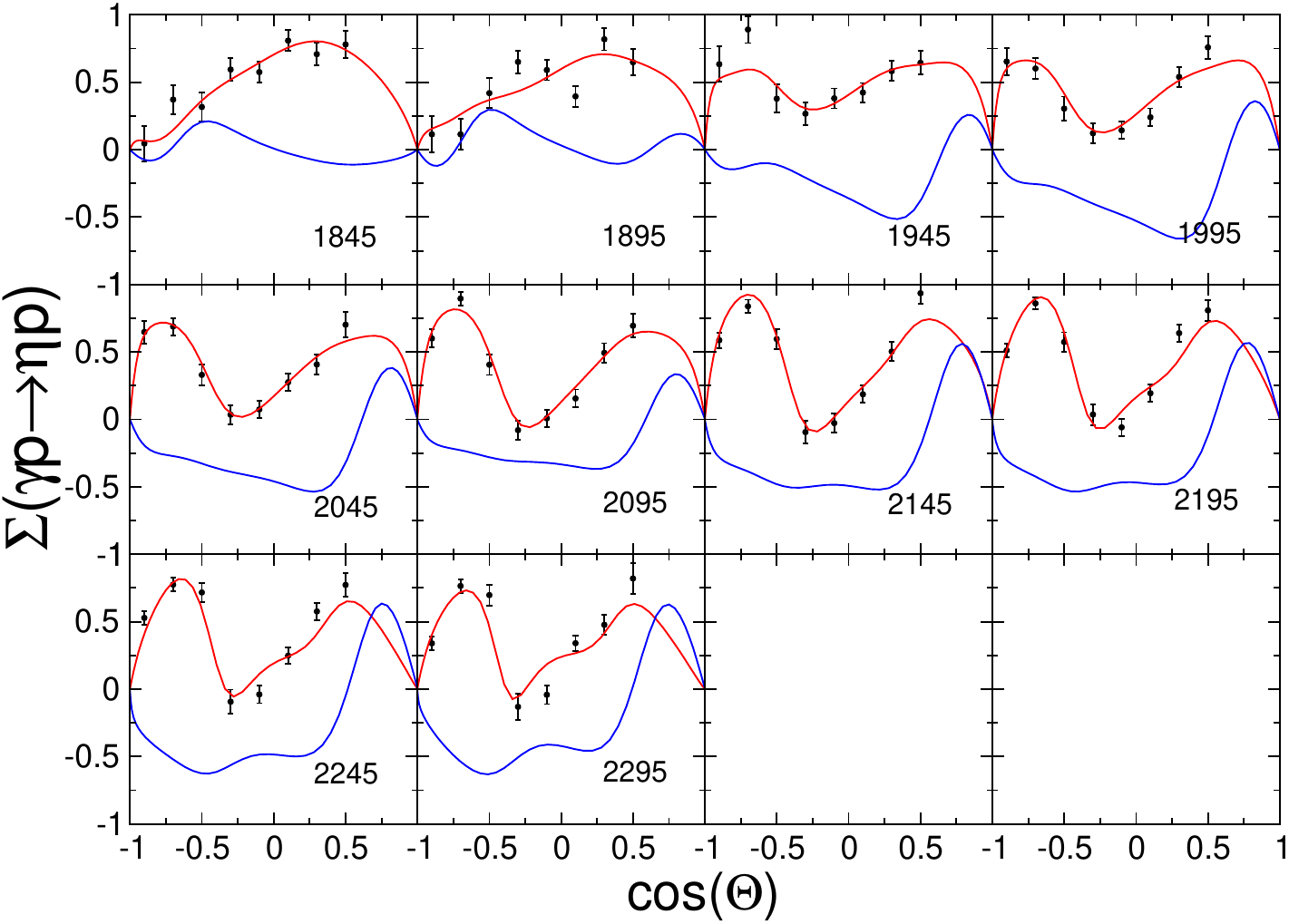} 
\end{center}
\caption{Current fit results (red) and the same solution without allowing $P_{13}$ poles to couple to $\eta N$ (blue) for  the newly included data sets from Ref.~\cite{LEPS2BGOegg:2022dop}. The numbers in each plot denote the center of mass energy in MeV.}
\label{fig:LEPS_data_without_P13}
\end{figure}

\begin{figure}[htbp]
\begin{center}
\includegraphics[width=1.\linewidth]{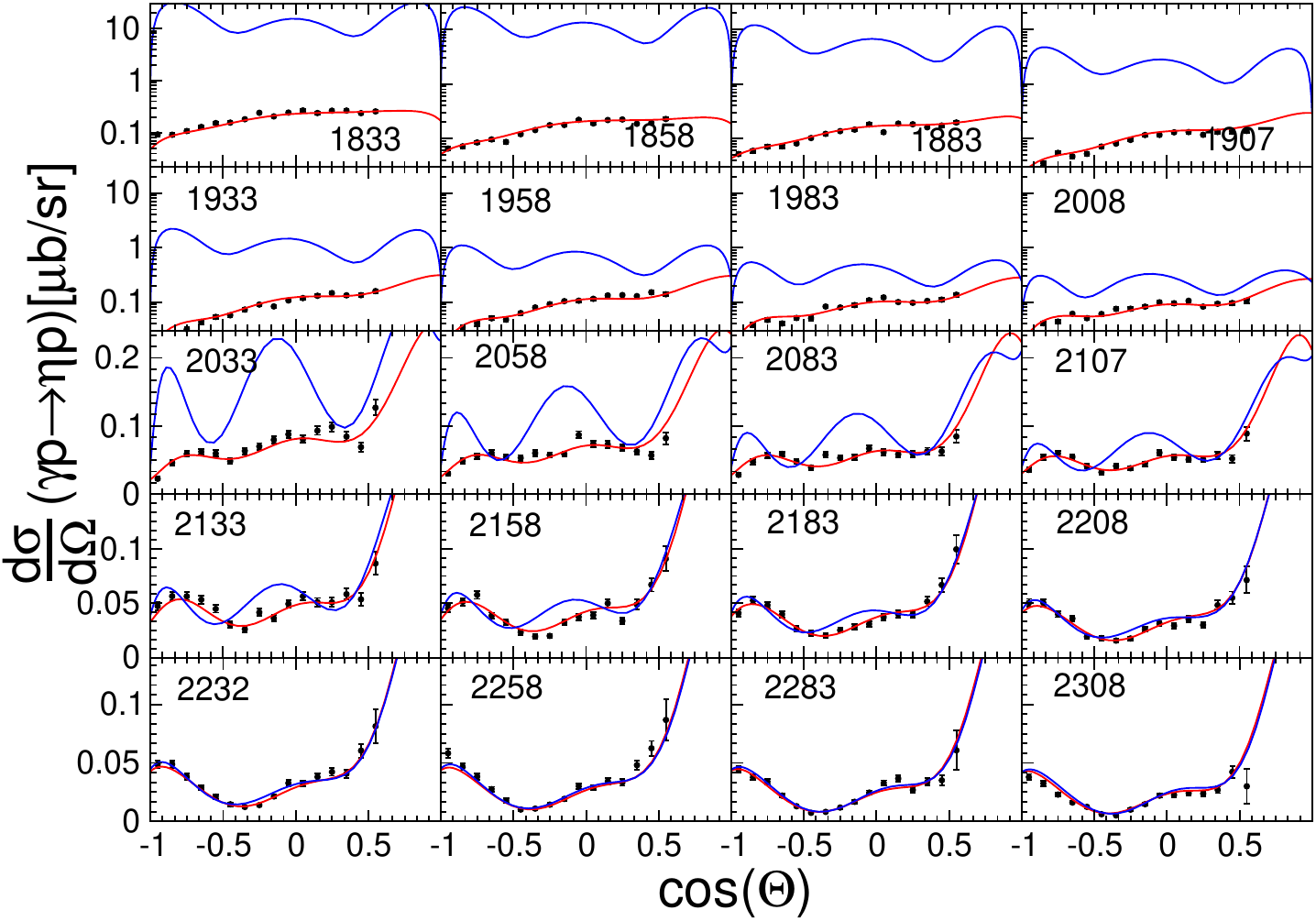} 
\includegraphics[width=1.\linewidth]{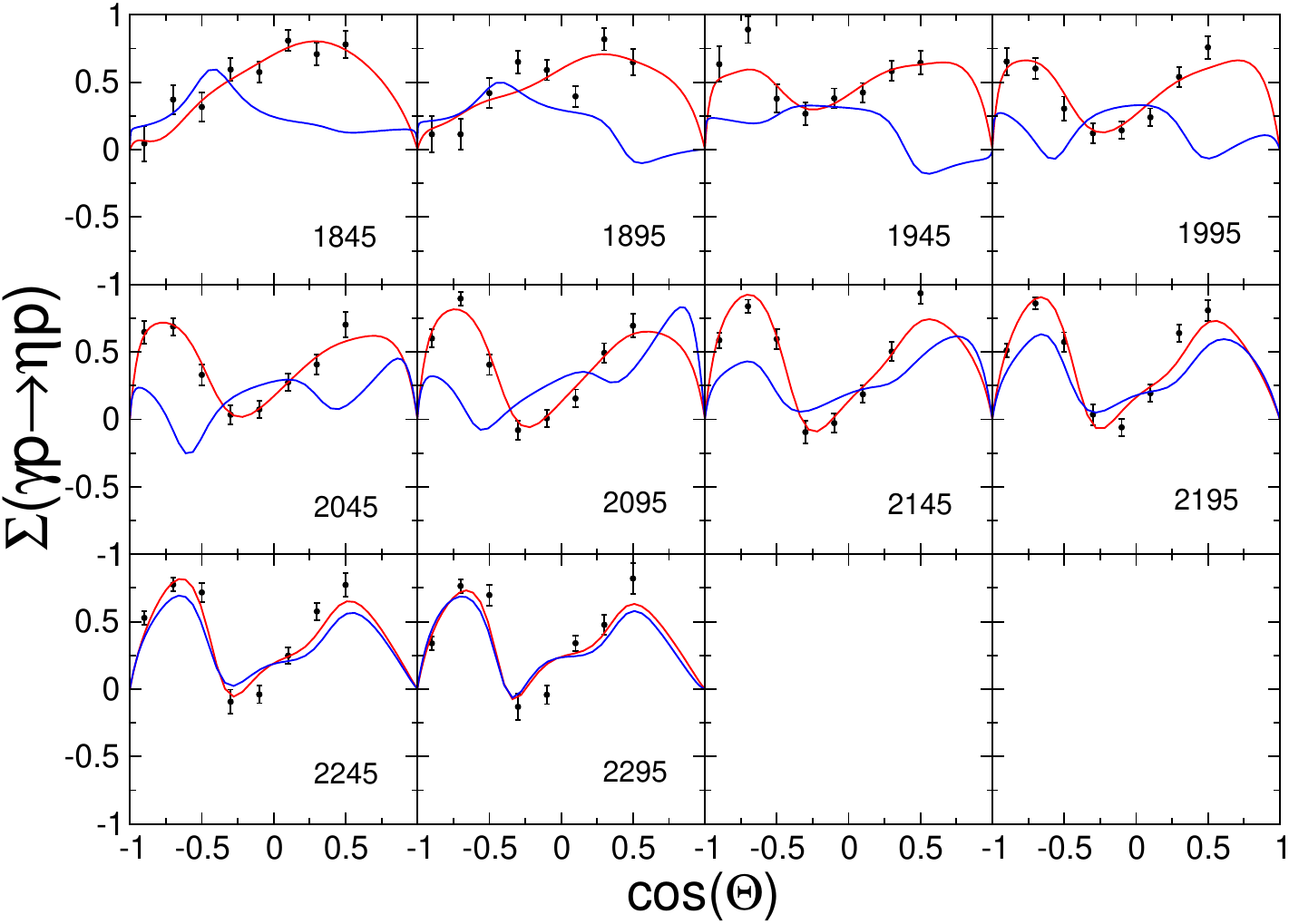} 
\end{center}
\caption{Current fit results (red) and the same solution without allowing the $F_{17}$ pole to couple to $\eta N$ (blue) for  the newly included data sets from Ref.~\cite{LEPS2BGOegg:2022dop}. Note that the scale for the observable ${d\sigma}/{d\Omega}$ in the first two rows is set to log-scale. The numbers in each plot denote the center of mass energy in MeV.}
\label{fig:LEPS_data_without_F17}
\end{figure}

For the two $P_{13}$ poles we observe a large impact on the $\eta p$ data sets that were newly included in the current study. This is shown in Fig.~\ref{fig:LEPS_data_without_P13}. We show analogously the effect of the $F_{17}$ pole $N(1990)7/2^+$ on the new $\eta p$-dataset in Fig.~\ref{fig:LEPS_data_without_F17}. Note that the scale for the observable d$\sigma/$d$\Omega$ in the first two rows is set to logarithmic scale such that the impact of the $N(1990)7/2^+$ can actually be seen. This large impact for the lowest energy  bins is explained by the pole position of the $N(1990)7/2^+$ at $1851$ MeV.

\bibliography{BIB}

\begin{thebibliography}{77}%
\makeatletter
\providecommand \@ifxundefined [1]{%
 \@ifx{#1\undefined}
}%
\providecommand \@ifnum [1]{%
 \ifnum #1\expandafter \@firstoftwo
 \else \expandafter \@secondoftwo
 \fi
}%
\providecommand \@ifx [1]{%
 \ifx #1\expandafter \@firstoftwo
 \else \expandafter \@secondoftwo
 \fi
}%
\providecommand \natexlab [1]{#1}%
\providecommand \enquote  [1]{``#1''}%
\providecommand \bibnamefont  [1]{#1}%
\providecommand \bibfnamefont [1]{#1}%
\providecommand \citenamefont [1]{#1}%
\providecommand \href@noop [0]{\@secondoftwo}%
\providecommand \href [0]{\begingroup \@sanitize@url \@href}%
\providecommand \@href[1]{\@@startlink{#1}\@@href}%
\providecommand \@@href[1]{\endgroup#1\@@endlink}%
\providecommand \@sanitize@url [0]{\catcode `\\12\catcode `\$12\catcode
  `\&12\catcode `\#12\catcode `\^12\catcode `\_12\catcode `\%12\relax}%
\providecommand \@@startlink[1]{}%
\providecommand \@@endlink[0]{}%
\providecommand \url  [0]{\begingroup\@sanitize@url \@url }%
\providecommand \@url [1]{\endgroup\@href {#1}{\urlprefix }}%
\providecommand \urlprefix  [0]{URL }%
\providecommand \Eprint [0]{\href }%
\providecommand \doibase [0]{http://dx.doi.org/}%
\providecommand \selectlanguage [0]{\@gobble}%
\providecommand \bibinfo  [0]{\@secondoftwo}%
\providecommand \bibfield  [0]{\@secondoftwo}%
\providecommand \translation [1]{[#1]}%
\providecommand \BibitemOpen [0]{}%
\providecommand \bibitemStop [0]{}%
\providecommand \bibitemNoStop [0]{.\EOS\space}%
\providecommand \EOS [0]{\spacefactor3000\relax}%
\providecommand \BibitemShut  [1]{\csname bibitem#1\endcsname}%
\let\auto@bib@innerbib\@empty
\bibitem [{\citenamefont {Gerasimov}(1965)}]{Gerasimov:1965et}%
  \BibitemOpen
  \bibfield  {author} {\bibinfo {author} {\bibfnamefont {S.~B.}\ \bibnamefont
  {Gerasimov}},\ }\bibfield  {title} {\enquote {\bibinfo {title} {{A Sum rule
  for magnetic moments and the damping of the nucleon magnetic moment in
  nuclei}},}\ }\href@noop {} {\bibfield  {journal} {\bibinfo  {journal} {Yad.
  Fiz.}\ }\textbf {\bibinfo {volume} {2}},\ \bibinfo {pages} {598--602}
  (\bibinfo {year} {1965})}\BibitemShut {NoStop}%
\bibitem [{\citenamefont {Drell}\ and\ \citenamefont
  {Hearn}(1966)}]{Drell:1966jv}%
  \BibitemOpen
  \bibfield  {author} {\bibinfo {author} {\bibfnamefont {S.~D.}\ \bibnamefont
  {Drell}}\ and\ \bibinfo {author} {\bibfnamefont {Anthony~C.}\ \bibnamefont
  {Hearn}},\ }\bibfield  {title} {\enquote {\bibinfo {title} {{Exact Sum Rule
  for Nucleon Magnetic Moments}},}\ }\href {\doibase
  10.1103/PhysRevLett.16.908} {\bibfield  {journal} {\bibinfo  {journal} {Phys.
  Rev. Lett.}\ }\textbf {\bibinfo {volume} {16}},\ \bibinfo {pages} {908--911}
  (\bibinfo {year} {1966})}\BibitemShut {NoStop}%
\bibitem [{\citenamefont {Drechsel}\ \emph {et~al.}(2001)\citenamefont
  {Drechsel}, \citenamefont {Kamalov},\ and\ \citenamefont
  {Tiator}}]{Drechsel:2000ct}%
  \BibitemOpen
  \bibfield  {author} {\bibinfo {author} {\bibfnamefont {D.}~\bibnamefont
  {Drechsel}}, \bibinfo {author} {\bibfnamefont {S.~S.}\ \bibnamefont
  {Kamalov}}, \ and\ \bibinfo {author} {\bibfnamefont {L.}~\bibnamefont
  {Tiator}},\ }\bibfield  {title} {\enquote {\bibinfo {title} {{The GDH sum
  rule and related integrals}},}\ }\href {\doibase 10.1103/PhysRevD.63.114010}
  {\bibfield  {journal} {\bibinfo  {journal} {Phys. Rev. D}\ }\textbf {\bibinfo
  {volume} {63}},\ \bibinfo {pages} {114010} (\bibinfo {year} {2001})},\
  \Eprint {http://arxiv.org/abs/hep-ph/0008306} {arXiv:hep-ph/0008306}
  \BibitemShut {NoStop}%
\bibitem [{\citenamefont {Drechsel}\ and\ \citenamefont
  {Tiator}(2004)}]{Drechsel:2004ki}%
  \BibitemOpen
  \bibfield  {author} {\bibinfo {author} {\bibfnamefont {Dieter}\ \bibnamefont
  {Drechsel}}\ and\ \bibinfo {author} {\bibfnamefont {Lothar}\ \bibnamefont
  {Tiator}},\ }\bibfield  {title} {\enquote {\bibinfo {title} {{The
  Gerasimov-Drell-Hearn sum rule and the spin structure of the nucleon}},}\
  }\href {\doibase 10.1146/annurev.nucl.54.070103.181159} {\bibfield  {journal}
  {\bibinfo  {journal} {Ann. Rev. Nucl. Part. Sci.}\ }\textbf {\bibinfo
  {volume} {54}},\ \bibinfo {pages} {69--114} (\bibinfo {year} {2004})},\
  \Eprint {http://arxiv.org/abs/nucl-th/0406059} {arXiv:nucl-th/0406059}
  \BibitemShut {NoStop}%
\bibitem [{\citenamefont {Schumacher}(2005)}]{Schumacher:2005an}%
  \BibitemOpen
  \bibfield  {author} {\bibinfo {author} {\bibfnamefont {Martin}\ \bibnamefont
  {Schumacher}},\ }\bibfield  {title} {\enquote {\bibinfo {title}
  {{Polarizability of the nucleon and Compton scattering}},}\ }\href {\doibase
  10.1016/j.ppnp.2005.01.033} {\bibfield  {journal} {\bibinfo  {journal} {Prog.
  Part. Nucl. Phys.}\ }\textbf {\bibinfo {volume} {55}},\ \bibinfo {pages}
  {567--646} (\bibinfo {year} {2005})},\ \Eprint
  {http://arxiv.org/abs/hep-ph/0501167} {arXiv:hep-ph/0501167} \BibitemShut
  {NoStop}%
\bibitem [{\citenamefont {Helbing}(2006)}]{Helbing:2006zp}%
  \BibitemOpen
  \bibfield  {author} {\bibinfo {author} {\bibfnamefont {Klaus}\ \bibnamefont
  {Helbing}},\ }\bibfield  {title} {\enquote {\bibinfo {title} {{The
  Gerasimov-Drell-Hearn Sum Rule}},}\ }\href {\doibase
  10.1016/j.ppnp.2005.09.003} {\bibfield  {journal} {\bibinfo  {journal} {Prog.
  Part. Nucl. Phys.}\ }\textbf {\bibinfo {volume} {57}},\ \bibinfo {pages}
  {405--469} (\bibinfo {year} {2006})},\ \Eprint
  {http://arxiv.org/abs/nucl-ex/0603021} {arXiv:nucl-ex/0603021} \BibitemShut
  {NoStop}%
\bibitem [{\citenamefont {Deur}\ \emph {et~al.}(2019)\citenamefont {Deur},
  \citenamefont {Brodsky},\ and\ \citenamefont {De~T\'eramond}}]{Deur:2018roz}%
  \BibitemOpen
  \bibfield  {author} {\bibinfo {author} {\bibfnamefont {Alexandre}\
  \bibnamefont {Deur}}, \bibinfo {author} {\bibfnamefont {Stanley~J.}\
  \bibnamefont {Brodsky}}, \ and\ \bibinfo {author} {\bibfnamefont {Guy~F.}\
  \bibnamefont {De~T\'eramond}},\ }\bibfield  {title} {\enquote {\bibinfo
  {title} {{The Spin Structure of the Nucleon}},}\ }\href {\doibase
  10.1088/1361-6633/ab0b8f} {\bibfield  {journal} {\bibinfo  {journal} {Rept.
  Prog. Phys.}\ }\textbf {\bibinfo {volume} {82}},\ \bibinfo {pages} {076201}
  (\bibinfo {year} {2019})},\ \Eprint {http://arxiv.org/abs/1807.05250}
  {arXiv:1807.05250 [hep-ph]} \BibitemShut {NoStop}%
\bibitem [{\citenamefont {Adhikari}\ \emph {et~al.}(2018)\citenamefont
  {Adhikari} \emph {et~al.}}]{CLAS:2017ozc}%
  \BibitemOpen
  \bibfield  {author} {\bibinfo {author} {\bibfnamefont {K.~P.}\ \bibnamefont
  {Adhikari}} \emph {et~al.} (\bibinfo {collaboration} {CLAS}),\ }\bibfield
  {title} {\enquote {\bibinfo {title} {{Measurement of the ${Q}^{2}$ Dependence
  of the Deuteron Spin Structure Function ${g}_{1}$ and its Moments at Low
  ${Q}^{2}$ with CLAS}},}\ }\href {\doibase 10.1103/PhysRevLett.120.062501}
  {\bibfield  {journal} {\bibinfo  {journal} {Phys. Rev. Lett.}\ }\textbf
  {\bibinfo {volume} {120}},\ \bibinfo {pages} {062501} (\bibinfo {year}
  {2018})},\ \Eprint {http://arxiv.org/abs/1711.01974} {arXiv:1711.01974
  [nucl-ex]} \BibitemShut {NoStop}%
\bibitem [{\citenamefont {Zheng}\ \emph {et~al.}(2021)\citenamefont {Zheng}
  \emph {et~al.}}]{CLAS:2021apd}%
  \BibitemOpen
  \bibfield  {author} {\bibinfo {author} {\bibfnamefont {X.}~\bibnamefont
  {Zheng}} \emph {et~al.} (\bibinfo {collaboration} {CLAS}),\ }\bibfield
  {title} {\enquote {\bibinfo {title} {{Measurement of the proton spin
  structure at long distances}},}\ }\href {\doibase 10.1038/s41567-021-01198-z}
  {\bibfield  {journal} {\bibinfo  {journal} {Nature Phys.}\ }\textbf {\bibinfo
  {volume} {17}},\ \bibinfo {pages} {736--741} (\bibinfo {year} {2021})},\
  \Eprint {http://arxiv.org/abs/2102.02658} {arXiv:2102.02658 [nucl-ex]}
  \BibitemShut {NoStop}%
\bibitem [{\citenamefont {Bernard}\ \emph {et~al.}(1993)\citenamefont
  {Bernard}, \citenamefont {Kaiser},\ and\ \citenamefont
  {Mei{\ss}ner}}]{Bernard:1992nz}%
  \BibitemOpen
  \bibfield  {author} {\bibinfo {author} {\bibfnamefont {V\'eronique}\
  \bibnamefont {Bernard}}, \bibinfo {author} {\bibfnamefont {Norbert}\
  \bibnamefont {Kaiser}}, \ and\ \bibinfo {author} {\bibfnamefont {Ulf-G.}\
  \bibnamefont {Mei{\ss}ner}},\ }\bibfield  {title} {\enquote {\bibinfo {title}
  {{Small momentum evolution of the extended Drell-Hearn-Gerasimov sum
  rule}},}\ }\href {\doibase 10.1103/PhysRevD.48.3062} {\bibfield  {journal}
  {\bibinfo  {journal} {Phys. Rev. D}\ }\textbf {\bibinfo {volume} {48}},\
  \bibinfo {pages} {3062--3069} (\bibinfo {year} {1993})},\ \Eprint
  {http://arxiv.org/abs/hep-ph/9212257} {arXiv:hep-ph/9212257} \BibitemShut
  {NoStop}%
\bibitem [{\citenamefont {Ji}\ \emph {et~al.}(2000)\citenamefont {Ji},
  \citenamefont {Kao},\ and\ \citenamefont {Osborne}}]{Ji:1999pd}%
  \BibitemOpen
  \bibfield  {author} {\bibinfo {author} {\bibfnamefont {Xiang-Dong}\
  \bibnamefont {Ji}}, \bibinfo {author} {\bibfnamefont {Chung-Wen}\
  \bibnamefont {Kao}}, \ and\ \bibinfo {author} {\bibfnamefont {Jonathan}\
  \bibnamefont {Osborne}},\ }\bibfield  {title} {\enquote {\bibinfo {title}
  {{Generalized Drell-Hearn-Gerasimov sum rule at order O(p**4) in chiral
  perturbation theory}},}\ }\href {\doibase 10.1016/S0370-2693(99)01365-9}
  {\bibfield  {journal} {\bibinfo  {journal} {Phys. Lett. B}\ }\textbf
  {\bibinfo {volume} {472}},\ \bibinfo {pages} {1--4} (\bibinfo {year}
  {2000})},\ \Eprint {http://arxiv.org/abs/hep-ph/9910256}
  {arXiv:hep-ph/9910256} \BibitemShut {NoStop}%
\bibitem [{\citenamefont {Bernard}\ \emph {et~al.}(2003)\citenamefont
  {Bernard}, \citenamefont {Hemmert},\ and\ \citenamefont
  {Mei{\ss}ner}}]{Bernard:2002pw}%
  \BibitemOpen
  \bibfield  {author} {\bibinfo {author} {\bibfnamefont {V\'eronique}\
  \bibnamefont {Bernard}}, \bibinfo {author} {\bibfnamefont {Thomas~R.}\
  \bibnamefont {Hemmert}}, \ and\ \bibinfo {author} {\bibfnamefont {Ulf-G.}\
  \bibnamefont {Mei{\ss}ner}},\ }\bibfield  {title} {\enquote {\bibinfo {title}
  {{Spin structure of the nucleon at low-energies}},}\ }\href {\doibase
  10.1103/PhysRevD.67.076008} {\bibfield  {journal} {\bibinfo  {journal} {Phys.
  Rev. D}\ }\textbf {\bibinfo {volume} {67}},\ \bibinfo {pages} {076008}
  (\bibinfo {year} {2003})},\ \Eprint {http://arxiv.org/abs/hep-ph/0212033}
  {arXiv:hep-ph/0212033} \BibitemShut {NoStop}%
\bibitem [{\citenamefont {Bernard}\ \emph {et~al.}(2013)\citenamefont
  {Bernard}, \citenamefont {Epelbaum}, \citenamefont {Krebs},\ and\
  \citenamefont {Mei{\ss}ner}}]{Bernard:2012hb}%
  \BibitemOpen
  \bibfield  {author} {\bibinfo {author} {\bibfnamefont {Veronique}\
  \bibnamefont {Bernard}}, \bibinfo {author} {\bibfnamefont {Evgeny}\
  \bibnamefont {Epelbaum}}, \bibinfo {author} {\bibfnamefont {Hermann}\
  \bibnamefont {Krebs}}, \ and\ \bibinfo {author} {\bibfnamefont {Ulf-G.}\
  \bibnamefont {Mei{\ss}ner}},\ }\bibfield  {title} {\enquote {\bibinfo {title}
  {{New insights into the spin structure of the nucleon}},}\ }\href {\doibase
  10.1103/PhysRevD.87.054032} {\bibfield  {journal} {\bibinfo  {journal} {Phys.
  Rev. D}\ }\textbf {\bibinfo {volume} {87}},\ \bibinfo {pages} {054032}
  (\bibinfo {year} {2013})},\ \Eprint {http://arxiv.org/abs/1209.2523}
  {arXiv:1209.2523 [hep-ph]} \BibitemShut {NoStop}%
\bibitem [{\citenamefont {Alarc\'on}\ \emph {et~al.}(2020)\citenamefont
  {Alarc\'on}, \citenamefont {Hagelstein}, \citenamefont {Lensky},\ and\
  \citenamefont {Pascalutsa}}]{Alarcon:2020icz}%
  \BibitemOpen
  \bibfield  {author} {\bibinfo {author} {\bibfnamefont {Jose~Manuel}\
  \bibnamefont {Alarc\'on}}, \bibinfo {author} {\bibfnamefont {Franziska}\
  \bibnamefont {Hagelstein}}, \bibinfo {author} {\bibfnamefont {Vadim}\
  \bibnamefont {Lensky}}, \ and\ \bibinfo {author} {\bibfnamefont {Vladimir}\
  \bibnamefont {Pascalutsa}},\ }\bibfield  {title} {\enquote {\bibinfo {title}
  {{Forward doubly-virtual Compton scattering off the nucleon in chiral
  perturbation theory: II. Spin polarizabilities and moments of polarized
  structure functions}},}\ }\href {\doibase 10.1103/PhysRevD.102.114026}
  {\bibfield  {journal} {\bibinfo  {journal} {Phys. Rev. D}\ }\textbf {\bibinfo
  {volume} {102}},\ \bibinfo {pages} {114026} (\bibinfo {year} {2020})},\
  \Eprint {http://arxiv.org/abs/2006.08626} {arXiv:2006.08626 [hep-ph]}
  \BibitemShut {NoStop}%
\bibitem [{\citenamefont {Ahrens}\ \emph {et~al.}(2001)\citenamefont {Ahrens}
  \emph {et~al.}}]{GDH:2001zzk}%
  \BibitemOpen
  \bibfield  {author} {\bibinfo {author} {\bibfnamefont {J.}~\bibnamefont
  {Ahrens}} \emph {et~al.} (\bibinfo {collaboration} {GDH, A2}),\ }\bibfield
  {title} {\enquote {\bibinfo {title} {{First measurement of the
  Gerasimov-Drell-Hearn integral for Hydrogen from 200 to 800 MeV}},}\ }\href
  {\doibase 10.1103/PhysRevLett.87.022003} {\bibfield  {journal} {\bibinfo
  {journal} {Phys. Rev. Lett.}\ }\textbf {\bibinfo {volume} {87}},\ \bibinfo
  {pages} {022003} (\bibinfo {year} {2001})},\ \Eprint
  {http://arxiv.org/abs/hep-ex/0105089} {arXiv:hep-ex/0105089} \BibitemShut
  {NoStop}%
\bibitem [{\citenamefont {Dutz}\ \emph {et~al.}(2003)\citenamefont {Dutz} \emph
  {et~al.}}]{GDH:2003xhc}%
  \BibitemOpen
  \bibfield  {author} {\bibinfo {author} {\bibfnamefont {H.}~\bibnamefont
  {Dutz}} \emph {et~al.} (\bibinfo {collaboration} {GDH}),\ }\bibfield  {title}
  {\enquote {\bibinfo {title} {{First measurement of the Gerasimov-Drell-Hearn
  sum rule for H-1 from 0.7-GeV to 1.8-GeV at ELSA}},}\ }\href {\doibase
  10.1103/PhysRevLett.91.192001} {\bibfield  {journal} {\bibinfo  {journal}
  {Phys. Rev. Lett.}\ }\textbf {\bibinfo {volume} {91}},\ \bibinfo {pages}
  {192001} (\bibinfo {year} {2003})}\BibitemShut {NoStop}%
\bibitem [{\citenamefont {Dutz}\ \emph {et~al.}(2004)\citenamefont {Dutz} \emph
  {et~al.}}]{Dutz:2004zz}%
  \BibitemOpen
  \bibfield  {author} {\bibinfo {author} {\bibfnamefont {H.}~\bibnamefont
  {Dutz}} \emph {et~al.},\ }\bibfield  {title} {\enquote {\bibinfo {title}
  {{Experimental Check of the Gerasimov-Drell-Hearn Sum Rule for H-1}},}\
  }\href {\doibase 10.1103/PhysRevLett.93.032003} {\bibfield  {journal}
  {\bibinfo  {journal} {Phys. Rev. Lett.}\ }\textbf {\bibinfo {volume} {93}},\
  \bibinfo {pages} {032003} (\bibinfo {year} {2004})}\BibitemShut {NoStop}%
\bibitem [{\citenamefont {Arndt}\ \emph {et~al.}(2005)\citenamefont {Arndt},
  \citenamefont {Briscoe}, \citenamefont {Strakovsky},\ and\ \citenamefont
  {Workman}}]{Arndt:2005wk}%
  \BibitemOpen
  \bibfield  {author} {\bibinfo {author} {\bibfnamefont {R.~A.}\ \bibnamefont
  {Arndt}}, \bibinfo {author} {\bibfnamefont {W.~J.}\ \bibnamefont {Briscoe}},
  \bibinfo {author} {\bibfnamefont {I.~I.}\ \bibnamefont {Strakovsky}}, \ and\
  \bibinfo {author} {\bibfnamefont {R.~L.}\ \bibnamefont {Workman}},\
  }\bibfield  {title} {\enquote {\bibinfo {title} {{Helicity-dependent
  photoabsorption cross sections on the nucleon}},}\ }\href {\doibase
  10.1103/PhysRevC.72.058203} {\bibfield  {journal} {\bibinfo  {journal} {Phys.
  Rev. C}\ }\textbf {\bibinfo {volume} {72}},\ \bibinfo {pages} {058203}
  (\bibinfo {year} {2005})},\ \Eprint {http://arxiv.org/abs/nucl-th/0508064}
  {arXiv:nucl-th/0508064} \BibitemShut {NoStop}%
\bibitem [{\citenamefont {Strakovsky}\ \emph {et~al.}(2022)\citenamefont
  {Strakovsky}, \citenamefont {\v{S}irca}, \citenamefont {Briscoe},
  \citenamefont {Deur}, \citenamefont {Schmidt},\ and\ \citenamefont
  {Workman}}]{Strakovsky:2022tvu}%
  \BibitemOpen
  \bibfield  {author} {\bibinfo {author} {\bibfnamefont {Igor}\ \bibnamefont
  {Strakovsky}}, \bibinfo {author} {\bibfnamefont {Simon}\ \bibnamefont
  {\v{S}irca}}, \bibinfo {author} {\bibfnamefont {William~J.}\ \bibnamefont
  {Briscoe}}, \bibinfo {author} {\bibfnamefont {Alexandre}\ \bibnamefont
  {Deur}}, \bibinfo {author} {\bibfnamefont {Axel}\ \bibnamefont {Schmidt}}, \
  and\ \bibinfo {author} {\bibfnamefont {Ron~L.}\ \bibnamefont {Workman}},\
  }\bibfield  {title} {\enquote {\bibinfo {title} {{Single-pion contribution to
  the Gerasimov-Drell-Hearn sum rule and related integrals}},}\ }\href
  {\doibase 10.1103/PhysRevC.105.045202} {\bibfield  {journal} {\bibinfo
  {journal} {Phys. Rev. C}\ }\textbf {\bibinfo {volume} {105}},\ \bibinfo
  {pages} {045202} (\bibinfo {year} {2022})},\ \Eprint
  {http://arxiv.org/abs/2201.06495} {arXiv:2201.06495 [nucl-th]} \BibitemShut
  {NoStop}%
\bibitem [{\citenamefont {Drechsel}\ \emph {et~al.}(2007)\citenamefont
  {Drechsel}, \citenamefont {Kamalov},\ and\ \citenamefont
  {Tiator}}]{Drechsel:2007if}%
  \BibitemOpen
  \bibfield  {author} {\bibinfo {author} {\bibfnamefont {D.}~\bibnamefont
  {Drechsel}}, \bibinfo {author} {\bibfnamefont {S.~S.}\ \bibnamefont
  {Kamalov}}, \ and\ \bibinfo {author} {\bibfnamefont {L.}~\bibnamefont
  {Tiator}},\ }\bibfield  {title} {\enquote {\bibinfo {title} {{Unitary Isobar
  Model - MAID2007}},}\ }\href {\doibase 10.1140/epja/i2007-10490-6} {\bibfield
   {journal} {\bibinfo  {journal} {Eur. Phys. J. A}\ }\textbf {\bibinfo
  {volume} {34}},\ \bibinfo {pages} {69--97} (\bibinfo {year} {2007})},\
  \Eprint {http://arxiv.org/abs/0710.0306} {arXiv:0710.0306 [nucl-th]}
  \BibitemShut {NoStop}%
\bibitem [{\citenamefont {Mart}\ and\ \citenamefont
  {Kholili}(2019)}]{Mart:2019fau}%
  \BibitemOpen
  \bibfield  {author} {\bibinfo {author} {\bibfnamefont {T.}~\bibnamefont
  {Mart}}\ and\ \bibinfo {author} {\bibfnamefont {M.~J.}\ \bibnamefont
  {Kholili}},\ }\bibfield  {title} {\enquote {\bibinfo {title} {{Partial wave
  analysis for $K\Sigma$ photoproduction on the nucleon valid from threshold up
  to $W$ = 2.8 GeV}},}\ }\href {\doibase 10.1088/1361-6471/ab34c6} {\bibfield
  {journal} {\bibinfo  {journal} {J. Phys. G}\ }\textbf {\bibinfo {volume}
  {46}},\ \bibinfo {pages} {105112} (\bibinfo {year} {2019})}\BibitemShut
  {NoStop}%
\bibitem [{\citenamefont {Anisovich}\ \emph {et~al.}(2012)\citenamefont
  {Anisovich}, \citenamefont {Beck}, \citenamefont {Klempt}, \citenamefont
  {Nikonov}, \citenamefont {Sarantsev},\ and\ \citenamefont
  {Thoma}}]{Anisovich:2011fc}%
  \BibitemOpen
  \bibfield  {author} {\bibinfo {author} {\bibfnamefont {A.~V.}\ \bibnamefont
  {Anisovich}}, \bibinfo {author} {\bibfnamefont {R.}~\bibnamefont {Beck}},
  \bibinfo {author} {\bibfnamefont {E.}~\bibnamefont {Klempt}}, \bibinfo
  {author} {\bibfnamefont {V.~A.}\ \bibnamefont {Nikonov}}, \bibinfo {author}
  {\bibfnamefont {A.~V.}\ \bibnamefont {Sarantsev}}, \ and\ \bibinfo {author}
  {\bibfnamefont {U.}~\bibnamefont {Thoma}},\ }\bibfield  {title} {\enquote
  {\bibinfo {title} {{Properties of baryon resonances from a multichannel
  partial wave analysis}},}\ }\href {\doibase 10.1140/epja/i2012-12015-8}
  {\bibfield  {journal} {\bibinfo  {journal} {Eur. Phys. J. A}\ }\textbf
  {\bibinfo {volume} {48}},\ \bibinfo {pages} {15} (\bibinfo {year} {2012})},\
  \Eprint {http://arxiv.org/abs/1112.4937} {arXiv:1112.4937 [hep-ph]}
  \BibitemShut {NoStop}%
\bibitem [{\citenamefont {M\"uller}\ \emph {et~al.}(2020)\citenamefont
  {M\"uller} \emph {et~al.}}]{CBELSATAPS:2019ylw}%
  \BibitemOpen
  \bibfield  {author} {\bibinfo {author} {\bibfnamefont {J.}~\bibnamefont
  {M\"uller}} \emph {et~al.} (\bibinfo {collaboration} {CBELSA/TAPS}),\
  }\bibfield  {title} {\enquote {\bibinfo {title} {{New data on $\vec{\gamma}
  \vec{p}\rightarrow \eta p$ with polarized photons and protons and their
  implications for $N^* \to N\eta$ decays}},}\ }\href {\doibase
  10.1016/j.physletb.2020.135323} {\bibfield  {journal} {\bibinfo  {journal}
  {Phys. Lett. B}\ }\textbf {\bibinfo {volume} {803}},\ \bibinfo {pages}
  {135323} (\bibinfo {year} {2020})},\ \Eprint
  {http://arxiv.org/abs/1909.08464} {arXiv:1909.08464 [nucl-ex]} \BibitemShut
  {NoStop}%
\bibitem [{\citenamefont {Sarantsev}\ \emph {et~al.}(2025)\citenamefont
  {Sarantsev}, \citenamefont {Klempt}, \citenamefont {Nikonov}, \citenamefont
  {Seifen}, \citenamefont {Thoma}, \citenamefont {Wunderlich}, \citenamefont
  {Achenbach}, \citenamefont {Burkert}, \citenamefont {Mokeev},\ and\
  \citenamefont {Crede}}]{Sarantsev:2025lik}%
  \BibitemOpen
  \bibfield  {author} {\bibinfo {author} {\bibfnamefont {A.~V.}\ \bibnamefont
  {Sarantsev}}, \bibinfo {author} {\bibfnamefont {E.}~\bibnamefont {Klempt}},
  \bibinfo {author} {\bibfnamefont {K.~V.}\ \bibnamefont {Nikonov}}, \bibinfo
  {author} {\bibfnamefont {T.}~\bibnamefont {Seifen}}, \bibinfo {author}
  {\bibfnamefont {U.}~\bibnamefont {Thoma}}, \bibinfo {author} {\bibfnamefont
  {Y.}~\bibnamefont {Wunderlich}}, \bibinfo {author} {\bibfnamefont
  {P.}~\bibnamefont {Achenbach}}, \bibinfo {author} {\bibfnamefont {V.~D.}\
  \bibnamefont {Burkert}}, \bibinfo {author} {\bibfnamefont {V.}~\bibnamefont
  {Mokeev}}, \ and\ \bibinfo {author} {\bibfnamefont {V.}~\bibnamefont
  {Crede}},\ }\bibfield  {title} {\enquote {\bibinfo {title} {{Decays of $N^*$
  and $\Delta^*$ resonances into $N\rho$, $\Delta\pi$, and $N\sigma$}},}\
  }\href@noop {} {\  (\bibinfo {year} {2025})},\ \Eprint
  {http://arxiv.org/abs/2503.16636} {arXiv:2503.16636 [nucl-th]} \BibitemShut
  {NoStop}%
\bibitem [{\citenamefont {Hunt}\ and\ \citenamefont
  {Manley}(2019)}]{Hunt:2018wqz}%
  \BibitemOpen
  \bibfield  {author} {\bibinfo {author} {\bibfnamefont {B.~C.}\ \bibnamefont
  {Hunt}}\ and\ \bibinfo {author} {\bibfnamefont {D.~M.}\ \bibnamefont
  {Manley}},\ }\bibfield  {title} {\enquote {\bibinfo {title} {{Updated
  determination of $N^*$ resonance parameters using a unitary, multichannel
  formalism}},}\ }\href {\doibase 10.1103/PhysRevC.99.055205} {\bibfield
  {journal} {\bibinfo  {journal} {Phys. Rev. C}\ }\textbf {\bibinfo {volume}
  {99}},\ \bibinfo {pages} {055205} (\bibinfo {year} {2019})},\ \Eprint
  {http://arxiv.org/abs/1810.13086} {arXiv:1810.13086 [nucl-ex]} \BibitemShut
  {NoStop}%
\bibitem [{\citenamefont {Kamano}\ \emph {et~al.}(2013)\citenamefont {Kamano},
  \citenamefont {Nakamura}, \citenamefont {Lee},\ and\ \citenamefont
  {Sato}}]{Kamano:2013iva}%
  \BibitemOpen
  \bibfield  {author} {\bibinfo {author} {\bibfnamefont {H.}~\bibnamefont
  {Kamano}}, \bibinfo {author} {\bibfnamefont {S.~X.}\ \bibnamefont
  {Nakamura}}, \bibinfo {author} {\bibfnamefont {T.~S.~H.}\ \bibnamefont
  {Lee}}, \ and\ \bibinfo {author} {\bibfnamefont {T.}~\bibnamefont {Sato}},\
  }\bibfield  {title} {\enquote {\bibinfo {title} {{Nucleon resonances within a
  dynamical coupled-channels model of $\pi N$ and $\gamma N$ reactions}},}\
  }\href {\doibase 10.1103/PhysRevC.88.035209} {\bibfield  {journal} {\bibinfo
  {journal} {Phys. Rev. C}\ }\textbf {\bibinfo {volume} {88}},\ \bibinfo
  {pages} {035209} (\bibinfo {year} {2013})},\ \Eprint
  {http://arxiv.org/abs/1305.4351} {arXiv:1305.4351 [nucl-th]} \BibitemShut
  {NoStop}%
\bibitem [{\citenamefont {Kamano}\ \emph {et~al.}(2016)\citenamefont {Kamano},
  \citenamefont {Nakamura}, \citenamefont {Lee},\ and\ \citenamefont
  {Sato}}]{Kamano:2016bgm}%
  \BibitemOpen
  \bibfield  {author} {\bibinfo {author} {\bibfnamefont {H.}~\bibnamefont
  {Kamano}}, \bibinfo {author} {\bibfnamefont {S.~X.}\ \bibnamefont
  {Nakamura}}, \bibinfo {author} {\bibfnamefont {T.~S.~H.}\ \bibnamefont
  {Lee}}, \ and\ \bibinfo {author} {\bibfnamefont {T.}~\bibnamefont {Sato}},\
  }\bibfield  {title} {\enquote {\bibinfo {title} {{Isospin decomposition of
  $\gamma N \to N^*$ transitions within a dynamical coupled-channels model}},}\
  }\href {\doibase 10.1103/PhysRevC.94.015201} {\bibfield  {journal} {\bibinfo
  {journal} {Phys. Rev. C}\ }\textbf {\bibinfo {volume} {94}},\ \bibinfo
  {pages} {015201} (\bibinfo {year} {2016})},\ \Eprint
  {http://arxiv.org/abs/1605.00363} {arXiv:1605.00363 [nucl-th]} \BibitemShut
  {NoStop}%
\bibitem [{\citenamefont {Rönchen}\ \emph {et~al.}(2013)\citenamefont
  {Rönchen}, \citenamefont {Döring}, \citenamefont {Huang}, \citenamefont
  {Haberzettl}, \citenamefont {Haidenbauer}, \citenamefont {Hanhart},
  \citenamefont {Krewald}, \citenamefont {Mei{\ss}ner},\ and\ \citenamefont
  {Nakayama}}]{Ronchen:2012eg}%
  \BibitemOpen
  \bibfield  {author} {\bibinfo {author} {\bibfnamefont {D.}~\bibnamefont
  {Rönchen}}, \bibinfo {author} {\bibfnamefont {M.}~\bibnamefont {Döring}},
  \bibinfo {author} {\bibfnamefont {F.}~\bibnamefont {Huang}}, \bibinfo
  {author} {\bibfnamefont {H.}~\bibnamefont {Haberzettl}}, \bibinfo {author}
  {\bibfnamefont {J.}~\bibnamefont {Haidenbauer}}, \bibinfo {author}
  {\bibfnamefont {C.}~\bibnamefont {Hanhart}}, \bibinfo {author} {\bibfnamefont
  {S.}~\bibnamefont {Krewald}}, \bibinfo {author} {\bibfnamefont {U.-G.}\
  \bibnamefont {Mei{\ss}ner}}, \ and\ \bibinfo {author} {\bibfnamefont
  {K.}~\bibnamefont {Nakayama}},\ }\bibfield  {title} {\enquote {\bibinfo
  {title} {{Coupled-channel dynamics in the reactions $\pi N\to\pi N, \eta N,
  K\Lambda, K\Sigma$}},}\ }\href {\doibase 10.1140/epja/i2013-13044-5}
  {\bibfield  {journal} {\bibinfo  {journal} {Eur. Phys. J. A}\ }\textbf
  {\bibinfo {volume} {49}},\ \bibinfo {pages} {44} (\bibinfo {year} {2013})},\
  \Eprint {http://arxiv.org/abs/1211.6998} {arXiv:1211.6998 [nucl-th]}
  \BibitemShut {NoStop}%
\bibitem [{\citenamefont {R\"onchen}\ \emph {et~al.}(2022)\citenamefont
  {R\"onchen}, \citenamefont {D\"oring}, \citenamefont {Mei\ss{}ner},\ and\
  \citenamefont {Shen}}]{Ronchen:2022hqk}%
  \BibitemOpen
  \bibfield  {author} {\bibinfo {author} {\bibfnamefont {Deborah}\ \bibnamefont
  {R\"onchen}}, \bibinfo {author} {\bibfnamefont {Michael}\ \bibnamefont
  {D\"oring}}, \bibinfo {author} {\bibfnamefont {Ulf-G.}\ \bibnamefont
  {Mei\ss{}ner}}, \ and\ \bibinfo {author} {\bibfnamefont {Chao-Wei}\
  \bibnamefont {Shen}},\ }\bibfield  {title} {\enquote {\bibinfo {title}
  {{Light baryon resonances from a coupled-channel study including $\mathbf
  {K\Sigma }$ photoproduction}},}\ }\href {\doibase
  10.1140/epja/s10050-022-00852-1} {\bibfield  {journal} {\bibinfo  {journal}
  {Eur. Phys. J. A}\ }\textbf {\bibinfo {volume} {58}},\ \bibinfo {pages} {229}
  (\bibinfo {year} {2022})},\ \Eprint {http://arxiv.org/abs/2208.00089}
  {arXiv:2208.00089 [nucl-th]} \BibitemShut {NoStop}%
\bibitem [{\citenamefont {D\"oring}\ \emph {et~al.}(2025)\citenamefont
  {D\"oring}, \citenamefont {Haidenbauer}, \citenamefont {Mai},\ and\
  \citenamefont {Sato}}]{Doring:2025sgb}%
  \BibitemOpen
  \bibfield  {author} {\bibinfo {author} {\bibfnamefont {Michael}\ \bibnamefont
  {D\"oring}}, \bibinfo {author} {\bibfnamefont {Johann}\ \bibnamefont
  {Haidenbauer}}, \bibinfo {author} {\bibfnamefont {Maxim}\ \bibnamefont
  {Mai}}, \ and\ \bibinfo {author} {\bibfnamefont {Toru}\ \bibnamefont
  {Sato}},\ }\bibfield  {title} {\enquote {\bibinfo {title} {{Dynamical
  coupled-channel models for hadron dynamics}},}\ }\href@noop {} {\  (\bibinfo
  {year} {2025})},\ \Eprint {http://arxiv.org/abs/2505.02745} {arXiv:2505.02745
  [nucl-th]} \BibitemShut {NoStop}%
\bibitem [{\citenamefont {Rönchen}\ \emph {et~al.}(2014)\citenamefont
  {Rönchen}, \citenamefont {Döring}, \citenamefont {Huang}, \citenamefont
  {Haberzettl}, \citenamefont {Haidenbauer}, \citenamefont {Hanhart},
  \citenamefont {Krewald}, \citenamefont {Mei\ss{}ner},\ and\ \citenamefont
  {Nakayama}}]{Ronchen:2014cna}%
  \BibitemOpen
  \bibfield  {author} {\bibinfo {author} {\bibfnamefont {D.}~\bibnamefont
  {Rönchen}}, \bibinfo {author} {\bibfnamefont {M.}~\bibnamefont {Döring}},
  \bibinfo {author} {\bibfnamefont {F.}~\bibnamefont {Huang}}, \bibinfo
  {author} {\bibfnamefont {H.}~\bibnamefont {Haberzettl}}, \bibinfo {author}
  {\bibfnamefont {J.}~\bibnamefont {Haidenbauer}}, \bibinfo {author}
  {\bibfnamefont {C.}~\bibnamefont {Hanhart}}, \bibinfo {author} {\bibfnamefont
  {S.}~\bibnamefont {Krewald}}, \bibinfo {author} {\bibfnamefont {U.-G.}\
  \bibnamefont {Mei\ss{}ner}}, \ and\ \bibinfo {author} {\bibfnamefont
  {K.}~\bibnamefont {Nakayama}},\ }\bibfield  {title} {\enquote {\bibinfo
  {title} {{Photocouplings at the Pole from Pion Photoproduction}},}\ }\href
  {\doibase 10.1140/epja/i2014-14101-3} {\bibfield  {journal} {\bibinfo
  {journal} {Eur. Phys. J. A}\ }\textbf {\bibinfo {volume} {50}},\ \bibinfo
  {pages} {101} (\bibinfo {year} {2014})},\ \bibinfo {note} {[Erratum:
  Eur.Phys.J.A 51, 63 (2015)]},\ \Eprint {http://arxiv.org/abs/1401.0634}
  {arXiv:1401.0634 [nucl-th]} \BibitemShut {NoStop}%
\bibitem [{\citenamefont {R\"onchen}\ \emph {et~al.}(2015)\citenamefont
  {R\"onchen}, \citenamefont {D\"oring}, \citenamefont {Haberzettl},
  \citenamefont {Haidenbauer}, \citenamefont {Mei\ss{}ner},\ and\ \citenamefont
  {Nakayama}}]{Ronchen:2015vfa}%
  \BibitemOpen
  \bibfield  {author} {\bibinfo {author} {\bibfnamefont {D.}~\bibnamefont
  {R\"onchen}}, \bibinfo {author} {\bibfnamefont {M.}~\bibnamefont {D\"oring}},
  \bibinfo {author} {\bibfnamefont {H.}~\bibnamefont {Haberzettl}}, \bibinfo
  {author} {\bibfnamefont {J.}~\bibnamefont {Haidenbauer}}, \bibinfo {author}
  {\bibfnamefont {U.-G.}\ \bibnamefont {Mei\ss{}ner}}, \ and\ \bibinfo {author}
  {\bibfnamefont {K.}~\bibnamefont {Nakayama}},\ }\bibfield  {title} {\enquote
  {\bibinfo {title} {{Eta photoproduction in a combined analysis of pion- and
  photon-induced reactions}},}\ }\href {\doibase 10.1140/epja/i2015-15070-7}
  {\bibfield  {journal} {\bibinfo  {journal} {Eur. Phys. J. A}\ }\textbf
  {\bibinfo {volume} {51}},\ \bibinfo {pages} {70} (\bibinfo {year} {2015})},\
  \Eprint {http://arxiv.org/abs/1504.01643} {arXiv:1504.01643 [nucl-th]}
  \BibitemShut {NoStop}%
\bibitem [{\citenamefont {Wang}\ \emph {et~al.}(2022)\citenamefont {Wang},
  \citenamefont {R\"onchen}, \citenamefont {Mei\ss{}ner}, \citenamefont {Lu},
  \citenamefont {Shen},\ and\ \citenamefont {Wu}}]{Wang:2022osj}%
  \BibitemOpen
  \bibfield  {author} {\bibinfo {author} {\bibfnamefont {Yu-Fei}\ \bibnamefont
  {Wang}}, \bibinfo {author} {\bibfnamefont {Deborah}\ \bibnamefont
  {R\"onchen}}, \bibinfo {author} {\bibfnamefont {Ulf-G.}\ \bibnamefont
  {Mei\ss{}ner}}, \bibinfo {author} {\bibfnamefont {Yu}~\bibnamefont {Lu}},
  \bibinfo {author} {\bibfnamefont {Chao-Wei}\ \bibnamefont {Shen}}, \ and\
  \bibinfo {author} {\bibfnamefont {Jia-Jun}\ \bibnamefont {Wu}},\ }\bibfield
  {title} {\enquote {\bibinfo {title} {{Reaction
  \ensuremath{\pi}N\textrightarrow{}\ensuremath{\omega}N in a dynamical
  coupled-channel approach}},}\ }\href {\doibase 10.1103/PhysRevD.106.094031}
  {\bibfield  {journal} {\bibinfo  {journal} {Phys. Rev. D}\ }\textbf {\bibinfo
  {volume} {106}},\ \bibinfo {pages} {094031} (\bibinfo {year} {2022})},\
  \Eprint {http://arxiv.org/abs/2208.03061} {arXiv:2208.03061 [nucl-th]}
  \BibitemShut {NoStop}%
\bibitem [{\citenamefont {Schütz}\ \emph {et~al.}(1998)\citenamefont
  {Schütz}, \citenamefont {Haidenbauer}, \citenamefont {Speth},\ and\
  \citenamefont {Durso}}]{Schutz:1998jx}%
  \BibitemOpen
  \bibfield  {author} {\bibinfo {author} {\bibfnamefont {C.}~\bibnamefont
  {Schütz}}, \bibinfo {author} {\bibfnamefont {J.}~\bibnamefont
  {Haidenbauer}}, \bibinfo {author} {\bibfnamefont {J.}~\bibnamefont {Speth}},
  \ and\ \bibinfo {author} {\bibfnamefont {J.~W.}\ \bibnamefont {Durso}},\
  }\bibfield  {title} {\enquote {\bibinfo {title} {{Extended coupled channels
  model for $\pi N$ scattering and the structure of N*(1440) and N*(1535)}},}\
  }\href {\doibase 10.1103/PhysRevC.57.1464} {\bibfield  {journal} {\bibinfo
  {journal} {Phys. Rev. C}\ }\textbf {\bibinfo {volume} {57}},\ \bibinfo
  {pages} {1464--1477} (\bibinfo {year} {1998})}\BibitemShut {NoStop}%
\bibitem [{\citenamefont {Krehl}\ \emph {et~al.}(2000)\citenamefont {Krehl},
  \citenamefont {Hanhart}, \citenamefont {Krewald},\ and\ \citenamefont
  {Speth}}]{Krehl:1999km}%
  \BibitemOpen
  \bibfield  {author} {\bibinfo {author} {\bibfnamefont {O.}~\bibnamefont
  {Krehl}}, \bibinfo {author} {\bibfnamefont {C.}~\bibnamefont {Hanhart}},
  \bibinfo {author} {\bibfnamefont {S.}~\bibnamefont {Krewald}}, \ and\
  \bibinfo {author} {\bibfnamefont {J.}~\bibnamefont {Speth}},\ }\bibfield
  {title} {\enquote {\bibinfo {title} {{What is the structure of the Roper
  resonance?}}}\ }\href {\doibase 10.1103/PhysRevC.62.025207} {\bibfield
  {journal} {\bibinfo  {journal} {Phys. Rev. C}\ }\textbf {\bibinfo {volume}
  {62}},\ \bibinfo {pages} {025207} (\bibinfo {year} {2000})},\ \Eprint
  {http://arxiv.org/abs/nucl-th/9911080} {arXiv:nucl-th/9911080} \BibitemShut
  {NoStop}%
\bibitem [{\citenamefont {Mai}\ \emph {et~al.}(2017)\citenamefont {Mai},
  \citenamefont {Hu}, \citenamefont {D{\"o}ring}, \citenamefont {Pilloni},\
  and\ \citenamefont {Szczepaniak}}]{Mai:2017vot}%
  \BibitemOpen
  \bibfield  {author} {\bibinfo {author} {\bibfnamefont {M.}~\bibnamefont
  {Mai}}, \bibinfo {author} {\bibfnamefont {B.}~\bibnamefont {Hu}}, \bibinfo
  {author} {\bibfnamefont {M.}~\bibnamefont {D{\"o}ring}}, \bibinfo {author}
  {\bibfnamefont {A.}~\bibnamefont {Pilloni}}, \ and\ \bibinfo {author}
  {\bibfnamefont {A.}~\bibnamefont {Szczepaniak}},\ }\bibfield  {title}
  {\enquote {\bibinfo {title} {{Three-body Unitarity with Isobars
  Revisited}},}\ }\href {\doibase 10.1140/epja/i2017-12368-4} {\bibfield
  {journal} {\bibinfo  {journal} {Eur. Phys. J. A}\ }\textbf {\bibinfo {volume}
  {53}},\ \bibinfo {pages} {177} (\bibinfo {year} {2017})},\ \Eprint
  {http://arxiv.org/abs/1706.06118} {arXiv:1706.06118 [nucl-th]} \BibitemShut
  {NoStop}%
\bibitem [{\citenamefont {Döring}\ \emph {et~al.}(2011)\citenamefont
  {Döring}, \citenamefont {Hanhart}, \citenamefont {Huang}, \citenamefont
  {Krewald}, \citenamefont {Mei\ss{}ner},\ and\ \citenamefont
  {Rönchen}}]{Doring:2010ap}%
  \BibitemOpen
  \bibfield  {author} {\bibinfo {author} {\bibfnamefont {M.}~\bibnamefont
  {Döring}}, \bibinfo {author} {\bibfnamefont {C.}~\bibnamefont {Hanhart}},
  \bibinfo {author} {\bibfnamefont {F.}~\bibnamefont {Huang}}, \bibinfo
  {author} {\bibfnamefont {S.}~\bibnamefont {Krewald}}, \bibinfo {author}
  {\bibfnamefont {U.-G.}\ \bibnamefont {Mei\ss{}ner}}, \ and\ \bibinfo {author}
  {\bibfnamefont {D.}~\bibnamefont {Rönchen}},\ }\bibfield  {title} {\enquote
  {\bibinfo {title} {{The reaction $\pi^+ p \to K^+\Sigma^+$ in a unitary
  coupled-channels model}},}\ }\href {\doibase 10.1016/j.nuclphysa.2010.12.010}
  {\bibfield  {journal} {\bibinfo  {journal} {Nucl. Phys. A}\ }\textbf
  {\bibinfo {volume} {851}},\ \bibinfo {pages} {58--98} (\bibinfo {year}
  {2011})},\ \Eprint {http://arxiv.org/abs/1009.3781} {arXiv:1009.3781
  [nucl-th]} \BibitemShut {NoStop}%
\bibitem [{\citenamefont {Wang}\ \emph
  {et~al.}(2024{\natexlab{a}})\citenamefont {Wang}, \citenamefont
  {Mei\ss{}ner}, \citenamefont {R\"onchen},\ and\ \citenamefont
  {Shen}}]{Wang:2023snv}%
  \BibitemOpen
  \bibfield  {author} {\bibinfo {author} {\bibfnamefont {Yu-Fei}\ \bibnamefont
  {Wang}}, \bibinfo {author} {\bibfnamefont {Ulf-G.}\ \bibnamefont
  {Mei\ss{}ner}}, \bibinfo {author} {\bibfnamefont {Deborah}\ \bibnamefont
  {R\"onchen}}, \ and\ \bibinfo {author} {\bibfnamefont {Chao-Wei}\
  \bibnamefont {Shen}},\ }\bibfield  {title} {\enquote {\bibinfo {title}
  {{Examination of the nature of the N* and \ensuremath{\Delta} resonances via
  coupled-channels dynamics}},}\ }\href {\doibase 10.1103/PhysRevC.109.015202}
  {\bibfield  {journal} {\bibinfo  {journal} {Phys. Rev. C}\ }\textbf {\bibinfo
  {volume} {109}},\ \bibinfo {pages} {015202} (\bibinfo {year}
  {2024}{\natexlab{a}})},\ \Eprint {http://arxiv.org/abs/2307.06799}
  {arXiv:2307.06799 [nucl-th]} \BibitemShut {NoStop}%
\bibitem [{\citenamefont {Wang}\ \emph {et~al.}(2025)\citenamefont {Wang},
  \citenamefont {Shen}, \citenamefont {R{\"o}nchen}, \citenamefont
  {Mei{\ss}ner}, \citenamefont {Zou},\ and\ \citenamefont
  {Huang}}]{Wang:2025ecf}%
  \BibitemOpen
  \bibfield  {author} {\bibinfo {author} {\bibfnamefont {Yu-Fei}\ \bibnamefont
  {Wang}}, \bibinfo {author} {\bibfnamefont {Chao-Wei}\ \bibnamefont {Shen}},
  \bibinfo {author} {\bibfnamefont {Deborah}\ \bibnamefont {R{\"o}nchen}},
  \bibinfo {author} {\bibfnamefont {Ulf-G.}\ \bibnamefont {Mei{\ss}ner}},
  \bibinfo {author} {\bibfnamefont {Bing-Song}\ \bibnamefont {Zou}}, \ and\
  \bibinfo {author} {\bibfnamefont {Fei}\ \bibnamefont {Huang}},\ }\bibfield
  {title} {\enquote {\bibinfo {title} {{The nature of the $P_c$ states from
  compositeness criteria}},}\ }\href@noop {} {\  (\bibinfo {year} {2025})},\
  \Eprint {http://arxiv.org/abs/2506.21858} {arXiv:2506.21858 [hep-ph]}
  \BibitemShut {NoStop}%
\bibitem [{\citenamefont {Mai}\ \emph {et~al.}(2021{\natexlab{a}})\citenamefont
  {Mai}, \citenamefont {D\"oring}, \citenamefont {Granados}, \citenamefont
  {Haberzettl}, \citenamefont {Mei\ss{}ner}, \citenamefont {R\"onchen},
  \citenamefont {Strakovsky},\ and\ \citenamefont {Workman}}]{Mai:2021vsw}%
  \BibitemOpen
  \bibfield  {author} {\bibinfo {author} {\bibfnamefont {Maxim}\ \bibnamefont
  {Mai}}, \bibinfo {author} {\bibfnamefont {Michael}\ \bibnamefont {D\"oring}},
  \bibinfo {author} {\bibfnamefont {Carlos}\ \bibnamefont {Granados}}, \bibinfo
  {author} {\bibfnamefont {Helmut}\ \bibnamefont {Haberzettl}}, \bibinfo
  {author} {\bibfnamefont {U.-G.}\ \bibnamefont {Mei\ss{}ner}}, \bibinfo
  {author} {\bibfnamefont {Deborah}\ \bibnamefont {R\"onchen}}, \bibinfo
  {author} {\bibfnamefont {Igor}\ \bibnamefont {Strakovsky}}, \ and\ \bibinfo
  {author} {\bibfnamefont {Ron}\ \bibnamefont {Workman}} (\bibinfo
  {collaboration} {J\"ulich-Bonn-Washington}),\ }\bibfield  {title} {\enquote
  {\bibinfo {title} {{J\"ulich-Bonn-Washington model for pion electroproduction
  multipoles}},}\ }\href {\doibase 10.1103/PhysRevC.103.065204} {\bibfield
  {journal} {\bibinfo  {journal} {Phys. Rev. C}\ }\textbf {\bibinfo {volume}
  {103}},\ \bibinfo {pages} {065204} (\bibinfo {year} {2021}{\natexlab{a}})},\
  \Eprint {http://arxiv.org/abs/2104.07312} {arXiv:2104.07312 [nucl-th]}
  \BibitemShut {NoStop}%
\bibitem [{\citenamefont {Mai}\ \emph {et~al.}(2021{\natexlab{b}})\citenamefont
  {Mai}, \citenamefont {D\"oring}, \citenamefont {Granados}, \citenamefont
  {Haberzettl}, \citenamefont {Hergenrather}, \citenamefont {Mei\ss{}ner},
  \citenamefont {R\"onchen}, \citenamefont {Strakovsky},\ and\ \citenamefont
  {Workman}}]{Mai:2021aui}%
  \BibitemOpen
  \bibfield  {author} {\bibinfo {author} {\bibfnamefont {Maxim}\ \bibnamefont
  {Mai}}, \bibinfo {author} {\bibfnamefont {Michael}\ \bibnamefont {D\"oring}},
  \bibinfo {author} {\bibfnamefont {Carlos}\ \bibnamefont {Granados}}, \bibinfo
  {author} {\bibfnamefont {Helmut}\ \bibnamefont {Haberzettl}}, \bibinfo
  {author} {\bibfnamefont {Jackson}\ \bibnamefont {Hergenrather}}, \bibinfo
  {author} {\bibfnamefont {U.-G.}\ \bibnamefont {Mei\ss{}ner}}, \bibinfo
  {author} {\bibfnamefont {Deborah}\ \bibnamefont {R\"onchen}}, \bibinfo
  {author} {\bibfnamefont {Igor}\ \bibnamefont {Strakovsky}}, \ and\ \bibinfo
  {author} {\bibfnamefont {Ron}\ \bibnamefont {Workman}} (\bibinfo
  {collaboration} {J\"ulich-Bonn-Washington}),\ }\bibfield  {title} {\enquote
  {\bibinfo {title} {{Coupled-channel analysis of pion- and
  eta-electroproduction with the J\"ulich-Bonn-Washington model}},}\ }\href
  {\doibase 10.1103/PhysRevC.106.015201} {\  (\bibinfo {year}
  {2021}{\natexlab{b}}),\ 10.1103/PhysRevC.106.015201},\ \Eprint
  {http://arxiv.org/abs/2111.04774} {arXiv:2111.04774 [nucl-th]} \BibitemShut
  {NoStop}%
\bibitem [{\citenamefont {Mai}\ \emph {et~al.}(2023)\citenamefont {Mai},
  \citenamefont {Hergenrather}, \citenamefont {D\"oring}, \citenamefont {Mart},
  \citenamefont {Mei\ss{}ner}, \citenamefont {R\"onchen},\ and\ \citenamefont
  {Workman}}]{Mai:2023cbp}%
  \BibitemOpen
  \bibfield  {author} {\bibinfo {author} {\bibfnamefont {M.}~\bibnamefont
  {Mai}}, \bibinfo {author} {\bibfnamefont {J.}~\bibnamefont {Hergenrather}},
  \bibinfo {author} {\bibfnamefont {M.}~\bibnamefont {D\"oring}}, \bibinfo
  {author} {\bibfnamefont {T.}~\bibnamefont {Mart}}, \bibinfo {author}
  {\bibfnamefont {Ulf-G.}\ \bibnamefont {Mei\ss{}ner}}, \bibinfo {author}
  {\bibfnamefont {D.}~\bibnamefont {R\"onchen}}, \ and\ \bibinfo {author}
  {\bibfnamefont {R.}~\bibnamefont {Workman}} (\bibinfo {collaboration}
  {J\"ulich\textendash{}Bonn\textendash{}Washington}),\ }\bibfield  {title}
  {\enquote {\bibinfo {title} {{Inclusion of $K\Lambda $ electroproduction data
  in a coupled channel analysis}},}\ }\href {\doibase
  10.1140/epja/s10050-023-01188-0} {\bibfield  {journal} {\bibinfo  {journal}
  {Eur. Phys. J. A}\ }\textbf {\bibinfo {volume} {59}},\ \bibinfo {pages} {286}
  (\bibinfo {year} {2023})},\ \Eprint {http://arxiv.org/abs/2307.10051}
  {arXiv:2307.10051 [nucl-th]} \BibitemShut {NoStop}%
\bibitem [{\citenamefont {Wang}\ \emph
  {et~al.}(2024{\natexlab{b}})\citenamefont {Wang}, \citenamefont {D\"oring},
  \citenamefont {Hergenrather}, \citenamefont {Mai}, \citenamefont {Mart},
  \citenamefont {Mei\ss{}ner}, \citenamefont {R\"onchen},\ and\ \citenamefont
  {Workman}}]{Wang:2024byt}%
  \BibitemOpen
  \bibfield  {author} {\bibinfo {author} {\bibfnamefont {Yu-Fei}\ \bibnamefont
  {Wang}}, \bibinfo {author} {\bibfnamefont {Michael}\ \bibnamefont
  {D\"oring}}, \bibinfo {author} {\bibfnamefont {Jackson}\ \bibnamefont
  {Hergenrather}}, \bibinfo {author} {\bibfnamefont {Maxim}\ \bibnamefont
  {Mai}}, \bibinfo {author} {\bibfnamefont {Terry}\ \bibnamefont {Mart}},
  \bibinfo {author} {\bibfnamefont {Ulf-G.}\ \bibnamefont {Mei\ss{}ner}},
  \bibinfo {author} {\bibfnamefont {Deborah}\ \bibnamefont {R\"onchen}}, \ and\
  \bibinfo {author} {\bibfnamefont {Ronald}\ \bibnamefont {Workman}} (\bibinfo
  {collaboration} {J\"ulich-Bonn-Washington}),\ }\bibfield  {title} {\enquote
  {\bibinfo {title} {{Global Data-Driven Determination of Baryon Transition
  Form Factors}},}\ }\href {\doibase 10.1103/PhysRevLett.133.101901} {\bibfield
   {journal} {\bibinfo  {journal} {Phys. Rev. Lett.}\ }\textbf {\bibinfo
  {volume} {133}},\ \bibinfo {pages} {101901} (\bibinfo {year}
  {2024}{\natexlab{b}})},\ \Eprint {http://arxiv.org/abs/2404.17444}
  {arXiv:2404.17444 [nucl-th]} \BibitemShut {NoStop}%
\bibitem [{\citenamefont {Huang}\ \emph {et~al.}(2012)\citenamefont {Huang},
  \citenamefont {Döring}, \citenamefont {Haberzettl}, \citenamefont
  {Haidenbauer}, \citenamefont {Hanhart}, \citenamefont {Krewald},
  \citenamefont {Mei\ss{}ner},\ and\ \citenamefont {Nakayama}}]{Huang:2011as}%
  \BibitemOpen
  \bibfield  {author} {\bibinfo {author} {\bibfnamefont {F.}~\bibnamefont
  {Huang}}, \bibinfo {author} {\bibfnamefont {M.}~\bibnamefont {Döring}},
  \bibinfo {author} {\bibfnamefont {H.}~\bibnamefont {Haberzettl}}, \bibinfo
  {author} {\bibfnamefont {J.}~\bibnamefont {Haidenbauer}}, \bibinfo {author}
  {\bibfnamefont {C.}~\bibnamefont {Hanhart}}, \bibinfo {author} {\bibfnamefont
  {S.}~\bibnamefont {Krewald}}, \bibinfo {author} {\bibfnamefont {U.-G.}\
  \bibnamefont {Mei\ss{}ner}}, \ and\ \bibinfo {author} {\bibfnamefont
  {K.}~\bibnamefont {Nakayama}},\ }\bibfield  {title} {\enquote {\bibinfo
  {title} {{Pion photoproduction in a dynamical coupled-channels model}},}\
  }\href {\doibase 10.1103/PhysRevC.85.054003} {\bibfield  {journal} {\bibinfo
  {journal} {Phys. Rev. C}\ }\textbf {\bibinfo {volume} {85}},\ \bibinfo
  {pages} {054003} (\bibinfo {year} {2012})},\ \Eprint
  {http://arxiv.org/abs/1110.3833} {arXiv:1110.3833 [nucl-th]} \BibitemShut
  {NoStop}%
\bibitem [{\citenamefont {{Figures representing the full fit result of this
  study, including a display of all data}}()}]{Juelichmodel:online}%
  \BibitemOpen
  \bibfield  {author} {\bibinfo {author} {\bibnamefont {{Figures representing
  the full fit result of this study, including a display of all data}}},\
  }\href@noop {} {}\bibinfo {howpublished}
  {\url{http://collaborations.fz-juelich.de/ikp/meson-baryon/juelich_amplitudes.html}}\BibitemShut
  {NoStop}%
\bibitem [{\citenamefont {Workman}\ \emph {et~al.}(2012)\citenamefont
  {Workman}, \citenamefont {Arndt}, \citenamefont {Briscoe}, \citenamefont
  {Paris},\ and\ \citenamefont {Strakovsky}}]{Workman:2012hx}%
  \BibitemOpen
  \bibfield  {author} {\bibinfo {author} {\bibfnamefont {R.~L.}\ \bibnamefont
  {Workman}}, \bibinfo {author} {\bibfnamefont {R.~A.}\ \bibnamefont {Arndt}},
  \bibinfo {author} {\bibfnamefont {W.~J.}\ \bibnamefont {Briscoe}}, \bibinfo
  {author} {\bibfnamefont {M.~W.}\ \bibnamefont {Paris}}, \ and\ \bibinfo
  {author} {\bibfnamefont {I.~I.}\ \bibnamefont {Strakovsky}},\ }\bibfield
  {title} {\enquote {\bibinfo {title} {{Parameterization dependence of T matrix
  poles and eigenphases from a fit to $\pi$N elastic scattering data}},}\
  }\href {\doibase 10.1103/PhysRevC.86.035202} {\bibfield  {journal} {\bibinfo
  {journal} {Phys. Rev. C}\ }\textbf {\bibinfo {volume} {86}},\ \bibinfo
  {pages} {035202} (\bibinfo {year} {2012})},\ \Eprint
  {http://arxiv.org/abs/1204.2277} {arXiv:1204.2277 [hep-ph]} \BibitemShut
  {NoStop}%
\bibitem [{\citenamefont {{SAID/GWU website}}()}]{SAID}%
  \BibitemOpen
  \bibfield  {author} {\bibinfo {author} {\bibnamefont {{SAID/GWU website}}},\
  }\href@noop {} {}\bibinfo {howpublished}
  {\url{http://gwdac.phys.gwu.edu}}\BibitemShut {NoStop}%
\bibitem [{\citenamefont {{Website of Bonn-Gatchina group with analysis
  results}}()}]{BnGa_web}%
  \BibitemOpen
  \bibfield  {author} {\bibinfo {author} {\bibnamefont {{Website of
  Bonn-Gatchina group with analysis results}}},\ }\href@noop {} {}\bibinfo
  {howpublished} {\url{https://pwa.hiskp.uni-bonn.de}}\BibitemShut {NoStop}%
\bibitem [{\citenamefont {Kim}\ \emph {et~al.}(2023)\citenamefont {Kim} \emph
  {et~al.}}]{CLAS:2023ddn}%
  \BibitemOpen
  \bibfield  {author} {\bibinfo {author} {\bibfnamefont {C.~W.}\ \bibnamefont
  {Kim}} \emph {et~al.} (\bibinfo {collaboration} {CLAS}),\ }\bibfield  {title}
  {\enquote {\bibinfo {title} {{Measurement of the helicity asymmetry ${\mathbb
  {E}}$ for the $\vec {\gamma }\vec {p} \rightarrow p \pi ^0$ reaction in the
  resonance region: The~CLAS~Collaboration}},}\ }\href {\doibase
  10.1140/epja/s10050-023-01123-3} {\bibfield  {journal} {\bibinfo  {journal}
  {Eur. Phys. J. A}\ }\textbf {\bibinfo {volume} {59}},\ \bibinfo {pages} {217}
  (\bibinfo {year} {2023})},\ \Eprint {http://arxiv.org/abs/2305.08616}
  {arXiv:2305.08616 [nucl-ex]} \BibitemShut {NoStop}%
\bibitem [{\citenamefont {Ahrens}\ \emph {et~al.}(2002)\citenamefont {Ahrens}
  \emph {et~al.}}]{GDH:2002pkk}%
  \BibitemOpen
  \bibfield  {author} {\bibinfo {author} {\bibfnamefont {J.}~\bibnamefont
  {Ahrens}} \emph {et~al.} (\bibinfo {collaboration} {GDH, A2}),\ }\bibfield
  {title} {\enquote {\bibinfo {title} {{The Helicity amplitudes A(1/2) and
  A(3/2) for the D(13)(1520) resonance obtained from the polarized-gamma
  polarized-p ---\ensuremath{>} p pi0 reaction}},}\ }\href {\doibase
  10.1103/PhysRevLett.88.232002} {\bibfield  {journal} {\bibinfo  {journal}
  {Phys. Rev. Lett.}\ }\textbf {\bibinfo {volume} {88}},\ \bibinfo {pages}
  {232002} (\bibinfo {year} {2002})},\ \Eprint
  {http://arxiv.org/abs/hep-ex/0203006} {arXiv:hep-ex/0203006} \BibitemShut
  {NoStop}%
\bibitem [{\citenamefont {Ahrens}\ \emph {et~al.}(2004)\citenamefont {Ahrens}
  \emph {et~al.}}]{GDH:2004ydy}%
  \BibitemOpen
  \bibfield  {author} {\bibinfo {author} {\bibfnamefont {J.}~\bibnamefont
  {Ahrens}} \emph {et~al.} (\bibinfo {collaboration} {GDH, A2}),\ }\bibfield
  {title} {\enquote {\bibinfo {title} {{Helicity dependence of the $\gamma p\to
  N \pi$ channels and multipole analysis in the $\Delta$ region}},}\ }\href
  {\doibase 10.1140/epja/i2003-10216-x} {\bibfield  {journal} {\bibinfo
  {journal} {Eur. Phys. J. A}\ }\textbf {\bibinfo {volume} {21}},\ \bibinfo
  {pages} {323--333} (\bibinfo {year} {2004})}\BibitemShut {NoStop}%
\bibitem [{\citenamefont {Ahrens}\ \emph {et~al.}(2006)\citenamefont {Ahrens}
  \emph {et~al.}}]{Ahrens:2006gp}%
  \BibitemOpen
  \bibfield  {author} {\bibinfo {author} {\bibfnamefont {J.}~\bibnamefont
  {Ahrens}} \emph {et~al.},\ }\bibfield  {title} {\enquote {\bibinfo {title}
  {{Measurement of the helicity dependence for the gamma p ---\ensuremath{>} n
  pi+ channel in the second resonance region}},}\ }\href {\doibase
  10.1103/PhysRevC.74.045204} {\bibfield  {journal} {\bibinfo  {journal} {Phys.
  Rev. C}\ }\textbf {\bibinfo {volume} {74}},\ \bibinfo {pages} {045204}
  (\bibinfo {year} {2006})}\BibitemShut {NoStop}%
\bibitem [{\citenamefont {Afzal}\ \emph {et~al.}(2024)\citenamefont {Afzal}
  \emph {et~al.}}]{A2:2024ydg}%
  \BibitemOpen
  \bibfield  {author} {\bibinfo {author} {\bibfnamefont {F.}~\bibnamefont
  {Afzal}} \emph {et~al.} (\bibinfo {collaboration} {A2}),\ }\bibfield  {title}
  {\enquote {\bibinfo {title} {{First Measurement Using Elliptically Polarized
  Photons of the Double-Polarization Observable E for
  \ensuremath{\gamma}p\textrightarrow{}p\ensuremath{\pi}0 and
  \ensuremath{\gamma}p\textrightarrow{}n\ensuremath{\pi}+}},}\ }\href {\doibase
  10.1103/PhysRevLett.132.121902} {\bibfield  {journal} {\bibinfo  {journal}
  {Phys. Rev. Lett.}\ }\textbf {\bibinfo {volume} {132}},\ \bibinfo {pages}
  {121902} (\bibinfo {year} {2024})},\ \Eprint
  {http://arxiv.org/abs/2402.05531} {arXiv:2402.05531 [nucl-ex]} \BibitemShut
  {NoStop}%
\bibitem [{\citenamefont {Gottschall}\ \emph {et~al.}(2014)\citenamefont
  {Gottschall} \emph {et~al.}}]{CBELSATAPS:2013btn}%
  \BibitemOpen
  \bibfield  {author} {\bibinfo {author} {\bibfnamefont {M.}~\bibnamefont
  {Gottschall}} \emph {et~al.} (\bibinfo {collaboration} {CBELSA/TAPS}),\
  }\bibfield  {title} {\enquote {\bibinfo {title} {{First measurement of the
  helicity asymmetry for $\gamma p\rightarrow p\pi^0$ in the resonance
  region}},}\ }\href {\doibase 10.1103/PhysRevLett.112.012003} {\bibfield
  {journal} {\bibinfo  {journal} {Phys. Rev. Lett.}\ }\textbf {\bibinfo
  {volume} {112}},\ \bibinfo {pages} {012003} (\bibinfo {year} {2014})},\
  \Eprint {http://arxiv.org/abs/1312.2187} {arXiv:1312.2187 [nucl-ex]}
  \BibitemShut {NoStop}%
\bibitem [{\citenamefont {Gottschall}\ \emph {et~al.}(2021)\citenamefont
  {Gottschall} \emph {et~al.}}]{CBELSATAPS:2019hhr}%
  \BibitemOpen
  \bibfield  {author} {\bibinfo {author} {\bibfnamefont {M.}~\bibnamefont
  {Gottschall}} \emph {et~al.} (\bibinfo {collaboration} {CBELSA/TAPS}),\
  }\bibfield  {title} {\enquote {\bibinfo {title} {{Measurement of the helicity
  asymmetry $E$ for the reaction $ \gamma p\rightarrow \pi ^0 p$}},}\ }\href
  {\doibase 10.1140/epja/s10050-020-00334-2} {\bibfield  {journal} {\bibinfo
  {journal} {Eur. Phys. J. A}\ }\textbf {\bibinfo {volume} {57}},\ \bibinfo
  {pages} {40} (\bibinfo {year} {2021})},\ \Eprint
  {http://arxiv.org/abs/1904.12560} {arXiv:1904.12560 [nucl-ex]} \BibitemShut
  {NoStop}%
\bibitem [{\citenamefont {Strauch}\ \emph {et~al.}(2015)\citenamefont {Strauch}
  \emph {et~al.}}]{CLAS:2015ykk}%
  \BibitemOpen
  \bibfield  {author} {\bibinfo {author} {\bibfnamefont {S.}~\bibnamefont
  {Strauch}} \emph {et~al.} (\bibinfo {collaboration} {CLAS}),\ }\bibfield
  {title} {\enquote {\bibinfo {title} {{First Measurement of the Polarization
  Observable E in the $\vec p(\vec \gamma,\pi^+)n$ Reaction up to 2.25 GeV}},}\
  }\href {\doibase 10.1016/j.physletb.2015.08.053} {\bibfield  {journal}
  {\bibinfo  {journal} {Phys. Lett. B}\ }\textbf {\bibinfo {volume} {750}},\
  \bibinfo {pages} {53--58} (\bibinfo {year} {2015})},\ \Eprint
  {http://arxiv.org/abs/1503.05163} {arXiv:1503.05163 [nucl-ex]} \BibitemShut
  {NoStop}%
\bibitem [{\citenamefont {Senderovich}\ \emph {et~al.}(2016)\citenamefont
  {Senderovich} \emph {et~al.}}]{CLAS:2015pjm}%
  \BibitemOpen
  \bibfield  {author} {\bibinfo {author} {\bibfnamefont {I.}~\bibnamefont
  {Senderovich}} \emph {et~al.} (\bibinfo {collaboration} {CLAS}),\ }\bibfield
  {title} {\enquote {\bibinfo {title} {{First measurement of the helicity
  asymmetry $E$ in $\eta$ photoproduction on the proton}},}\ }\href {\doibase
  10.1016/j.physletb.2016.01.044} {\bibfield  {journal} {\bibinfo  {journal}
  {Phys. Lett. B}\ }\textbf {\bibinfo {volume} {755}},\ \bibinfo {pages}
  {64--69} (\bibinfo {year} {2016})},\ \Eprint
  {http://arxiv.org/abs/1507.00325} {arXiv:1507.00325 [nucl-ex]} \BibitemShut
  {NoStop}%
\bibitem [{\citenamefont {Zachariou}\ \emph {et~al.}(2021)\citenamefont
  {Zachariou} \emph {et~al.}}]{CLAS:2021udy}%
  \BibitemOpen
  \bibfield  {author} {\bibinfo {author} {\bibfnamefont {N.}~\bibnamefont
  {Zachariou}} \emph {et~al.} (\bibinfo {collaboration} {CLAS}),\ }\bibfield
  {title} {\enquote {\bibinfo {title} {{Double polarisation observable $\mathbb
  G$ for single pion photoproduction from the proton}},}\ }\href {\doibase
  10.1016/j.physletb.2021.136304} {\bibfield  {journal} {\bibinfo  {journal}
  {Phys. Lett. B}\ }\textbf {\bibinfo {volume} {817}},\ \bibinfo {pages}
  {136304} (\bibinfo {year} {2021})}\BibitemShut {NoStop}%
\bibitem [{\citenamefont {Mornacchi}\ \emph {et~al.}(2024)\citenamefont
  {Mornacchi} \emph {et~al.}}]{A2CollaborationatMAMI:2023twj}%
  \BibitemOpen
  \bibfield  {author} {\bibinfo {author} {\bibfnamefont {E.}~\bibnamefont
  {Mornacchi}} \emph {et~al.} (\bibinfo {collaboration} {A2 Collaboration at
  MAMI}),\ }\bibfield  {title} {\enquote {\bibinfo {title} {{Evaluation of the
  E2/M1 ratio in the N\textrightarrow{}\ensuremath{\Delta}(1232) transition
  from the \ensuremath{\gamma}p\textrightarrow{}p\ensuremath{\pi}0
  reaction}},}\ }\href {\doibase 10.1103/PhysRevC.109.055201} {\bibfield
  {journal} {\bibinfo  {journal} {Phys. Rev. C}\ }\textbf {\bibinfo {volume}
  {109}},\ \bibinfo {pages} {055201} (\bibinfo {year} {2024})},\ \Eprint
  {http://arxiv.org/abs/2312.08211} {arXiv:2312.08211 [nucl-ex]} \BibitemShut
  {NoStop}%
\bibitem [{\citenamefont {Hashimoto}\ \emph {et~al.}(2022)\citenamefont
  {Hashimoto} \emph {et~al.}}]{LEPS2BGOegg:2022dop}%
  \BibitemOpen
  \bibfield  {author} {\bibinfo {author} {\bibfnamefont {T.}~\bibnamefont
  {Hashimoto}} \emph {et~al.} (\bibinfo {collaboration} {LEPS2/BGOegg, BGOegg,
  LEPS2}),\ }\bibfield  {title} {\enquote {\bibinfo {title} {{Differential
  cross sections and photon beam asymmetries of $\eta$ photoproduction on the
  proton at $E_\gamma$ = 1.3-2.4 GeV}},}\ }\href {\doibase
  10.1103/PhysRevC.106.035201} {\bibfield  {journal} {\bibinfo  {journal}
  {Phys. Rev. C}\ }\textbf {\bibinfo {volume} {106}},\ \bibinfo {pages}
  {035201} (\bibinfo {year} {2022})},\ \Eprint
  {http://arxiv.org/abs/2202.13688} {arXiv:2202.13688 [nucl-ex]} \BibitemShut
  {NoStop}%
\bibitem [{\citenamefont {Clark}\ \emph {et~al.}(2025)\citenamefont {Clark}
  \emph {et~al.}}]{CLAS:2024bzi}%
  \BibitemOpen
  \bibfield  {author} {\bibinfo {author} {\bibfnamefont {L.}~\bibnamefont
  {Clark}} \emph {et~al.} (\bibinfo {collaboration} {CLAS}),\ }\bibfield
  {title} {\enquote {\bibinfo {title} {{Photoproduction of the
  \ensuremath{\Sigma}+ hyperon using linearly polarized photons with CLAS}},}\
  }\href {\doibase 10.1103/PhysRevC.111.025204} {\bibfield  {journal} {\bibinfo
   {journal} {Phys. Rev. C}\ }\textbf {\bibinfo {volume} {111}},\ \bibinfo
  {pages} {025204} (\bibinfo {year} {2025})},\ \Eprint
  {http://arxiv.org/abs/2404.19404} {arXiv:2404.19404 [nucl-ex]} \BibitemShut
  {NoStop}%
\bibitem [{\citenamefont {James}\ and\ \citenamefont
  {Roos}(1975)}]{James:1975dr}%
  \BibitemOpen
  \bibfield  {author} {\bibinfo {author} {\bibfnamefont {F.}~\bibnamefont
  {James}}\ and\ \bibinfo {author} {\bibfnamefont {M.}~\bibnamefont {Roos}},\
  }\bibfield  {title} {\enquote {\bibinfo {title} {{Minuit: A System for
  Function Minimization and Analysis of the Parameter Errors and
  Correlations}},}\ }\href {\doibase 10.1016/0010-4655(75)90039-9} {\bibfield
  {journal} {\bibinfo  {journal} {Comput. Phys. Commun.}\ }\textbf {\bibinfo
  {volume} {10}},\ \bibinfo {pages} {343--367} (\bibinfo {year}
  {1975})}\BibitemShut {NoStop}%
\bibitem [{\citenamefont {{J\"{u}lich Supercomputing Centre}}(2021)}]{JURECA}%
  \BibitemOpen
  \bibfield  {author} {\bibinfo {author} {\bibnamefont {{J\"{u}lich
  Supercomputing Centre}}},\ }\bibfield  {title} {\enquote {\bibinfo {title}
  {{JURECA: Data Centric and Booster Modules implementing the Modular
  Supercomputing Architecture at J\"{u}lich Supercomputing Centre}},}\ }\href
  {\doibase 10.17815/jlsrf-7-182} {\bibfield  {journal} {\bibinfo  {journal}
  {Journal of large-scale research facilities}\ }\textbf {\bibinfo {volume}
  {7}} (\bibinfo {year} {2021}),\ 10.17815/jlsrf-7-182}\BibitemShut {NoStop}%
\bibitem [{\citenamefont {Tibshirani}(1996)}]{Tibshirani:1996fxl}%
  \BibitemOpen
  \bibfield  {author} {\bibinfo {author} {\bibfnamefont {Robert}\ \bibnamefont
  {Tibshirani}},\ }\bibfield  {title} {\enquote {\bibinfo {title} {{Regression
  Shrinkage and Selection Via the Lasso}},}\ }\href {\doibase
  10.1111/j.2517-6161.1996.tb02080.x} {\bibfield  {journal} {\bibinfo
  {journal} {J. Roy. Statist. Soc. B}\ }\textbf {\bibinfo {volume} {58}},\
  \bibinfo {pages} {267--288} (\bibinfo {year} {1996})}\BibitemShut {NoStop}%
\bibitem [{\citenamefont {Landay}\ \emph {et~al.}(2017)\citenamefont {Landay},
  \citenamefont {D\"oring}, \citenamefont {Fern\'andez-Ram\'\i{}rez},
  \citenamefont {Hu},\ and\ \citenamefont {Molina}}]{Landay:2016cjw}%
  \BibitemOpen
  \bibfield  {author} {\bibinfo {author} {\bibfnamefont {J.}~\bibnamefont
  {Landay}}, \bibinfo {author} {\bibfnamefont {M.}~\bibnamefont {D\"oring}},
  \bibinfo {author} {\bibfnamefont {C.}~\bibnamefont
  {Fern\'andez-Ram\'\i{}rez}}, \bibinfo {author} {\bibfnamefont
  {B.}~\bibnamefont {Hu}}, \ and\ \bibinfo {author} {\bibfnamefont
  {R.}~\bibnamefont {Molina}},\ }\bibfield  {title} {\enquote {\bibinfo {title}
  {{Model Selection for Pion Photoproduction}},}\ }\href {\doibase
  10.1103/PhysRevC.95.015203} {\bibfield  {journal} {\bibinfo  {journal} {Phys.
  Rev. C}\ }\textbf {\bibinfo {volume} {95}},\ \bibinfo {pages} {015203}
  (\bibinfo {year} {2017})},\ \Eprint {http://arxiv.org/abs/1610.07547}
  {arXiv:1610.07547 [nucl-th]} \BibitemShut {NoStop}%
\bibitem [{\citenamefont {Landay}\ \emph {et~al.}(2019)\citenamefont {Landay},
  \citenamefont {Mai}, \citenamefont {D\"oring}, \citenamefont {Haberzettl},\
  and\ \citenamefont {Nakayama}}]{Landay:2018wgf}%
  \BibitemOpen
  \bibfield  {author} {\bibinfo {author} {\bibfnamefont {J.}~\bibnamefont
  {Landay}}, \bibinfo {author} {\bibfnamefont {M.}~\bibnamefont {Mai}},
  \bibinfo {author} {\bibfnamefont {M.}~\bibnamefont {D\"oring}}, \bibinfo
  {author} {\bibfnamefont {H.}~\bibnamefont {Haberzettl}}, \ and\ \bibinfo
  {author} {\bibfnamefont {K.}~\bibnamefont {Nakayama}},\ }\bibfield  {title}
  {\enquote {\bibinfo {title} {{Towards the Minimal Spectrum of Excited
  Baryons}},}\ }\href {\doibase 10.1103/PhysRevD.99.016001} {\bibfield
  {journal} {\bibinfo  {journal} {Phys. Rev. D}\ }\textbf {\bibinfo {volume}
  {99}},\ \bibinfo {pages} {016001} (\bibinfo {year} {2019})},\ \Eprint
  {http://arxiv.org/abs/1810.00075} {arXiv:1810.00075 [nucl-th]} \BibitemShut
  {NoStop}%
\bibitem [{\citenamefont {D\"oring}\ \emph {et~al.}(2016)\citenamefont
  {D\"oring}, \citenamefont {Revier}, \citenamefont {R\"onchen},\ and\
  \citenamefont {Workman}}]{Doring:2016snk}%
  \BibitemOpen
  \bibfield  {author} {\bibinfo {author} {\bibfnamefont {M.}~\bibnamefont
  {D\"oring}}, \bibinfo {author} {\bibfnamefont {J.}~\bibnamefont {Revier}},
  \bibinfo {author} {\bibfnamefont {D.}~\bibnamefont {R\"onchen}}, \ and\
  \bibinfo {author} {\bibfnamefont {R.~L.}\ \bibnamefont {Workman}},\
  }\bibfield  {title} {\enquote {\bibinfo {title} {{Correlations of $\pi$N
  partial waves for multireaction analyses}},}\ }\href {\doibase
  10.1103/PhysRevC.93.065205} {\bibfield  {journal} {\bibinfo  {journal} {Phys.
  Rev. C}\ }\textbf {\bibinfo {volume} {93}},\ \bibinfo {pages} {065205}
  (\bibinfo {year} {2016})},\ \Eprint {http://arxiv.org/abs/1603.07265}
  {arXiv:1603.07265 [nucl-th]} \BibitemShut {NoStop}%
\bibitem [{\citenamefont {Arndt}\ \emph {et~al.}(1995)\citenamefont {Arndt},
  \citenamefont {Strakovsky}, \citenamefont {Workman},\ and\ \citenamefont
  {Pavan}}]{Arndt:1995bj}%
  \BibitemOpen
  \bibfield  {author} {\bibinfo {author} {\bibfnamefont {Richard~A.}\
  \bibnamefont {Arndt}}, \bibinfo {author} {\bibfnamefont {Igor~I.}\
  \bibnamefont {Strakovsky}}, \bibinfo {author} {\bibfnamefont {Ron~L.}\
  \bibnamefont {Workman}}, \ and\ \bibinfo {author} {\bibfnamefont
  {Marcello~M.}\ \bibnamefont {Pavan}},\ }\bibfield  {title} {\enquote
  {\bibinfo {title} {{Updated analysis of pi N elastic scattering data to
  2.1-GeV: The Baryon spectrum}},}\ }\href {\doibase 10.1103/PhysRevC.52.2120}
  {\bibfield  {journal} {\bibinfo  {journal} {Phys. Rev. C}\ }\textbf {\bibinfo
  {volume} {52}},\ \bibinfo {pages} {2120--2130} (\bibinfo {year} {1995})},\
  \Eprint {http://arxiv.org/abs/nucl-th/9505040} {arXiv:nucl-th/9505040}
  \BibitemShut {NoStop}%
\bibitem [{\citenamefont {Briscoe}\ \emph {et~al.}(2020)\citenamefont
  {Briscoe}, \citenamefont {Kudryavtsev}, \citenamefont {Strakovsky},
  \citenamefont {Tarasov},\ and\ \citenamefont {Workman}}]{Briscoe:2020qat}%
  \BibitemOpen
  \bibfield  {author} {\bibinfo {author} {\bibfnamefont {W.~J.}\ \bibnamefont
  {Briscoe}}, \bibinfo {author} {\bibfnamefont {A.~E.}\ \bibnamefont
  {Kudryavtsev}}, \bibinfo {author} {\bibfnamefont {I.~I.}\ \bibnamefont
  {Strakovsky}}, \bibinfo {author} {\bibfnamefont {V.~E.}\ \bibnamefont
  {Tarasov}}, \ and\ \bibinfo {author} {\bibfnamefont {R.~L.}\ \bibnamefont
  {Workman}},\ }\bibfield  {title} {\enquote {\bibinfo {title} {{Threshold $\pi
  ^-$ photoproduction on the neutron}},}\ }\href {\doibase
  10.1140/epja/s10050-020-00221-w} {\bibfield  {journal} {\bibinfo  {journal}
  {Eur. Phys. J. A}\ }\textbf {\bibinfo {volume} {56}},\ \bibinfo {pages} {218}
  (\bibinfo {year} {2020})},\ \Eprint {http://arxiv.org/abs/2004.01742}
  {arXiv:2004.01742 [nucl-th]} \BibitemShut {NoStop}%
\bibitem [{\citenamefont {Navas}\ \emph {et~al.}(2024)\citenamefont {Navas}
  \emph {et~al.}}]{ParticleDataGroup:2024cfk}%
  \BibitemOpen
  \bibfield  {author} {\bibinfo {author} {\bibfnamefont {S.}~\bibnamefont
  {Navas}} \emph {et~al.} (\bibinfo {collaboration} {Particle Data Group}),\
  }\bibfield  {title} {\enquote {\bibinfo {title} {{Review of particle
  physics}},}\ }\href {\doibase 10.1103/PhysRevD.110.030001} {\bibfield
  {journal} {\bibinfo  {journal} {Phys. Rev. D}\ }\textbf {\bibinfo {volume}
  {110}},\ \bibinfo {pages} {030001} (\bibinfo {year} {2024})}\BibitemShut
  {NoStop}%
\bibitem [{\citenamefont {R\"onchen}\ \emph {et~al.}(2018)\citenamefont
  {R\"onchen}, \citenamefont {D\"oring},\ and\ \citenamefont
  {Mei\ss{}ner}}]{Ronchen:2018ury}%
  \BibitemOpen
  \bibfield  {author} {\bibinfo {author} {\bibfnamefont {D.}~\bibnamefont
  {R\"onchen}}, \bibinfo {author} {\bibfnamefont {M.}~\bibnamefont {D\"oring}},
  \ and\ \bibinfo {author} {\bibfnamefont {U.~G}\ \bibnamefont {Mei\ss{}ner}},\
  }\bibfield  {title} {\enquote {\bibinfo {title} {{The impact of
  $K^{+}\Lambda$ photoproduction on the resonance spectrum}},}\ }\href
  {\doibase 10.1140/epja/i2018-12541-3} {\bibfield  {journal} {\bibinfo
  {journal} {Eur. Phys. J. A}\ }\textbf {\bibinfo {volume} {54}},\ \bibinfo
  {pages} {110} (\bibinfo {year} {2018})},\ \Eprint
  {http://arxiv.org/abs/1801.10458} {arXiv:1801.10458 [nucl-th]} \BibitemShut
  {NoStop}%
\bibitem [{\citenamefont {Döring}\ \emph {et~al.}(2009)\citenamefont
  {Döring}, \citenamefont {Hanhart}, \citenamefont {Huang}, \citenamefont
  {Krewald},\ and\ \citenamefont {Mei\ss{}ner}}]{Doring:2009yv}%
  \BibitemOpen
  \bibfield  {author} {\bibinfo {author} {\bibfnamefont {M.}~\bibnamefont
  {Döring}}, \bibinfo {author} {\bibfnamefont {C.}~\bibnamefont {Hanhart}},
  \bibinfo {author} {\bibfnamefont {F.}~\bibnamefont {Huang}}, \bibinfo
  {author} {\bibfnamefont {S.}~\bibnamefont {Krewald}}, \ and\ \bibinfo
  {author} {\bibfnamefont {U.-G.}\ \bibnamefont {Mei\ss{}ner}},\ }\bibfield
  {title} {\enquote {\bibinfo {title} {{Analytic properties of the scattering
  amplitude and resonances parameters in a meson exchange model}},}\ }\href
  {\doibase 10.1016/j.nuclphysa.2009.08.010} {\bibfield  {journal} {\bibinfo
  {journal} {Nucl. Phys. A}\ }\textbf {\bibinfo {volume} {829}},\ \bibinfo
  {pages} {170--209} (\bibinfo {year} {2009})},\ \Eprint
  {http://arxiv.org/abs/0903.4337} {arXiv:0903.4337 [nucl-th]} \BibitemShut
  {NoStop}%
\bibitem [{\citenamefont {Heuser}\ \emph {et~al.}(2024)\citenamefont {Heuser},
  \citenamefont {Chanturia}, \citenamefont {Guo}, \citenamefont {Hanhart},
  \citenamefont {Hoferichter},\ and\ \citenamefont {Kubis}}]{Heuser:2024biq}%
  \BibitemOpen
  \bibfield  {author} {\bibinfo {author} {\bibfnamefont {L.~A.}\ \bibnamefont
  {Heuser}}, \bibinfo {author} {\bibfnamefont {G.}~\bibnamefont {Chanturia}},
  \bibinfo {author} {\bibfnamefont {F.~K.}\ \bibnamefont {Guo}}, \bibinfo
  {author} {\bibfnamefont {C.}~\bibnamefont {Hanhart}}, \bibinfo {author}
  {\bibfnamefont {M.}~\bibnamefont {Hoferichter}}, \ and\ \bibinfo {author}
  {\bibfnamefont {B.}~\bibnamefont {Kubis}},\ }\bibfield  {title} {\enquote
  {\bibinfo {title} {{From pole parameters to line shapes and branching
  ratios}},}\ }\href {\doibase 10.1140/epjc/s10052-024-12884-6} {\bibfield
  {journal} {\bibinfo  {journal} {Eur. Phys. J. C}\ }\textbf {\bibinfo {volume}
  {84}},\ \bibinfo {pages} {599} (\bibinfo {year} {2024})},\ \Eprint
  {http://arxiv.org/abs/2403.15539} {arXiv:2403.15539 [hep-ph]} \BibitemShut
  {NoStop}%
\bibitem [{\citenamefont {Tiator}\ \emph {et~al.}(2018)\citenamefont {Tiator},
  \citenamefont {Gorchtein}, \citenamefont {Kashevarov}, \citenamefont
  {Nikonov}, \citenamefont {Ostrick}, \citenamefont
  {Had{\v{z}}imehmedovi{\'c}}, \citenamefont {Omerovi{\'c}}, \citenamefont
  {Osmanovi{\'c}}, \citenamefont {Stahov},\ and\ \citenamefont
  {{\v{S}}varc}}]{Tiator:2018heh}%
  \BibitemOpen
  \bibfield  {author} {\bibinfo {author} {\bibfnamefont {L.}~\bibnamefont
  {Tiator}}, \bibinfo {author} {\bibfnamefont {M.}~\bibnamefont {Gorchtein}},
  \bibinfo {author} {\bibfnamefont {V.~L.}\ \bibnamefont {Kashevarov}},
  \bibinfo {author} {\bibfnamefont {K.}~\bibnamefont {Nikonov}}, \bibinfo
  {author} {\bibfnamefont {M.}~\bibnamefont {Ostrick}}, \bibinfo {author}
  {\bibfnamefont {M.}~\bibnamefont {Had{\v{z}}imehmedovi{\'c}}}, \bibinfo
  {author} {\bibfnamefont {R.}~\bibnamefont {Omerovi{\'c}}}, \bibinfo {author}
  {\bibfnamefont {H.}~\bibnamefont {Osmanovi{\'c}}}, \bibinfo {author}
  {\bibfnamefont {J.}~\bibnamefont {Stahov}}, \ and\ \bibinfo {author}
  {\bibfnamefont {A.}~\bibnamefont {{\v{S}}varc}},\ }\bibfield  {title}
  {\enquote {\bibinfo {title} {{Eta and Etaprime Photoproduction on the Nucleon
  with the Isobar Model EtaMAID2018}},}\ }\href {\doibase
  10.1140/epja/i2018-12643-x} {\bibfield  {journal} {\bibinfo  {journal} {Eur.
  Phys. J. A}\ }\textbf {\bibinfo {volume} {54}},\ \bibinfo {pages} {210}
  (\bibinfo {year} {2018})},\ \Eprint {http://arxiv.org/abs/1807.04525}
  {arXiv:1807.04525 [nucl-th]} \BibitemShut {NoStop}%
\bibitem [{\citenamefont {Zhao}\ \emph {et~al.}(2002)\citenamefont {Zhao},
  \citenamefont {Al-Khalili},\ and\ \citenamefont {Bennhold}}]{Zhao:2002wf}%
  \BibitemOpen
  \bibfield  {author} {\bibinfo {author} {\bibfnamefont {Qiang}\ \bibnamefont
  {Zhao}}, \bibinfo {author} {\bibfnamefont {J.~S.}\ \bibnamefont
  {Al-Khalili}}, \ and\ \bibinfo {author} {\bibfnamefont {C.}~\bibnamefont
  {Bennhold}},\ }\bibfield  {title} {\enquote {\bibinfo {title} {{Contributions
  of vector meson photoproduction to GDH sum rule}},}\ }\href {\doibase
  10.1103/PhysRevC.65.032201} {\bibfield  {journal} {\bibinfo  {journal} {Phys.
  Rev. C}\ }\textbf {\bibinfo {volume} {65}},\ \bibinfo {pages} {032201}
  (\bibinfo {year} {2002})},\ \Eprint {http://arxiv.org/abs/nucl-th/0201002}
  {arXiv:nucl-th/0201002} \BibitemShut {NoStop}%
\bibitem [{\citenamefont {Ahrens}\ \emph {et~al.}(2003)\citenamefont {Ahrens}
  \emph {et~al.}}]{GDH:2003fjc}%
  \BibitemOpen
  \bibfield  {author} {\bibinfo {author} {\bibfnamefont {J}~\bibnamefont
  {Ahrens}} \emph {et~al.} (\bibinfo {collaboration} {GDH, A2}),\ }\bibfield
  {title} {\enquote {\bibinfo {title} {{Helicity dependence of the gamma(pol.)
  p(pol.) --\ensuremath{>} n pi+ pi0 reaction in the second resonance
  region}},}\ }\href {\doibase 10.1016/S0370-2693(02)03008-3} {\bibfield
  {journal} {\bibinfo  {journal} {Phys. Lett. B}\ }\textbf {\bibinfo {volume}
  {551}},\ \bibinfo {pages} {49--55} (\bibinfo {year} {2003})}\BibitemShut
  {NoStop}%
\bibitem [{\citenamefont {Hirata}\ \emph {et~al.}(2003)\citenamefont {Hirata},
  \citenamefont {Katagiri},\ and\ \citenamefont {Takaki}}]{Hirata:2002tp}%
  \BibitemOpen
  \bibfield  {author} {\bibinfo {author} {\bibfnamefont {Michihiro}\
  \bibnamefont {Hirata}}, \bibinfo {author} {\bibfnamefont {Nobuhiko}\
  \bibnamefont {Katagiri}}, \ and\ \bibinfo {author} {\bibfnamefont {Takashi}\
  \bibnamefont {Takaki}},\ }\bibfield  {title} {\enquote {\bibinfo {title} {{Pi
  N N coupling and two pion photoproduction on the nucleon}},}\ }\href
  {\doibase 10.1103/PhysRevC.67.034601} {\bibfield  {journal} {\bibinfo
  {journal} {Phys. Rev. C}\ }\textbf {\bibinfo {volume} {67}},\ \bibinfo
  {pages} {034601} (\bibinfo {year} {2003})},\ \Eprint
  {http://arxiv.org/abs/nucl-th/0210063} {arXiv:nucl-th/0210063} \BibitemShut
  {NoStop}%
\end{thebibliography}%

\end{document}